\documentclass[12pt]{article}
\pdfoutput=1
\usepackage{commands}
\usepackage{putex} 
\usepackage[all,cmtip]{xy}
\usepackage{fancybox}
\usepackage{subcaption}
\usepackage{graphicx}
\usepackage{cite}
\usepackage{physics}
\usepackage{mciteplus}
\usepackage{tensor}
\usepackage{skak}
\usepackage{slashed}
\usepackage{esint}
\usepackage{amsmath}
\usepackage{comment}
\usepackage{multirow}
\usepackage{calc}
\usepackage{footmisc}
\usepackage{mathtools,amssymb}
\usepackage{tikz}
\usepackage{pgfplots}
\pgfplotsset{compat=1.18}
\usetikzlibrary{decorations.markings,decorations.pathmorphing}
\usetikzlibrary{arrows.meta}
\DeclareFontFamily{OT1}{pzc}{}
\DeclareFontShape{OT1}{pzc}{m}{it}{<-> s * [1.10] pzcmi7t}{}
\DeclareMathAlphabet{\mathpzc}{OT1}{pzc}{m}{it}

\def\({\left(}
\def\){\right)}

\usepackage{tikz-cd}
\usepackage[compat=1.0.0]{tikz-feynman}
\makeatletter
\DeclareFontFamily{OMX}{MnSymbolE}{}
\DeclareSymbolFont{MnLargeSymbols}{OMX}{MnSymbolE}{m}{n}
\SetSymbolFont{MnLargeSymbols}{bold}{OMX}{MnSymbolE}{b}{n}
\DeclareFontShape{OMX}{MnSymbolE}{m}{n}{
    <-6>  MnSymbolE5
   <6-7>  MnSymbolE6
   <7-8>  MnSymbolE7
   <8-9>  MnSymbolE8
   <9-10> MnSymbolE9
  <10-12> MnSymbolE10
  <12->   MnSymbolE12
}{}
\DeclareFontShape{OMX}{MnSymbolE}{b}{n}{
    <-6>  MnSymbolE-Bold5
   <6-7>  MnSymbolE-Bold6
   <7-8>  MnSymbolE-Bold7
   <8-9>  MnSymbolE-Bold8
   <9-10> MnSymbolE-Bold9
  <10-12> MnSymbolE-Bold10
  <12->   MnSymbolE-Bold12
}{}

\let\llangle\@undefined
\let\rrangle\@undefined
\DeclareMathDelimiter{\llangle}{\mathopen}%
                     {MnLargeSymbols}{'164}{MnLargeSymbols}{'164}
\DeclareMathDelimiter{\rrangle}{\mathclose}%
                     {MnLargeSymbols}{'171}{MnLargeSymbols}{'171}
\makeatother

\usepackage{latexsym}
\usepackage{stmaryrd}

\usepackage{cite}
\usepackage{framed}
\definecolor{shadecolor}{rgb}{0.95,0.95,0.97}
\definecolor{refkey}{rgb}{0.5,0.5,0}
\definecolor{labelkey}{rgb}{0.5,0.5,0}
\definecolor{citekey}{rgb}{0.5,0.5,0}
\definecolor{darkgreen}{rgb}{0,0.5,0}
\definecolor{darkblue}{cmyk}{0.9,0.9,0,0}
\definecolor{darkred}{rgb}{0.6,0,0.3}
\colorlet{mydarkblue}{blue!50!black}
\colorlet{myred}{red!65!black}
\usepackage{ifpdf}
\ifpdf
\usepackage{graphicx} 
\usepackage{epsfig}
\usepackage[setpagesize=false,pagebackref=false, linktocpage, bookmarksopen=true, colorlinks=true, linkcolor=blue,citecolor=blue,urlcolor=blue]{hyperref}

\else
\usepackage{graphicx}

\fi

\renewcommand{\Im}{{\rm Im}}
\renewcommand{\Re}{{\rm Re}}

\def\eqref#1{(\ref{#1})}
\def\comma{\,,}
\def\period{\,.}
\def\figref#1{Figure \ref{#1}}

\def\XXint#1#2#3{{\setbox0=\hbox{$#1{#2#3}{\int}$}
		\vcenter{\hbox{$#2#3$}}\kern-.5\wd0}}

\newcommand{\beq}{\begin{equation}}
\newcommand{\eeq}{\end{equation}}

\def\nullify#1{}

\makeatletter
\def\section{\@startsection {section}{1}{\z@}{-3.5ex plus -1ex minus 
		-.2ex}{2.3ex plus .2ex}{\large\bf}}
\makeatother
\makeatletter
\def\subsection{\@startsection {subsection}{1}{\z@}{-3.5ex plus -1ex minus 
		-.2ex}{2.3ex plus .2ex}{\normalsize\bf}}
\makeatother

\begin{document}

	\preprint{DESY-24-010\\
CERN-TH-2024-014}
	
	\institution{DESY}{Deutsches Elektronen-Synchrotron DESY, Notkestr. 85, 22607 Hamburg, Germany}
 	\institution{CERN}{Department of Theoretical Physics, CERN, 1211 Meyrin, Switzerland}
	
	\title{
	2d QCD and Integrability\\
       \LARGE	 Part II: Generalized QCD 
 	}
	\authors{Federico Ambrosino\worksat{\DESY} and Shota Komatsu\worksat{\CERN} }
	
		\abstract
	{We extend the study of integrable structures and analyticity of the spectrum in large $N_c$ QCD$_2$ to a broad class of theories called the generalized QCD, which are given by the  Lagrangian $\mathcal{L}\propto {\rm tr}\,B\wedge F- {\rm tr}\,V(B)$ coupled to quarks in the fundamental representation. We recast the Bethe-Salpeter equation for the meson spectrum into a TQ-Baxter equation and determine a transfer matrix in a closed form for any given polynomial $V(B)$. Using an associated Fredholm equation, we numerically study the analytic structures of the spectrum  as a function of the coefficients of $V(B)$. We determine the region of couplings where the theory admits a positive and discrete spectrum of mesons. Furthermore, we uncover a multi-sheeted structure with infinitely many multi-critical points, where several mesons become simultaneously massless. Lastly, we illustrate that this structure persists in the large-representation limit of the generalized QCD with the SU(2) gauge group.
}
	\date{\today}

	\maketitle
	
	\tableofcontents

\section{Introduction}
Understanding strong-coupling dynamics of gauge theory is one of the most important challenges in modern physics with diverse theoretical and phenomenological implications. In particular, essential mechanisms of how the Yang-Mills theory exhibits color confinement through the formation of chromoelectric fluxtubes and how the dynamics of the fluxtubes controls physical observables, such as the meson and glueball spectrum, still remain elusive.

To make progress on this front, in this paper we continue our exploration of toy models that are amenable to analytical methods but exhibit qualitative features similar to more realistic, yet complicated examples of four-dimensional gauge theories. Specifically, we study a class of theories in $1+1$ dimensions called {\it generalized QCD} in the large $N_c$ limit. These are generalizations of the two-dimensional QCD (also known as the 't Hooft model) and are described by the {\it generalized Yang-Mills}, with Lagrangian given by $\mathcal{L}\propto {\rm tr}\, B\wedge F -{\rm tr} \, V(B)$, coupled to quarks in the fundamental representation of the gauge group.

In our previous paper \cite{Ambrosino:2023dik}, building on the results of \cite{Fateev_2009}, we demonstrated that the integral equation that determines the spectrum of mesons in the  't Hooft model can be reformulated into a finite difference equation that takes the form of the TQ-Baxter equations --- the equations that determine the spectrum of integrable systems. This reformulation also enables asymptotic expansions of the mass spectrum and the wavefunctions.

As with the 't Hooft model, for a class of potentials $V(B)$ the generalized QCD at large $N_c$ has stable mesonic bound states with a discrete spectrum.
Thanks to the simplification that occurs at large $N_c$, the mass spectrum can be computed explicitly by solving integral equations that generalize the one for the 't Hooft model. In this paper, we show that these generalized equations can also be recast into the TQ-Baxter equation, which turns out to be remarkably simple and universal:
\beq
Q(\nu+2i)+Q(\nu-2i)-2 Q(\nu)=T(\nu)Q(\nu)\comma
\eeq
where the dependence on the potential $V(B)$ is contained entirely in the transfer matrix $T(\nu)$, which can be computed in a closed form for any given potential. See \eqref{eq:TQBaxter} for details. As in the 't Hooft model, this reformulation potentially enables systematic asymptotic expansions of the mass spectrum and the wavefunctions although we leave its detailed study for the future work.

Furthermore, the reformulation in terms of the TQ-Baxter equation and an associated inhomogeneous Fredholm equation allows us to explore detailed analytic structures of the meson spectrum as a function of parameters of the theory. A similar study for the 't Hooft model revealed an intricate structure of the spectrum in the complex quark-mass plane \cite{Ambrosino:2023dik,Fateev_2009}: it has a multi-sheeted structure with infinitely many branch points, each of which signals a tachyonic instability of one of the mesons. The first of such branch points corresponds to a massless quark point where the chiral symmetry becomes exact. The IR phase of that point is described by a CFT with the $U(1)_N$ current algebra \cite{Delmastro:2021otj,Delmastro:2022prj}. This suggests the correspondence between the branch points in the complex mass plane and actual {\it critical points} of the theory. In this paper, we extend these results to the generalized QCD$_2$. A novelty in the present case is that we can also study the dependence on the coefficients of the potential $V(B)$. We found that the theory develops instability for a certain range of parameters and the onset of the instability is signaled by ``critical points'' at which one of the mesons becomes massless (and beyond which it becomes tachyonic). The connection to the inhomogeneous Fredholm equation provides a simple and systematic criterion for the stability, bypassing the case-by-case numerical analysis performed in \cite{Douglas_1994}. In addition, we found that, when the coefficients of the potential are appropriately tuned, several mesons can become simultaneously massless, realizing ``multi-critical points".  As in the 't Hooft model, we conjecture that these ``critical points" correspond to actual critical points of the theory. Unlike the standard QCD$_2$, the IR phases of the generalized QCD$_2$ have not been sufficiently explored in the literature and our results strongly motivate such analysis.

These findings extend and strengthen a surprising connection between QCD$_2$ and integrable systems, initially found in \cite{Fateev_2009} and further explored in \cite{Ambrosino:2023dik,Litvinov:2024riz}. For theories with fundamental quarks studied in this paper, one might argue that this is simply a technical improvement since it is basically a rewriting of the already-known integral equation (although there are numerous advantages in doing so as we emphasized above). However such a  rewriting could be truly beneficial for theories with quarks in the adjoint representation of the gauge group (often referred to as {\it adjoint QCD}$_2$). There have been renewed interest in these theories since they could serve as toy models of confining fluxtubes with nontrivial dynamics \cite{Dalley:1992yy,Kutasov:1993gq,Boorstein:1993nd,Bhanot:1993xp,Demeterfi:1993rs,Smilga:1994hc,Lenz:1994du,Katz:2013qua,Katz:2014uoa,Cherman:2019hbq,Komargodski:2020mxz,Dempsey:2021xpf,Popov:2022vud,Dempsey_2023,Dempsey:2023fvm,Dempsey:2024ofo}. Furthermore, there is the possibility that the confining fluxtube becomes integrable at a certain point in the parameter space \cite{Dubovsky:2018dlk,Donahue:2019adv,Donahue:2019fgn,Donahue:2022jxu,Asrat:2022aov}. However, unlike theories with fundamental quarks, there is no simple integral equation that determines the spectrum of these theories. One possible strategy to overcome this difficulty may be to {\it assume} integrability and try to write down a consistent TQ-Baxter equation directly, based on basic properties of the theory such as the symmetry and also on educated guesses. Such a strategy worked quite well for $\mathcal{N}=4$ supersymmetric Yang-Mills theory in four dimensions \cite{Beisert:2010jr} and it may not be so far-fetched to hope for the same for the adjoint QCD$_2$. To achieve this, it would be important to understand better the connection between TQ-equations and physics of QCD-like theories in two dimensions. This is one of the main reasons for studying generalized QCD$_2$ in this paper. Also, the deformation of the Yang-Mills Lagrangian by a potential $V(B)$ may be important for finding an aforementioned integrable point of the adjoint QCD$_2$.

We also perform a similar analysis for the {\it large representation limit} of generalized Yang-Mills with the SU(2) gauge group coupled to quarks in the spin $J$ representation of SU(2). This is the limit introduced recently for the 't Hooft model in \cite{Kaushal:2023ezo}, in which $J$ is sent to infinity keeping $\lambda_J \equiv g_{\rm YM}^2 J$ fixed. A similar double-scaling limit was studied extensively in the context of large charge limits of conformal field theory (see e.g~\cite{Bourget:2018obm,Arias-Tamargo:2019xld,Arias-Tamargo:2019kfr,Badel:2019oxl,Watanabe:2019pdh,Giombi:2020enj,Caetano:2023zwe} and references therein for the large charge double scaling limits and \cite{Cuomo:2022xgw,Aharony:2022ntz,Aharony:2023amq,Rodriguez-Gomez:2022xwm,Beccaria:2022bcr,Rodriguez-Gomez:2022gif,CarrenoBolla:2023sos} for the large representation limits of line defects) since it simplifies the dynamics of interacting conformal field theories and makes it accessible through analytic methods while still maintaining important qualitative features of the original theory. As in those examples, it was found in \cite{Kaushal:2023ezo} that the dynamics of the 't Hooft model simplifies in the large representation limit and the meson spectrum can be determined by {\it exactly the same} integral equation as in the large $N_c$ limit. We extend this result to the generalized QCD$_2$. Our result also provides evidence that the structures found in the large $N_c$ limit (the connection to integrability and the analytic structure in the complex quark-mass plane) may persist even at finite $N_c$.

{\bf The outline of the paper} is the following: In {\bf Section} \ref{sec:thoofteq}, we review the generalized QCD$_2$ and the derivation of the integral equation determining the mass spectrum of the mesons. In {\bf Section} \ref{sec:Baxter}, we rewrite the integral equation into the TQ-Baxter equation. In {\bf Section} \ref{analiticalpr} we analyze the analytical structure of the spectrum of mesons as a function for generic $V(B)$ and complex quark masses, and 
we study in detail the example of a cubic potential $V(B)$ that demonstrates all the relevant features of the generic case. In {\bf Section} \ref{sec:largespin}, we generalize the analysis to the large representation limit of the generalized QCD$_2$ with gauge group $SU(2)$. Finally we discuss future directions in {\bf Section} \ref{sec:conclusion}. 

\paragraph{\small Note Added:} The first version of this preprint contained a subsection on the asymptotic expansion of the spectrum generalizing \cite{Fateev_2009}. However, although the result reproduces the numerical data rather well, the argument employed there relied on the use of conjectural identities (generalizing (2.17) of \cite{Fateev_2009}), whose validity has not been fully tested. Thus, upon further consideration, we decided to remove the subsection, and mention only the resulting formula as a curious observation leaving the further scrutiny for future works. We thank the anonymous referee of our first paper \cite{Ambrosino:2023dik}, whose questions led to this reconsideration.

\section{Generalized QCD in the large N limit}
\label{sec:thoofteq}
Generalized QCD$_2$ is described by the Lagrangian\cite{Douglas_1994}: 
\be    \label{lagrangian}
\cL =\frac{N}{8\pi} {\rm tr} \, B \wedge F -  {\rm tr} \,  \frac{N}{4\pi} g^2 \, V(B) + {\rm tr} \, \overline{\Psi} \(i \gamma^\mu D_\mu -m\)\Psi \comma
\ee
where both the field strength $F = \dd{A}$ and the $B$-field are in the adjoint representation of $SU(N)$, for simplicity we consider a single flavor of quarks $\Psi$ in the fundamental representation\footnote{The generalization to multi-flavors is analogous to the one for the  't Hooft model \cite{THOOFT1974461}. }. Classical power counting implies that \textit{any} potential for the scalar field $B$ is allowed in two dimensions. In this work, we take the potential for the $B$-field to be given by the formal power series\footnote{This is not the most generic form of a self-interaction term: also power of traces would be allowed. }:
\be \label{potential}
V(B)= \sum_{n=2}^{\infty} v_n \,  {\rm tr} \, B^n\period
\ee
Integrating out the field $B$ gives: $\dd{A}= 2 g^2\, V'(B)$, therefore, by fixing $V(B) = \frac{1}{8\pi} B^2$, the usual Yang-Mills Lagrangian is retrieved. While one could expect that the theory could admit a discrete and positive spectrum of mesons when the potential $V(B)$ is bounded from below, one of the results of this paper is that this is not a sufficient condition. This is due due to instabilities produced by the odd powers in \eqref{potential}. Instead, the more conservative choice $v_{2n+1}=0, v_{2n} >0$, guarantees positivity and discreteness of the spectrum. At the same time, one of the surprising facts we illustrate in this paper is that also for unbounded-from below potentials, there exist regions of the parameter space where the theory admits a positive spectrum in the large $N$ limit.  

\paragraph{Lightcone gauge.}
As for the standard 't Hooft model \cite{THOOFT1974461}, upon going to lightcone coordinates\footnote{The conventions on lightcone are the same of \S 2.1 of \cite{Ambrosino:2023dik}. } $x_\pm = \frac{1}{\sqrt{2}} (x_0 \pm x_1)$, and fixing lightcone gauge $A_-=0, A_+ = A$, the lagrangian presentation \eqref{lagrangian} simplifies  and reduces to:
\be\label{lagrangianlightcone}
\cL =\frac{N}{8\pi} {\rm tr} \, B \partial_- A -  {\rm tr} \,  \frac{N}{4\pi} g^2 \, \sum_{n=2}^{\infty} v_n {\rm tr} B^n  + {\rm tr} \, \overline{\Psi}_a (i \gamma_\mu \partial^\mu\, - m_a -  \gamma_- A) \Psi_a \period 
\ee
Here, we can regard all the $ v_n {\rm tr }\, B^n$ terms as interaction vertices, so the only Feynman rules are:
\be 
\begin{tikzpicture} [baseline={(current bounding box.center)}]
\feynmandiagram [horizontal=a to b] {
 a -- [fermion] b
};
\end{tikzpicture}  = \frac{p_-}{2p_+p_- - m^2 + i \epsilon}\comma \qquad \begin{tikzpicture}[baseline={(current bounding box.center)}]
\begin{feynman}
\vertex   (a1)     ;
\vertex[below=1 of a1] (a2);
\vertex[below right =0.5 and 1 of a1] (a3);
\vertex[right =0.6 of a3] (a4);
\vertex[right =0.6 of a4] (a5);
\diagram* {
(a1)--[fermion] (a3)-- [fermion] (a2),
(a3) --[photon] (a4),
(a4)-- [scalar] (a5),
};
\end{feynman}
\end{tikzpicture} =  \frac{-8\pi}{k_-}\comma \qquad  \begin{tikzpicture}[baseline={(current bounding box.center)}]
\begin{feynman}
\vertex[very thick,dot] (a0) {};
\vertex[right = 0.5 of a0] (at);
\vertex[above=1 of a0] (au)      ;
\vertex[below=1 of a0] (ad);
\vertex[below = -0.51 of at] (ar) {$\vdots$};
\vertex[left = 1 of a0] (al);
\vertex[above right =0.7 and 0.7 of a0] (aur);
\vertex[above left =0.7 and 0.7 of a0] (aul);
\vertex[below left =0.7 and 0.7 of a0] (adl);
\vertex[below right =0.7 and 0.7 of a0] (adr);
\diagram* {
(a0)--[scalar] (au),
(a0)--[scalar] (ad),
(a0)--[scalar] (al),
(a0)--[scalar,opacity=0] (ar),
(a0)--[scalar] (aur),
(a0)--[scalar] (aul),
(a0)--[scalar] (adr),
(a0)--[scalar] (adl),
};
\end{feynman}
\end{tikzpicture} =  \frac{i n v_n}{4\pi} \period
\ee
Henceforth in an arbitrary Feynman diagram, any propagator from a $B^n$ vertex must be connected with a quark line. In the large $N$ limit, the only Feynman diagrams that survive are the planar ones; an example of a planar contribution to the quark self energy is reported in \figref{qseplanar}. For the case of the 't Hooft model \cite{THOOFT1974461} this revealed to be a crucial simplification and allowed us to obtain exact expression for the fermion propagators and for the equation determining the mesons' spectrum. Remarkably, as illustrated in \cite{Douglas_1994} similar results are also available for its generalized version. 
\begin{figure}
\be \begin{split}
\Gamma(p,q) &= \begin{tikzpicture}[baseline={(current bounding box.center)}]
\begin{feynman}
\vertex (a0) {};
\vertex[right= 0.7cm of a0] (a1l);
\vertex[right= 0.7cm of a1l] (a1);
\vertex[right= 0.7cm of a1] (a1r);
\vertex[right= 0.5cm of a1r] (a2);
\vertex[above= 1cm of a1,dot] (a1t) {};
\diagram* {
(a0)--[fermion] (a1l)--[fermion] (a1)--[fermion] (a1r)--[fermion] (a2),
(a1l)--[scalar] (a1t),
(a1)--[scalar] (a1t)-- [scalar] (a1r),
};
\end{feynman}
\end{tikzpicture} + \begin{tikzpicture}[baseline={(current bounding box.center)}]
\begin{feynman}
\vertex (a0);
\vertex[right= 0.8cm of a0] (a2);
\vertex[right= -0.3cm of a2] (a2l);
\vertex[right= 0.3cm of a2] (a2r);
\vertex[right= 0.5cm of a2r] (a3);
\vertex[above= 1cm of a2,dot] (a2t) {};
\diagram* {
(a0)--[fermion]  (a2l)  -- [fermion] (a2r)--[fermion] (a3), 
(a2l)--[scalar] (a2t) -- [scalar] (a2r),
};
\end{feynman}
\end{tikzpicture} + 
\begin{tikzpicture}[baseline={(current bounding box.center)}]
\begin{feynman}
\vertex (a0);
\vertex[right = 1.0 cm of a0] (a3);
\vertex[right= -0.2cm of a3] (a3l);
\vertex[right = -0.4cm of a3l ] (a3ll);
\vertex[right= 0.2cm of a3] (a3r);
\vertex[right = 0.4cm of a3r ] (a3rr);
\vertex[above = 1cm of a3, dot ] (a3t) {};
\vertex[right= 0.5 cm of a3rr] (a4) {} ;
\diagram* {
(a0) --[fermion] (a3ll) --[fermion] (a3l) -- [fermion] (a3r) --[fermion] (a3rr),
(a3ll)--[scalar] (a3t) -- [scalar] (a3l),
(a3rr)--[scalar] (a3t) -- [scalar] (a3r),
(a3rr)-- [fermion] (a4),
};
\end{feynman}
\end{tikzpicture} +\hspace{0.3cm} \cdots\hspace{0.3cm}
= 2 g^2 \sum_{n=2}^{\infty} (-i)^n n v_n \, I_n \comma
\end{split}\ee 
\caption{Planar diagram contributions to the quark self-energy. The \textit{master integral} $I_n$ is reported in \eqref{masterintegral}}
\label{qseplanar}\end{figure}

\subsection{Bound state equation}
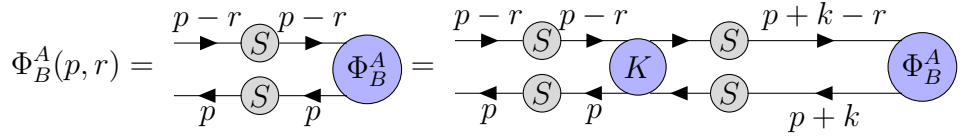
\begin{figure}
    \centering
\begin{align*}
\Phi^A_{B  }(p,r) = \begin{tikzpicture}[baseline={([yshift=-.5ex]current bounding box.center)}]
\begin{feynman}[every blob={/tikz/fill=blue!30,/tikz/inner sep=2pt}]
\vertex[blob] (m) {$\Phi^A_B$};
\vertex[above left=0.35cm  and 0.27cm of m] (m1);
\vertex[below left = 0.35cm and 0.27cm of m] (m2);
\vertex[ left= 0.9cm of m1] (a);
\vertex[small,draw,circle,inner sep=0pt,minimum size=0.5cm,fill=gray!30, left = 0.9cm of m1] {$S$};
\vertex[ left= 0.5cm of a] (a0);
\vertex[ left= 0.9cm of a0] (a01);
\vertex[ left= 0.9cm of m2] (b);
\vertex[small,draw,circle,inner sep=0pt,minimum size=0.5cm,fill=gray!30, left = 0.9cm of m2] {$S$};
\vertex[ left= 0.5cm of b] (b0);
\vertex[ left= 0.9cm of b0] (b01);
\diagram* [edges=fermion]{
(a)--[edge label= $p-r$](m1),
(a01)--[edge label= $p-r$](a0),
(m2)--[edge label= $p$](b),
(b0)--[edge label= $p$](b01),
};
\end{feynman}
\end{tikzpicture}  = \begin{tikzpicture}[baseline={([yshift=-.5ex]current bounding box.center)}]
\begin{feynman}[every blob={/tikz/fill=blue!30,/tikz/inner sep=2pt}]
\vertex (a0);
\vertex[above =0.35cm of a0] (a1);
\vertex[below = 0.35cm of a0](b1);
\vertex[right = 0.9cm of a1] (a11);
\vertex[small,draw,circle,inner sep=0pt,minimum size=0.5cm, fill=gray!30,right = 0.9cm of a1] {$S$};
\vertex[right = 0.5cm of a11] (a12);
\vertex[right = 0.9cm of a12] (a2);
\vertex[right = 0.9cm of b1] (b11);
\vertex[small,draw,circle,inner sep=0pt,minimum size=0.5cm, fill=gray!30,right = 0.9cm of b1] {$S$};
\vertex[right = 0.5cm of b11] (b12);
\vertex[right = 0.9cm of b12] (b2);
\vertex[right = 0.29cm of a2] (a3);
\vertex[right = 0.29cm of b2] (b3);
\vertex[right = 0.8cm of a3] (a4);
\vertex[right = 0.8cm of b3] (b4);
\vertex[small,draw,circle,inner sep=0pt,minimum size=0.5cm, fill=gray!30,right = 0.8cm of a3] {$S$};
\vertex[small,draw,circle,inner sep=0pt,minimum size=0.5cm,fill=gray!30, right = 0.8cm of b3] {$S$};
\vertex[right = 0.5 cm of a4] (a5);
\vertex[right = 0.5 cm of b4] (b5);
\vertex[blob, right = 5.75 cm of a0] (m) {$\Phi^A_B$};
\vertex[blob, right = 2.05cm of a0] (m2) {$K$};
\vertex[above left=0.35cm  and 0.27cm of m] (a6);
\vertex[below left = 0.35cm and 0.27cm of m] (b6);
\diagram* [edges=fermion]{
(a1)--[edge label= $p-r$](a11),
(a12)--[edge label= $p-r$](a2),
(a3)--(a4),
(b2)--[edge label= $p$](b12),
(b11)--[edge label= $p$](b1),
(b4)--(b3),
(a5)--[edge label= $p+k-r$](a6),
(b6)--[edge label= $p+k$](b5),
};
\end{feynman} \end{tikzpicture}
\end{align*}   
\caption{Diagrammatic representation of Bethe-Salpeter equation \eqref{bseq}. }
    \label{fig:bs}
\end{figure}

The analog of 't Hooft equation for generalized QCD is obtained, as in the 't Hooft model, upon computing the 2PI kernel:
\be \begin{split}
K(p,q,p',q') &= \begin{tikzpicture}[baseline={(current bounding box.center)}]
\begin{feynman}
\vertex (a0);
\vertex[above = 2cm of a0] (b0); ;
\vertex[right= 1.3cm of a0] (a1);
\vertex[above = 1cm of a1, dot] (a1t) {};
\vertex[above= 2cm of a1] (b1);
\vertex[right= -0.3cm of a1] (a1l);
\vertex[right= 0.3cm of a1] (a1r);
\vertex[right= 1cm of a1r] (a2);
\vertex[above= 2cm of a2] (b2);
\diagram* {
(a0)--[fermion, edge label' = { $p$ }] (a1l) -- [fermion] (a1r) --[fermion, edge label' = { $p'$ }](a2) ,
(b0) --[fermion, edge label = {$p-q$}] (b1),
(a1l)--[scalar] (a1t) -- [scalar](a1r),
(a1t)--[scalar] (b1)--[fermion, edge label = {$p'-q'$}] (b2),
};
\end{feynman}
\end{tikzpicture} + 
\begin{tikzpicture}[baseline={(current bounding box.center)}]
\begin{feynman}
\vertex (a0);
\vertex[above = 2cm of a0] (b0);
\vertex[right= 1.2cm of b0] (b2);
\vertex[left = 0.7cm of b2] (b2l);
\vertex[right= 0.7cm of b2] (b2r);
\vertex[below= 2cm of b2] (a2);
\vertex[above= 1cm of a2,dot] (b2t) {};
\vertex[left = 0.7cm of a2] (a2l);
\vertex[right= 0.7cm of a2] (a2r);
\vertex[right = 0.5cm of a2r] (a3);
\vertex[above = 2cm of a3] (b3);
\diagram* {
(b0)--[fermion] (b2l)--[fermion] (b2)--[fermion](b2r)--[fermion](b3),
(b2l)--[scalar](b2t)--[scalar](b2),
(b2t)--[scalar](b2r),
(a0)--[fermion] (a2l)--[fermion] (a2)--[fermion](a2r)--[fermion](a3),
(a2l)--[scalar](b2t)--[scalar](a2),
(b2t)--[scalar](a2r),
};
\end{feynman}
\end{tikzpicture} + 
\begin{tikzpicture}[baseline={(current bounding box.center)}]
\begin{feynman}
\vertex (a0);
\vertex[above = 2cm of a0] (b0);
\vertex[right = 1.1 cm of a0] (a3);
\vertex[right= -0.2cm of a3] (a3l);
\vertex[right = -0.4cm of a3l ] (a3ll);
\vertex[right= 0.2cm of a3] (a3r);
\vertex[right = 0.4cm of a3r ] (a3rr);
\vertex[above = 1cm of a3, dot ] (a3t) {};
\vertex[above = 2cm of a3 ] (b3);
\vertex[right = 0.25cm of b3 ] (b3r);
\vertex[left = 0.25cm of b3 ] (b3l);
\vertex[right =2.2 cm of a0] (a4) ;
\vertex[above = 2 cm of a4] (b4) ;
\diagram* {
(a0) --[fermion] (a3ll) --[fermion] (a3l) -- [fermion] (a3r) --[fermion] (a3rr)--[fermion](a4),
(a3ll)--[scalar] (a3t) -- [scalar] (a3l),
(a3rr)--[scalar] (a3t) -- [scalar] (a3r),
(b3l)--[scalar](a3t)--[scalar](b3r),
(b0)--[fermion] (b3l)-- [fermion](b3r)--[fermion] (b4),
};
\end{feynman}
\end{tikzpicture} + \hspace{0.7cm} \cdots 
\\
&= 4g^2 \sum_{n=2}^{\infty} (-i)^n n v_n \sum_{l=1}^n I_l(q,q')I_{n-l}(p'-q',p-q) \comma\end{split}
\ee

that is expressed in terms of the following {\it master integrals} $I_n$:\be \label{masterintegral}
I_n (p,q)= \int\dd[n-1]{k} \frac{1}{p_- - k_1} \, \sgn {k_1} \,  \frac{1}{k_1 - k_2} \, \sgn {k_2} \cdots \frac{1}{k_{n-2} - k_{n-1}}  \,\sgn{k_{n-1}}\, \frac{1}{k_{n-1} - q_-}  \comma
\ee
whose computation is technically challenging because of their IR divergences. We refer to \cite{Douglas_1994} for a thorough discussion. 
Following the same steps of \cite{THOOFT1974461} (also reviewed in \cite{Ambrosino:2023dik}), from the homogeneous part of the  4-point Dyson equation we obtain the {\it Bethe-Salpeter} equation:
\be \label{bseq1}
\Phi(p,r) = S(p)\, S(p-r) \int\dd[2]{k} K(k,p,r)\,\Phi(p+k,r)  \comma
\ee
schematically illustrated in \figref{fig:bs}, that in this model takes the form:
\be \label{bseq}
\left[ 2p_+ - m^2 \( \frac{1}{p_- } + \frac{1}{p_--q_- } \) \right]\phi(p,p-q) = \(\Gamma(q) + \Gamma(p-q) \)\phi(q) + \int_0^{q_-} \dd{k_-} K(k,p;q,q) \phi(k)
\ee
In this model, \eqref{bseq} can be expressed in the following contour-integral form \cite{Douglas_1994}:
\be
    \begin{split}
     \bigg[   2\pi^2 \lambda - (\alpha + 1)\, & \( \frac{1}{x} + \frac{1}{1-x} \) \bigg] \phi(x) =\\ 
     =-2\pi \oint_{\gamma_\infty} \frac{\dd{z}}{2\pi i} V'(z) \Bigg[& iz \frac{z^2 + \pi^2}{(z^2 - \pi^2)^2}\, \frac{2}{\pi} \arctan(\ts \frac{\pi}{iz}) \(\frac{1}{x} + \frac{1}{1-x}\) \phi(x) \, + \\
     &+ 2 \frac{z^2}{(z^2 - \pi^2)^2} \fint_0^1\dd{y} \frac{1}{(x-y)^2} \exp(\frac{2}{\pi}\arctan(\ts\frac{\pi}{iz})\, \log(\ts\frac{x(1-y)}{y(1-x)} )) \phi(y)
     \Bigg]\comma
    \end{split}
\ee
where: $x=  p_-/q_-$ is the momentum fraction, $\alpha = \pi m^2/g^2 -1 $, and $\phi(x)$ is the reduced meson wavefunction: $\phi(x) = \int\dd{p_+} \Phi(p,r)$ $(\Phi(p,r) = \bra{\Omega} \overline{\psi}\psi \ket{\Phi})$. 
The contour integral is along a curve $\gamma_\infty$ enclosing all the singularities but the one at $z = \infty$. This, as 't Hooft equation, is an eigenproblem in $x\in [0,1]$ for the mesons wavefunctions $\phi(x)$ and masses $\mu^2$.  
\paragraph{Explicit form. }
We found closed-form expressions for the result of the contour integrations. The first term in the square bracket can be readily evaluated:
\be\begin{split}
\label{alphandef}
\alpha_n &= 2\pi \oint \frac{\dd{z}}{2\pi i}\, i n \,z^{n}  \frac{z^2 + \pi^2}{(z^2 - \pi^2)^2}\, \frac{2}{\pi} \arctan(\ts\frac{\pi}{iz}) =\\  &=\begin{cases}
    0 &\quad n \, \text{odd}\\
    2\pi^{n-1}\, n^2 \(-1 + {\rm H}(\frac{n-1}{2}) + \log( 4) \)  &\quad n \, \text{even}
\end{cases} \comma
\end{split}
\ee
where $ \rm{H}(n)$ are the Harmonic numbers\footnote{The combination $\({\rm H}(\frac{n-1}{2}) + \log(4) \)$ is a rational number for $n \in 2\bZ $ .  }. 

Denoting $m = n-2$ and employing the variable:
\be
   \Lambda :=  \log(\ts \frac{x(1-y)}{y(1-x)})\comma
\ee
the second term is expressed as\footnote{In practice, rather than using the explicit expression we exhibit in the text, it is more convenient to directly evaluate the integral below by taking the residue at infinity for any specific $n$. }:
\be
\begin{split}
\label{pndef}
    &p_n(\Lambda) =  2\pi  \oint \frac{\dd{z}}{2\pi i}\,  \frac{ n\,z^{n+1}}{(z^2 - \pi^2)^2} \exp(\Lambda\,\frac{2}{\pi}\arctan(\frac{\pi}{iz}) ) = \\
    &=   4\pi n \sum_{q = 0}^{\lfloor\frac{m}{2}\rfloor}\sum^{\lceil\frac{m}{2}\rceil -1 -q}_{p=0}\frac{(-2 i)^{m-2p-2q}(q+1)\pi^{2q+2p}}{(m-2q-2p)!} \Lambda^{m-2q-2p} \kappa(p)\comma \\
    &\kappa(p) = \sum_{k_0,\cdots, k_p = 0}^{2p} \binom{2p}{k_0, \,\cdots, k_p} \prod^p_{t=2}\(\frac{1}{2t+1}\)^{k_t}\, \delta\({\textstyle\sum^p_{l=0} k_l -2p}\)\, \delta\({\textstyle\sum^p_{l=1} l\, k_l - p}\) \period
    \end{split}
\ee
From this analytical expression\footnote{This was also observed empirically in \cite{Douglas_1994}.}, it is evident that for any $n$ even (odd)  $p_n(\Lambda)$ is a polynomial with real (imaginary) coefficients of degree $n-2$ in $\Lambda$ containing only even (odd) powers thereof. The imaginary coefficients of the odd-terms guarantee simultaneously the hermitianity of the equation and the following parity of the eigenfunctions $\phi_n(x) = (-1)^n\phi_n(1-x)$. 

Using the formulae above, the generalized 't Hooft equation is:
\be
\label{thoofteq}
    \mu^2 \phi(x) =  \(\alpha + 1 - \sum^\infty_{n=1} v_{2n}\, \alpha_{2n}\) \(\frac{1}{x} + \frac{1}{1-x}\) \phi(x) - \fint_0^1\dd{y}\,\frac{\sum_{n=2}^{\infty} v_n\, p_n(\Lambda)}{(x-y)^2} \, \phi(y) \period
\ee
We report here the first few terms as computed from the expressions above:
\be\label{firstterms}\begin{split}
\sum^\infty_{n=1} v_{2n}\, \alpha_{2n} &= 8 \pi  v_2 +\frac{160}{3}  \pi^3 v_4 + \frac{744}{5}  \pi^5 v_6 + \frac{31616}{105} \pi^7 v_8 + \frac{32440}{63}  \pi^9 v_{10} + \cdots \\
   \sum_{n=2}^{\infty} v_n \, p_n(\Lambda) &= 8\pi v_2  -24 i\pi \, v_3\, \Lambda + 32\pi\,v_4\,\(-\Lambda^2 + \pi^2\) - \frac{40 i \pi}{3}v_5\, \(-2\Lambda^3 + 7 \pi^2\Lambda \) + \cdots
\end{split}\ee
As expected, by setting $v_n = \frac{1}{8\pi}\, \delta_{2,n}$ we recover the familiar 't Hooft equation of QCD$_2$:
\be
\label{thoofteqym}
    \mu^2 \phi(x) =  \alpha \(\frac{1}{x} + \frac{1}{1-x}\) \phi(x) - \fint_0^1\dd{y}\,\frac{1}{(x-y)^2} \, \phi(y) \period
\ee
A direct comparison between \eqref{thoofteq} and \eqref{thoofteqym} illustrates that the self-interaction potential for the $B$-field at the level of the mesons effectively {\it renormalizes} the mass of the quarks $\alpha$, to:
 \be\label{tildealphadef} \alpha  \to \tilde{\alpha} =  \alpha + 1 - \sum v_{2n} \alpha_{2n}
 \comma \ee 
and modifies the kernel (that is an  effective inter-quark potential) by adding logarithmic corrections. 
\section{Generalized 't Hooft equation as a TQ-system}\label{sec:Baxter}

\subsection[$\nu$-space representation]{\boldmath $\nu$-space representation}
Following the approach initiated for the 't Hooft model \cite{Fateev_2009}, we perform a Fourier transform with respect to the rapidity variables $\theta = \frac{1}{2}\log\frac{x}{1-x}$:  
\be
    \phi(y) = \infint \frac{\dd{\nu}}{2\pi } \(\frac{x}{1-x}\)^{-i \nu/2} \Psi(\nu)\comma \qquad  \Psi(\nu) = \int_{0}^{1} \frac{\dd{x}}{2 x(1-x) } \(\frac{x}{1-x}\)^{i \nu/2} \phi(x)\period
\ee
In $\nu$-space, the generalized 't Hooft equation \eqref{thoofteq} takes the form ($2\pi^2 \lambda = \mu^2$):
\be
\label{eq:eignuspace}
   \lambda \infint \dd{\nu'} \frac{\pi (\nu - \nu')\,\Psi(\nu') }{2 \sinh(\frac{\pi}{2}(\nu - \nu') )} = \underbrace{\left[ \frac{2\tilde{\alpha}}{\pi} + \sum_{n=2}^{\infty} v_n\,  p_n (\mathcal{I})   \right]}_{=: f(\nu)}
\Psi (\nu) \comma
\ee
where the polynomial $p_n\(\cI \)$ is obtained by replacing in $p_n(\Lambda)$ of \eqref{pndef} powers of $\Lambda^n$ with $\cI_{n+2}$:
\be\label{integrals}
   \cI_{n+2} =  \(\frac{\(\frac{x}{1-x}\)^{\frac{i \nu }{2}}}{\frac{2}{\pi}(x-1) x}\)^{-1} \fint_0^1\dd{y} \frac{\(\frac{y}{1-y}\)^{\frac{i \nu }{2}} \(\log \(\frac{y}{1-y}\)-\log \(\frac{x}{1-x}\)\)^{n}}{(x-y)^2}\period
\ee
E.g.\ the first few terms are (cf.\ \eqref{firstterms}):
\be
   \sum_{n=2}^{\infty} v_n \, p_n(\cI) = 8\pi v_2\,\cI_2  -24\pi i\, v_3\, \cI_3 + 32\pi\,v_4\,\(-\cI_4 + \pi^2 \cI_2\) - \frac{40 i \pi}{3}v_5\, \(-2\cI_5 + 7 \pi^2\cI_3 \) + \cdots \period 
\ee
Computing the principal parts integrals in \eqref{integrals} is made straightforward by the following observation\footnote{If one neglects the principal part integration, this follows by iteration from the definition of $\cI_n$:
\be 
\dv{\nu}\cI_{n+2} = \frac{i}{2}\(\frac{\(\frac{x}{1-x}\)^{\frac{i \nu }{2}}}{\frac{2}{\pi}(x-1) x}\)^{-1} \fint_0^1\dd{y} \(\log \(\ts\frac{y}{(1-y)}\) - \log\( \ts\frac{x}{(1-x)}\)\) \frac{\(\frac{y}{1-y}\)^{\frac{i \nu }{2}} \log^n \(\ts\frac{y(1-x)}{x(1-y)}\)}{(x-y)^2} = \frac{i}{2} \cI_{n+3}\period
\ee A more thorough proof is provided in Appendix \ref{app:rec} where we compute them as Fourier transform of distributions. }:
\be\label{recintegral}
    \cI_{n+2} = \(\frac{2}{i}\)^n \dv[n]{\nu} \cI_2 \comma \qquad \cI_2 = \nu \coth(\frac{\pi \nu}{2})\comma 
\ee as $\cI_2$ coincides with the integral computed in eq.\ (3.3) of \cite{Ambrosino:2023dik} for the standard QCD$_2$.  We list a few of them to give a flavor of the resulting expressions:\small
\be
\begin{split}
\cI_3 = \frac{i   (\pi  \nu -\sinh (\pi  \nu ))}{\sinh ^2\(\frac{\pi  \nu }{2}\)} \comma  \cI_4 = 2 \pi   \frac{2-\pi  \nu  \coth \(\frac{\pi  \nu }{2}\)} {\sinh^2\(\frac{\pi  \nu }{2}\)} \comma \cI_5 = 2 i \pi ^2 \frac{3 \sinh (\pi  \nu ) - \pi  \nu  (\cosh (\pi  \nu )+2)}{\sinh^4\(\frac{\pi \nu }{2}\) }
\end{split}
\ee
\normalsize
Henceforth, we can straightforwardly compute the function $f(\nu)$ given any potential $V(B)$ expressed via a polynomial form\footnote{In $\nu$-space the hermitianity of \eqref{eq:eignuspace} for $v_i \in \bR$ is evident from the fact that in the products $v_{2n+1}\, \mathcal{I}_{2m+1}$ are always real. }:
\be \label{fnu}
f(\nu) = \alpha +1 + \sum_{n=1}^{\infty} v_{2n} \alpha_{2n} +  \sum_{n=2}^{\infty} v_n p_n\(\cI\) \period
\ee As an example, taken  $V(B)$ to be a quartic polynomial, the resulting expression for $f(\nu)$ is:
\be \begin{split}
f(\nu) = \frac{2}{\pi} \(\alpha+1 -8\pi v_2 - \frac{160\pi^3}{3} v_4 \) + \frac{8\pi}{\tanh\(\frac{\pi\nu}{2} \)} \(  v_2\nu+4 \pi^2 v_4\nu -6 v_3\)& \\+ \frac{8\pi^2}{\sinh^2\(\frac{\pi\nu}{2}\)} \(3 v_3 \nu +8 \pi  v_4 \nu  \coth\(\frac{\pi\nu}{2}\)-16v_4\)&
\end{split}\ee
\subsection{Discreteness of the spectrum and instabilities}
\label{discreteness}
Although for any potential there is an associated bound state equation \eqref{eq:eignuspace}, arbitrary choices of $V(B)$ will not always lead to positive mass spectrum for the mesons\footnote{
This is a generalization of the condition $\alpha > -1$ in 't Hooft model that ensures a stable positive spectrum of mesons. In QCD$_2$ this condition corresponds to imposing reality of the quark masses: $ \pi \frac{m^2}{g^2} - 1 = \alpha$.  }. This was firstly empirically observed for the case of a cubic potential in \cite{Douglas_1994}, where,  upon numerical direct solution of \eqref{thoofteq}, it was found that for high values of $v_3$ the model exhibits tachyonic bound states. 
 Below we use the reformulation of the eigenproblem \eqref{thoofteq} in $\nu$-space in order to analyze the instablities of the generalized 't Hooft model and illustrates a rich and interesting analytical structure. 

Exactly as discussed for the 't Hooft model \cite{Fateev_2009}, the generalized 't Hooft equation in $\nu$-space takes the form of a homogeneous Fredholm equation of the second kind:
\be \label{fred}
   \phi(\nu) =\lambda \infint\dd{\nu'} \kappa(\nu,\nu')\,\phi(\nu'), \qquad  \kappa(\nu,\nu') := \frac{\pi (\nu - \nu')}{2 \sinh(\frac{\pi}{2}(\nu - \nu') )} \frac{1}{\sqrt{f(\nu)\, f(\nu')}}\; .
\ee
where $f(\nu)$ is defined in \eqref{eq:eignuspace}. The finiteness of the $L^2(\bR^2)$ of $\kappa(\nu,\nu')$, guaranteed by $f(\nu\in \bR)\neq 0$, ensures that this self-adjoint kernel defines a Hilbert-Schmidt integral operator with discrete and non-degenerate spectrum. Since from \eqref{eq:eignuspace} follows that $\Psi(\nu)$ has poles at the location of zeros of $f(\nu)$, when the latter has zeros on the real axis, the eigenfunctions are not normalizable and do not correspond to physical bound states.  

Henceforth, upon studying the zeros of $f(\nu)$ in the complex plane of the couplings, one determines the regions where the generalized QCD produces a physical confining theory. We defer to section \ref{analiticalpr} a detailed discussion of the rich analytical structure of generalized QCD as function of the (complexified) 
couplings.  

We conclude this subsection by remarking that the region of coupling space where the meson spectrum is positive and real does not correspond to imposing that $V(B)$ is bounded from below as one could expect. Even potentials that are unbounded from below, such as polynomials with odd leading powers, surprisingly admit physical regions in the coupling space where the spectrum is real and discrete. A prototypical example of this feature, is the one of a cubic $V(B)$, analyzed thoroughly in section \ref{cubic}.  Moreover, even for potentials that are bounded from below, for large values of any of the odd couplings $v_{2n+1}$, the spectrum develops instabilities, and becomes tachyonic at large values thereof. Indeed, already at the classical level, a potential with odd couplings does not have their minimum at $B \equiv 0$ and tachyonic modes are ultimately just a manifestation of the instability of the vacuum at the origin. Naturally, this classical analysis receives corrections from quantum effects: from the discussion in section \ref{analiticalpr}, one can verify that even for a potential of the form $V(B) = v_2 B^2 + v_3 B^3 + v_4 B^4$, even though $B=0$ is never the true minimum for $v_3 <0$, there is a critical value $\overline{v_3}$ (depending on $v_2$ and $v_4$) such that for any $\abs{v_3} < \overline{v_3}$ the model has a real and discrete spectrum of mesons. 
\paragraph{Normalization and discrete symmetries.}
Having identified where the model admits a discrete meson spectrum, in this paragraph we restrict to such coupling range. The normalization of the eigenfunction in $\nu$-space is fixed by (cf.\ eq.\ 3.11 of \cite{Ambrosino:2023dik}):
\be
    \lambda_n\,\delta_{nm} =  \lambda_n \int_0^1\dd{x} \phi_n(x)\,\phi^*_m(x) =\int_{-\infty}^{+\infty} \frac{\dd{\mu}}{2\pi} \frac{f(\mu)}{2\pi}\, \Psi_n(\mu)\, \Psi_m^*(\mu) \period
\ee
Convergence of this norm imposes that eigenfunctions decay at least polynomially fast on the real axis.
Furthermore, since the function $f(\nu)$ enjoys the discrete symmetry: \be f(\nu) = \Check{f}(-\nu)  \equiv f(-\nu) \eval_{v_{2n+1}\to - v_{2n+1}} \comma \ee
the solutions $\Psi_n$ will be also eigenfunction thereof:\be \label{parity}
\Psi(\nu) = (-1)^n \, \Check{\Psi}(-\nu)\comma \qquad \Check{\Psi}(\nu) \equiv \Psi(\nu)\eval_{v_{2n+1} \to - v_{2n+1}} \period
\ee
This discrete symmetry is a direct generalization of the parity symmetry $f(\nu)\to f(-\nu)$ of 't Hooft model, and we consequently still refer to that as \textit{parity}. 
\subsection{TQ-equation}
In the following, we choose the generalized YM potential $V(B)$ such that $f(\nu \in \bR) \neq 0$. According to the discussion of the previous subsection \ref{discreteness}, this ensures that the model admits a physical discrete spectrum of mesons with real masses and normalizable eigenfunction. 
In particular, the right hand side $f(\nu)\Psi(\nu)$ of \eqref{eq:eignuspace} does not have any singularities in the strip\footnote{The discussion of this subsection follows closely the approach of section 3.2 of \cite{Ambrosino:2023dik}, and we refer the reader to that paper for details concerning all the derivations. }  $\abs{\Im \, \nu} < 2$ where we can analytically continue both sides of \eqref{eq:eignuspace}. Then, following \cite{Ambrosino:2023dik, Fateev_2009}, we introduce the {\it $Q$-functions}:
\be\label{Qdef}
    Q(\nu) = \sinh(\frac{\pi\nu}{2}) \left[ \frac{2\tilde{\alpha}}{\pi} + \sum_{n=2}^{\infty} v_n\,  p_n (\mathcal{I})  \right]\, \Psi(\nu) = \sinh(\frac{\pi\nu}{2}) \, f(\nu)\, \Psi(\nu) \comma 
\ee
that are defined to be regular in the strip, to decay less-than exponentially at $\abs{\nu}\to \infty$. Moreover, these $Q$-functions satisfy the related integral equation:
\be
\label{inteqq}
       Q(\nu) = \lambda\sinh(\frac\pi 2 \nu) \int_{-\infty}^{+\infty} \dd{\nu'} \frac{\pi (\nu - \nu') }{2 \sinh(\frac{\pi}{2}(\nu - \nu') )} \frac{Q(\nu')}{\sinh(\frac \pi 2 \nu')f(\nu')} \comma
\ee
together with the \textit{quantization} conditions \be \label{quantizations}Q(0) = Q(\pm 2i) = 0\period \ee This is completely analogous to the case of 't Hooft model \cite{Fateev_2009}; in particular, one can show that $\Psi(\nu)$ is a solution of \eqref{eq:eignuspace} decaying polynomially at infinity if and only if the associated $Q(\nu)$ (through \eqref{Qdef}), analytical in $\abs{\Im \, \nu} <2 $, solves the following TQ-Baxter equation:
\be\label{eq:TQBaxter}
\boxed{        Q(\nu + 2i) + Q(\nu - 2i) -2 Q(\nu) = \frac{-4\pi \lambda}{ f(\nu) } \, Q(\nu) =\frac{-4\pi \lambda\, Q(\nu) }{ \frac{2\tilde{\alpha}}{\pi} + \sum\limits_{n=2}^{\infty} v_n\,  p_n (\mathcal{I}) } } \comma
\ee
subject to the quantization conditions \eqref{quantizations}, and decaying less-than-exponentially at infinity in the whole strip. 

Thus, for any given generalized Yang-Mills potential, the only modification to  the TQ-Baxter equation is contained inside the single function $f(\nu)$, computed in closed form (through equations \eqref{fnu}, \eqref{pndef},\eqref{alphandef}) for given any polynomial $V(B)$.
\paragraph{Asymptotic decay.}
The asymptotic decay of the eigenfunctions can be further refined upon studying the large-$\nu$ limit of \eqref{eq:TQBaxter}. Assuming the asymptotic behavior $Q(\nu) \asymp  e^{\pi k \nu} q(\nu)$, with  $q(\nu)$ periodic by $2i$,  and bounded by any exponential, \eqref{eq:TQBaxter} is:
\be\label{eq:asymptq}
      0 =  Q(\nu + 2i) + Q(\nu - 2i) -2 Q(\nu) + \frac{4\pi \lambda}{ 8\pi \nu \sum\limits_{n=1}^{\infty}n^2 v_{2n} } \, Q(\nu) \asymp -4 e^{\pi  k \nu } \sin ^2(\pi  k) q(\nu ) \comma
\ee
from which one deduces the stronger conditions of exponential decay of $\Psi(\nu)$ at $\abs {\Re \nu} \to \infty$ in the strip $(-2i,2i)$, that refines the polynomial decay guaranteed by square integrability\footnote{This result could have been imported directly from \cite{Ambrosino:2023dik} given that in eq. (3.32) the r.h.s.\ does not depend on $\lambda$. }.

\subsection{Inhomogenous extension}
The TQ-system \eqref{eq:TQBaxter} admits solutions with the given decay at infinity and quantization only at the discrete spectrum $\lambda = \lambda_n$ of the integral operator \eqref{fred}. The TQ-Baxter equation is a simultaneous problem for $Q$ and $\lambda$ and,  as such, it is difficult to develop analytic approaches to the spectrum. However, in \cite{Fateev_2009} a notable conceptual and practical simplification specific to this problem was obtained upon considering an auxiliary \textit{inhomogeneous} problem, obtained by modifying \eqref{eq:eignuspace} and \eqref{quantizations}, so that \eqref{eq:TQBaxter} admits solutions for {\it any} $\lambda\in \bR$. This turned out to be extremely useful as it allowed one to study the TQ-equation in a two-step procedure: we first solve the simpler inhomogeneous problem, and then use this solutions to obtain the discrete spectrum of the homogeneous problem. This ultimately led to a fastly-convergent asymptotic analytic expansion for the spectral data \cite{Ambrosino:2023dik,Fateev_2009,Litvinov:2024riz}]. We refer the readers to those papers for a more thorough discussion. 

The extension to the inhomogeneous problem is possible even in the generalized QCDs. We do so by:
\begin{align}
\label{eq:extfred}
        \frac {F(\nu)}{\sqrt{f(\nu)}} &=    \phi(\nu|\lambda) - \lambda \int_{-\infty}^{\infty}\dd{\nu'} \kappa(\nu,\nu') \phi(\nu'|\lambda),\\
\label{inhomogeneousterm}
    F_+(\nu) &= \frac{\nu}{\sinh(\frac \pi 2 \nu)}, \qquad F_-(\nu) = \frac{ 1}{\sinh(\frac \pi 2 \nu)}\period
\end{align}
The inhomogeneous terms \eqref{inhomogeneousterm} are the same as in \cite{Ambrosino:2023dik,Fateev_2009} and have definite parity under the discrete symmetry  \eqref{parity}. Each of those drives a unique solution of the inhomogeneous integral equation depending on the continuous parameter $\lambda$:
\be
\label{eq:exteig}
   F_\pm(\nu) = f(\nu)\,\Psi_\pm(\nu|\lambda)  - \lambda \fint_{-\infty}^{+\infty} \dd{\nu'}\frac{\pi (\nu - \nu') }{2 \sinh(\frac{\pi}{2}(\nu - \nu') )}\,\Psi_\pm(\nu'|\lambda) \comma
\ee
whose solutions inherits definite parity:
\be 
\Psi_\pm (\pm \nu) = \pm \Check{\Psi}_\pm (\nu) \period
\ee
The \textit{inhomogeneous} Q-function defined by \eqref{Qdef}:
\be\label{inhqdef}
Q_\pm(\nu|\lambda)= \sinh(\frac{\pi\nu}{2})\,f(\nu) \, \Psi_\pm(\nu|\lambda) = \mp Q_\pm(-\nu|\lambda)
\ee
are also shown to satisfy the same TQ-equation of the homogeneous problem \eqref{eq:TQBaxter}:
\be\label{tqinhom}
Q_\pm(\nu+2i|\lambda) + Q_\pm(\nu-2i|\lambda)- 2Q_\pm(\nu|\lambda) = \frac{-4\pi\lambda }{f(\nu) } Q_\pm(\nu|\lambda)\comma
\ee 
but at any values of $\lambda \in \bR$, upon assuming the same  asymptotic, but subject to modified quantization conditions\footnote{Note that this changes the analytical properties of the solutions $\Psi(\nu|\lambda)$ compared to $\Psi_n(\nu)$, allowing for poles at the boundary of the strip.}:
\be\label{quantinho}
Q_+(\pm 2i) = \pm 2i  \comma \qquad Q_{+}(0)=0\comma\qquad  Q_-(\pm 2i) =Q_{-}(0)=  1   \period
\ee
The Wronskian constructed with the two solutions $Q_\pm(\nu|\lambda)$ results to be a constant:
\be\label{eq:defwron}
W(\nu|\lambda) = Q_+(\nu + i|\lambda)Q_-(\nu - i|\lambda) - Q_+(\nu - i|\lambda)Q_-(\nu + i|\lambda) = W(i|\lambda) = 2i \period
\ee

\subsection{Comments on asymptotic expansions}
Having reformulated the bound state problem into the TQ equation, one can try to generalize the approaches in \cite{Fateev_2009} to obtain the analytic asymptotic expansion of the spectrum and wave functions.

\begin{figure}[t]
    \centering
    \includegraphics{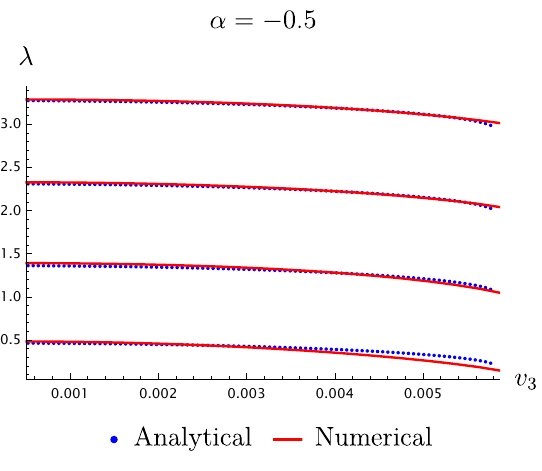}
    \caption{Comparison between the numerical eigenvalues and the analytical expression \eqref{specanalytical}, at $\alpha = -0.5$. For this value of the mass, the expected critical point is $v_3^{(\rm II)} \simeq -0.0058$.}
    \label{fig:numspectrum-0.5}
\end{figure}

For the cubic potential given by
\be
    V(B) = \frac{1}{8\pi} B^2 + v_3 B^3\comma \qquad f(\nu,\alpha,v_3) = \frac{2\alpha}{ \pi}  +  \nu  \coth \(\frac{\pi  \nu }{2}\)+ 24 \pi v_3 \frac{\pi \nu - \sinh(\pi\nu)}{\sinh^2\(\frac{\pi\nu}{2}\)}\period
\ee
we followed the argument in \cite{Fateev_2009} to arrive at the following conjectural formula for the spectrum:
\be
\begin{split}
\label{specanalytical}
   & n = 2 \lambda_n  - \frac{2 \alpha  \log (2\lambda_n)}{\pi ^2} + \frac{\alpha^2}{\pi ^4 \lambda_n } - \frac{3}{4} - \frac{2 \alpha  \log (4\pi e^{\gamma_E})}{\pi ^2} +\frac{1}{\pi^3}\,\cI(\alpha,v_3)\\&+\frac{1}{2\pi^6\lambda^2_n}\(\alpha^3 + (-1)^n\pi^2 (\alpha+1) + (-1)^n v_3^2 \left(C_1 + \frac{288 \log ^2(\lambda_n)}{\pi ^2}+ C_2\log (\lambda_n ) \right) \) + \mathcal{O}(\lambda^{-3}_n)\comma
\end{split}
\end{equation}
where the constants $C_1$ and $C_2$, and the integral $\mathcal{I}(\alpha,v_3)$ are given in appendix \ref{app:spectraldet}. However, the derivation relies on generalizations of the identity (2.17) in \cite{Fateev_2009}, whose validity has not bee fully tested. Nonetheless, as shown in Figure \ref{fig:numspectrum-0.5}, \eqref{specanalytical} approximates the spectrum rather well (even for low-lying states). That said, sans other evidence, we only mention it as a curious observation, leaving further scrutiny for the future work.

\begin{figure}
\begin{subfigure}{0.3\linewidth}
            \centering
    \includegraphics[width = \textwidth]{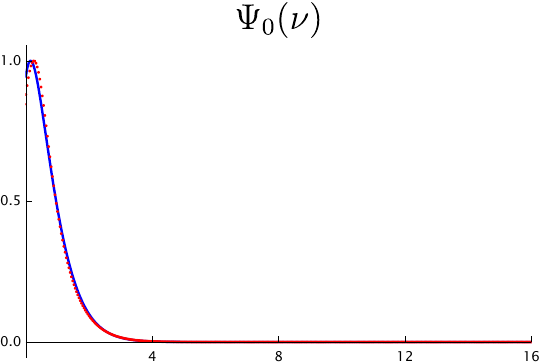}
\end{subfigure}\hfill
\begin{subfigure}{0.3\linewidth}
            \centering
    \includegraphics[width = \textwidth]{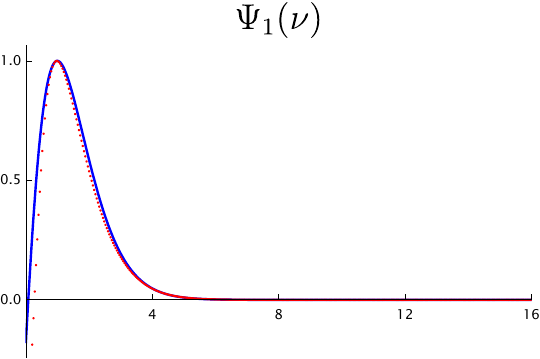}
\end{subfigure}\hfill
\begin{subfigure}{0.3\linewidth}
            \centering
    \includegraphics[width = \textwidth]{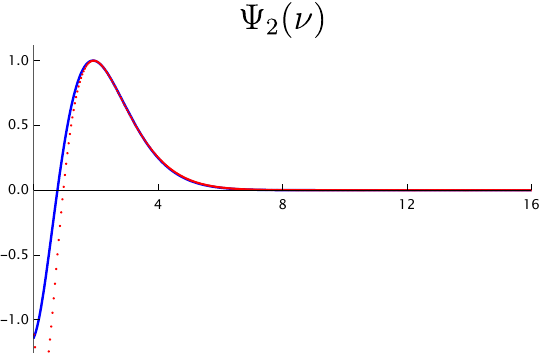}
\end{subfigure}
\caption{Comparison between the asymptotic expression for the wavefunction and the numerical solutions for  $\alpha = 0.5$, $v_3 = 0.0026$.}
\label{generanavsnum}
\end{figure}

One can also write down the approximation for the wave function by constructing the solution to the TQ-equation directly as follows:
\small
\be 
\begin{split}
&\Psi_n(\nu) = \frac{2 \sinh \(\frac{\pi  \nu }{2}\)  }{48 \pi^2 v_3+\sinh (\pi  \nu )} \, \exp(\frac{i \pi  \lambda  (\cosh (\pi  \nu )-1)}{48 \pi ^2v_3+\sinh (\pi  \nu )}) \\ 
& \times {\rm M}\(1 +i \frac{ 2 \(\cosh (\pi  \nu )-1) \alpha  +48 v_3 \pi ^3 \nu +\pi  (\nu -48 v_3 \pi ) \sinh (\pi  \nu )\)}{2 \pi  \(48 \pi ^2 v_3+\sinh (\pi  \nu )\)}, 2 , -\frac{2 i \pi  \lambda  (\cosh (\pi  \nu )-1)}{48 \pi ^2v_3+\sinh (\pi  \nu )}\)
\end{split}\ee
\normalsize
This approximaton is even cruder than the one performed in \cite{Fateev_2009} for the 't Hooft model since the resulting wave function does not have a required analyticity in the complex plane. However, on the real axis, it approximates well the wave functions computed numerically (see Figure \ref{generanavsnum}).

\section{Analytical structure of the spectrum and critical points}
\label{analiticalpr}
The simple form of 't Hooft equation in $\nu$-variables \eqref{eq:eignuspace} allows for a straightforward qualitative analysis of the main features of the model in the complexified space of couplings. In particular, the analytical structure of the function $f(\nu)$ completely determines the analytical structure of the eignevalues $\lambda_n(\alpha,\bs{v})$ as functions of the complexified couplings. In what follows, in order to emphasize the dependence on the complexified couplings, we denote $f(\nu)$ by $f(\nu,\alpha,\bs{v})$.

As already observed in section \ref{discreteness}, the zeros of $f(\nu,\alpha,\boldsymbol{v})$ coincide with the poles of the eigenfunction $\Psi_n(\nu)$, so that whenever they do not lie on the real line, the model will admit a discrete spectrum of mesons with normalizable eingenfunctions. Among the zeros of $f(\nu,\alpha,\bs{v})$ the higher order zeros, where also the derivatives thereof vanish are special and distinct points. 
Already in the case of the 't Hooft model \cite{Ambrosino:2023dik}, which can be recovered by setting all the $v_k$ to zero,  we found that the location of double zeros of $f(\nu,\alpha)$ in the complex plane on $\nu$ determines infinitely many square-root branch points at some specific values $\alpha_k\in\mathbb{C}$ where of one of the mesons becomes massless. At finite $N_c$ these critical points are likely to become non-unitary interacting CFTs. This interesting structure extends and enriches greatly in the generalized Yang-Mills theory with polynomial potentials.  In this section we describe the general features of the analytical structure, and of the type of criticality that we encounter in this very broad class of models by varying the polynomial potential $V(B)$. The reader can refer to the concrete example of a cubic self-interaction that we discuss thoroughly in section \ref{cubic},  that illustrates all the main features of the generic case. 

\paragraph{$\mathcal{PT}$-symmetry and reality of the spectrum.}
It is important to note that physical choices of $V(B)$ shall not be limited to real values for the coefficients.  Since the $B$-field is odd under $\cP\cT$-symmetry (or equivalently,  charge-conjugation), potentials with imaginary odd couplings, $v_{2n+1}\in i \mathbb{R}$, provide examples of $\cP\cT$-symmetric quantum field theories. It is by now well-established that non-hermitian field theories enjoying unbroken $\cP\cT$-symmetry, can have real spectra; we refer the reader to the classical papers \cite{Bender_1998,Bender_2005,Bender:2004sa} (and references therein) for a thorough discussion on the topic. However, $\mathcal{PT}$-symmetry can also be spontaneously broken, in which case the spectrum contains complex conjugate pairs, and in generalized QCD both cases happen  (recently also observed in the context of non-unitary minimal models in \cite{Lencses:2022ira,Lencses:2023evr,Lencses:2024wib}), see for instance ection \ref{tricrit}.

\subsection{Multicritical points from collision-of-roots}
\begin{figure}
    \centering
    \begin{subfigure}{0.4\textwidth}
    \centering
          \includegraphics[width = \textwidth]{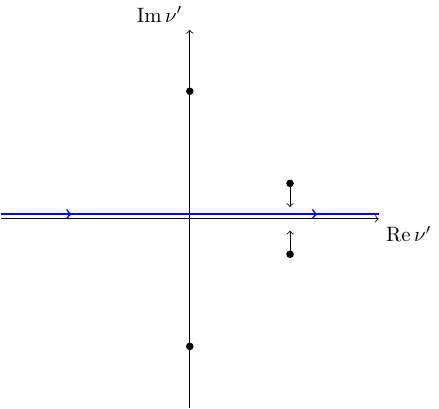}
    \caption{Trapped contour: singular point.}
    \label{pinching}
    \end{subfigure}
    \begin{subfigure}{0.4\textwidth}
    \centering
          \includegraphics[width = \textwidth]{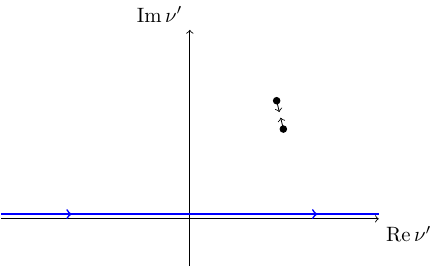}
    \caption{No singularity on this sheet. }  
    \label{notpinch}
    \end{subfigure}
    \caption{Collision-of-roots points points may or may not lead to singularities.}
    \label{fig:collisionoofroots}
\end{figure}

By tuning the parameters of the potential $V(B)$, one can make $f(\nu,\alpha,\bs{v})$ to have zeros of higher order\footnote{For the physical potential, the order of the zeros is bounded from above by $\deg V(B)$. }. As we explain below, a $n$-th order zero corresponds to a $n$-th root-branch point of the eigenvalues, if it is not possible to deform the contour of integration to avoid the pinching singularity. At each of these $n$-th branch-points the eigenvalues behave as $\lambda \sim (\alpha - \alpha^{(n)})^{1/n}$, (and similarly as functions of all the other couplings), and correspond to multicritical points where exactly $n-1$ mesons become simultaneously massless.
\paragraph{Collision-of-roots point and branch points.}
In general, roots of $f(\nu,\alpha,\bs{v})$, do not correspond to branch-point singularities of the spectrum, but only to poles of $\Psi(\nu)$. This is due to the fact that if the colliding roots are away from the integration contour, the integral eigenproblem \eqref{eq:eignuspace} is still well-defined. On the other hand, if two roots collide in such a way that the contour of integration is trapped in between, the spectrum will develop a branch cut\footnote{This is very well known in the context of scattering amplitudes where many singularities of these kind arise, cfr.\ e.g.\ \cite{Eden:1966dnq}.}. Only the collision-of-roots of type  \figref{pinching} would lead to a singular point of $\lambda$, while the one of \figref{notpinch} is not a singularity on this sheet of the complex plane since the contour is not trapped by the collision. However it will be a singularity on another sheet of the complex plane as we will shortly discuss.
\paragraph{Type of singular points.}
 The precise type of singularity associated to this critical point is inferred by studying the monodromy of the zeros of the function $f$ as we meromorphically  continue the parameter $\alpha$ along a closed curve around any of the values of the critical values $\alpha^{(n)}$. If $\nu^{(n)}$ is the position of the $n$-th order zeros at $\alpha = \alpha^{(n)}$, as we move in the parameter space to $\alpha\neq \alpha^{(n)}$, the degeneracy of the roots will be generically split and we will be left with exactly $n$ non degenerates roots  $\nu_1\cdots \nu_n$. Then it is easy to see that upon the analytical continuation of $\alpha$ along a closed curve winding around $\alpha^{(n)}$ a single time, $f(\nu,\alpha,\bs{v})$ encompass a monodromy that permutes the roots by: $\nu_{i}\to \nu_{i+1}$. Because of this non-trivial monodromy, each root of these root may or may not 
 cross the integration contour, if this happens, then the integral eigenproblem will differ by the respective residue terms. 
 
 The simplest case is the one of a square-root cut, as in \figref{fig:collisionoofroots}: in this case after a $2\pi$ analytical continuation  $\alpha\to \widetilde{\alpha}$, the eigenproblem will also undergo a non-trivial monodromy transformation, that act by adding two further residue terms:
 \be 
 \label{boundstatein2sheet}
 \begin{split}
f(\nu) \Psi(\nu) &=  \widetilde\lambda(\alpha) \int_{-\infty}^{+\infty}\dd{\nu'} \frac{\pi (\nu - \nu')}{ 2 \sinh{\frac\pi 2 (\nu-\nu')}} \Psi(\nu')\\
 &+  \widetilde{\lambda}(\widetilde{\alpha}) \text{Res} \left( \frac{\pi (\nu - \nu')}{ 2 \sinh{\frac\pi 2 (\nu-\nu')}} \Psi(\nu';\alpha_0, \bs{v_0}) , \nu' = \nu_1\right)\\
 &-    \widetilde{\lambda}(\widetilde\alpha) \text{Res} \left( \frac{\pi (\nu - \nu')}{ 2 \sinh{\frac\pi 2 (\nu-\nu')}} \Psi(\nu';\alpha_0, \bs{v_0}) , \nu' = \nu_2\right) \period
\end{split} \ee
This indicates $\alpha^{(\rm II)}$ is a square-root branch point of the eigenvalues. This point coincides with the boundary of the allowed region for having a positive spectrum of mesons, and indeed we find a corresponding critical point where one of the meson becomes massless upon direct solution (analytical or numerical) of the spectral problem. The critical point that we found by solving asymptotically the spectrum in   \figref{fig:numspectrum-0.5}, is exactly of this nature. 
\begin{figure}
    \centering
    \begin{subfigure}{0.4\textwidth}
          \includegraphics[width = \textwidth]{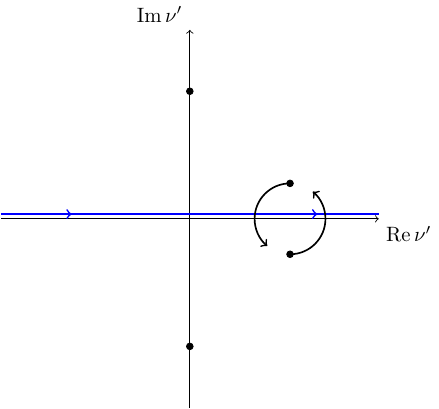}
    \end{subfigure}
    \begin{subfigure}{0.4\textwidth}
          \includegraphics[width = \textwidth]{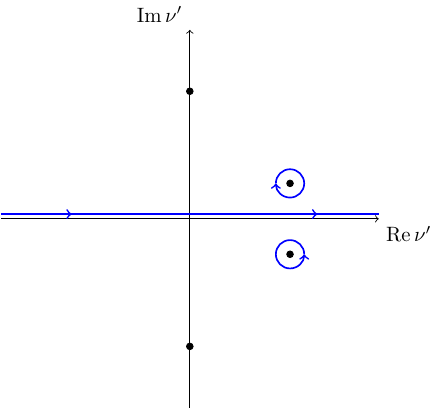}
    \end{subfigure}
    \caption{Complex plane before and after the a $2\pi$ analytical continuation with non-trivial monodromy. }
    \label{fig:monodromy}
\end{figure}
To illustrate a more complicated case, let us examine the case of a zero of order 3. As we thoroughly explain in section \ref{cubic} the typical situation is the one of figure \figref{mono3pt1.pdf}, where the contour cannot be continuously deformed to avoid a pinching-of-roots singularity at some  special values of the couplings $\alpha^{(\rm III)},v^{(\rm III)}_3$. Then, it is easy to see that the monodromy of $\alpha$ under a $6\pi$ monodromy along a curve encircling  the collision-of-root point  $\alpha^{(\rm III)}$ is the one described in \figref{mono3pt2.pdf}, \figref{mono3pt3.pdf} and \figref{mono3pt4.pdf}, i.e.\ a third order zero leads to a cubic-root branch cut of the spectrum and a trictritical point where two of the meson masses turn simultaneously to zero (cfr.\ \figref{tricriticalhermite}).
\begin{figure}
    \centering
    \begin{subfigure}{0.4\textwidth}
                  \includegraphics[width = \textwidth]{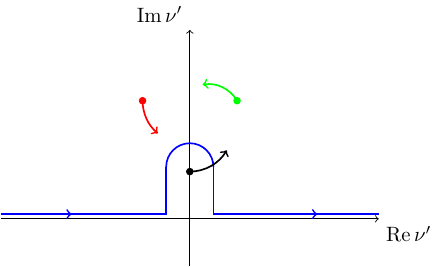}
                  \caption{Roots in the first sheet.}
                  \label{mono3pt1.pdf}
    \end{subfigure}
        \begin{subfigure}{0.4\textwidth}
                  \includegraphics[width = \textwidth]{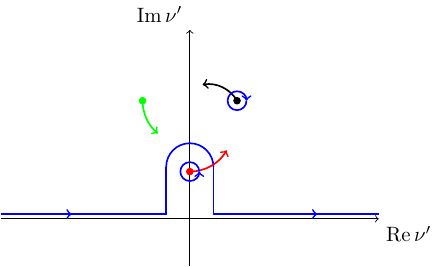}
                 \caption{$2\pi$ monodromy.}
                      \label{mono3pt2.pdf}
    \end{subfigure}\\
    \centering
        \begin{subfigure}{0.4\textwidth}
                  \includegraphics[width = \textwidth]{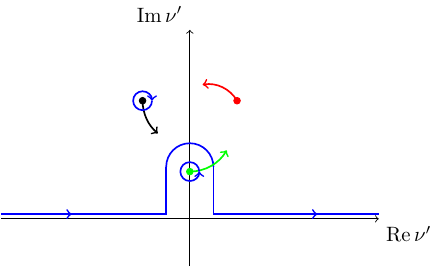}                 \caption{$4\pi$ monodromy.}
                                    \label{mono3pt3.pdf}
    \end{subfigure}
        \begin{subfigure}{0.4\textwidth}
                  \includegraphics[width = \textwidth]{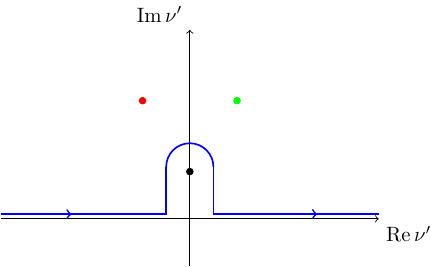}
                  \caption{$6\pi$ monodromy.}
                                   \label{mono3pt4.pdf}

    \end{subfigure}
    \caption{Monodromy of the roots around a 3rd order zero in the complex plane.}
    \label{fig:3rdorderroot}
\end{figure}
It is straightforward from the graphical representation in \figref{fig:3rdorderroot} to write down, on each of the three sheets represented there, the corresponding bound state equation as we have done in \eqref{boundstatein2sheet}. The generalization to arbitrary order of root is completely straightforward.

\paragraph{Geometry of the phase space.}
For any given polynomial $V(B)$ of degree $n$, the singularities of type $p$-th root will form a codimension $n-p$ locus in the phase space of the theory given that they generically solve a system of $p$ equations. In particular, $n$-th-root branch points form a set of infinitely many isolated points in the space of couplings,  $(n-1)$-th-root branch points form set of  lines that may intersect at $n$-th root branch points, and so on. The typical structure of the first sheet of the phase space is the one in \figref{fig:tricriticalpoints}. Each of the several connected components of these loci, lives in a unique sheet of the complex plane. On a given sheet, only the collision-of-roots which pinches the contour leads to singularities, while the others are singular points on other sheets, as we discuss more later. 
These critical loci correspond to the boundary of regions where the spectrum of mesons is positive and discrete. At the special points in the space of couplings, several  mesons become massless (as in \figref{realspectrumreal} and \figref{fig:numericalf3}).

\subsection{Critical points beyond the first sheet}
\begin{figure}
    \centering
    \includegraphics{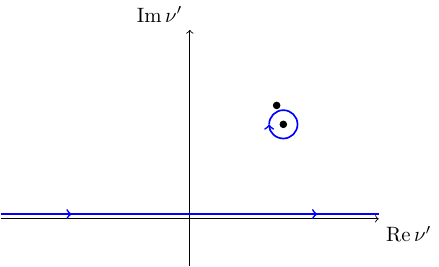}
    \caption{Collision-of-roots away from the contour are singularities in higher-sheet of the complex $\nu'$-plane}
    \label{fig:secondsheet2}
\end{figure}

As anticipated already in the previous subsection, the singularities on the first sheet are only one of the many connected components forming the singular loci of the parameter space. Indeed, while collision-of-roots happening away from the integration contour are not singularity on the first sheet, they are singularities on other sheets of the complex plane: As we analytically continue beyond the cuts on the first sheet, the contour of integration for the integral equation picks up additional contributions from poles that crossed the contour (see \eqref{boundstatein2sheet} and  \figref{fig:3rdorderroot}). After this happens, if these poles collide with other poles on the complex plane, they lead to singularities on the corresponding sheet. See e.g.~\figref{fig:secondsheet2} (and  the corresponding situation on the first sheet \figref{notpinch}). 

By comparing them with the analytical prediction  or with the numerics, one can check that those singularities correspond to critical points where one of the higher mesons (instead of the lowest one) becomes tachyonic and triggers a critical phase transition.

\subsection{A case study: cubic potential}
\label{cubic}
In this section, we analyze the spectrum and phases of generalized QCD with a cubic potential: $V(B) = v_2\,B^2 + v_3\, B^3$ that is never bounded from below except for $v_3 = 0$.
The latter is a purely quadratic potential and is the  simplest example  of a gYM potential $V(B) = v_2\,B^2$. In this case we have
\be
f(\nu,\alpha,v_2) = \frac{2(\alpha +1 - 8\pi v_2) }{\pi }+8 \pi  v_2\,\nu  \coth \(\frac{\pi  \nu }{2}\) = 8\pi  v_2 \(\bar{\alpha} + \nu\coth(\frac{\pi\nu}{2})  \)\period
\ee
Clearly, the problem depends only on the single variable $ \bar{\alpha} = \frac{\alpha -8 \pi  v_2+1}{8 \pi  v_2}$ and the entire structure is equivalent to the usual QCD$_2$ analyzed in \cite{Ambrosino:2023dik,Fateev_2009}, upon the replacement $\bar{\alpha}\to \alpha$: the spectrum has a square root branch cut in $\bar{\alpha} = -1$ for any values of $v_2$, and the energy levels are real and positive  for $v_2>0 (<0)$ for $\bar{\alpha} >-1 (<-1)$ . 

For $v_3\neq 0$, only two of the three couplings $(\alpha,v_2,v_3)$ are physically relevant and we can always canonically normalize $v_2 = \frac{1}{8\pi}$. From \eqref{fnu}, we obtain
\be 
\label{alphaf3n}
f(\nu,\alpha,v_3) = \frac{2\alpha}{ \pi}  +  \nu  \coth \(\frac{\pi  \nu }{2}\)+ 24 \pi v_3 \frac{\pi \nu - \sinh(\pi\nu)}{\sinh^2\(\frac{\pi\nu}{2}\)}\period
\ee
To identify regions in the parameter space where this model exhibits a real and positive spectrum of mesons, below we systematically study zeros of $f(\nu)$ in the complex space of the couplings $(\alpha, v_3)$. 
\subsubsection{Real couplings}
\label{realcouplings}
\begin{figure}[htbp]
  \centering
  \begin{subfigure}{0.3\linewidth}
    \includegraphics[width=\linewidth]{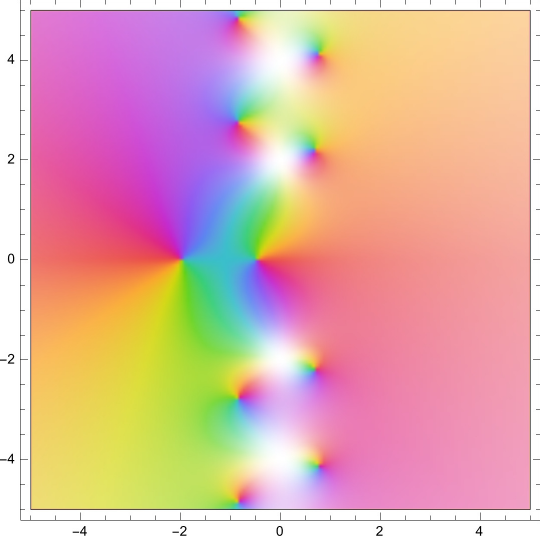}
    \caption{$v_3 = -0.0156856$}
    \label{sub:twozeros1}
  \end{subfigure}
  \hfill
  \begin{subfigure}{0.3\linewidth}
    \includegraphics[width=\linewidth]{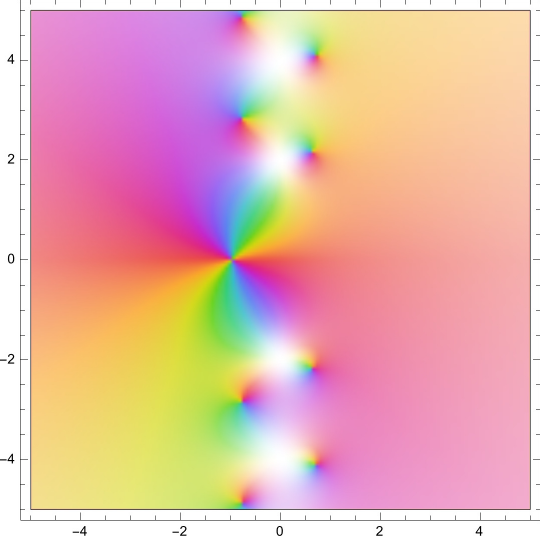}
    \caption{$v_3 =-v_3^{({\rm II})}(0.5)$}
        \label{sub:dege1}
  \end{subfigure}
  \hfill
  \begin{subfigure}{0.3\linewidth}
    \includegraphics[width=\linewidth]{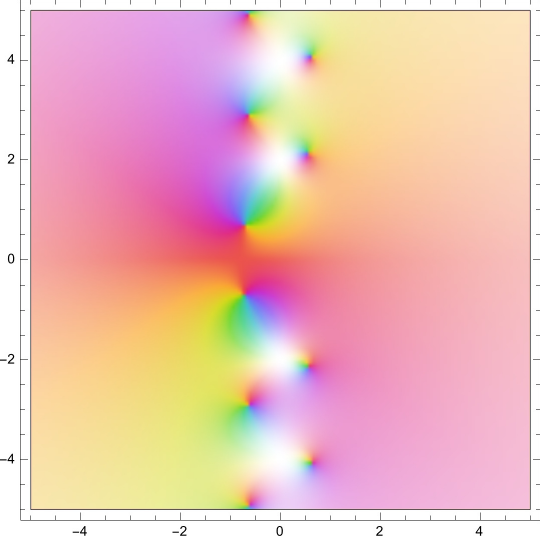}
    \caption{$v_3 =-0.00784281  $  }
  \end{subfigure}
\par\bigskip
  \centering
  \begin{subfigure}{0.3\linewidth}
    \includegraphics[width=\linewidth]{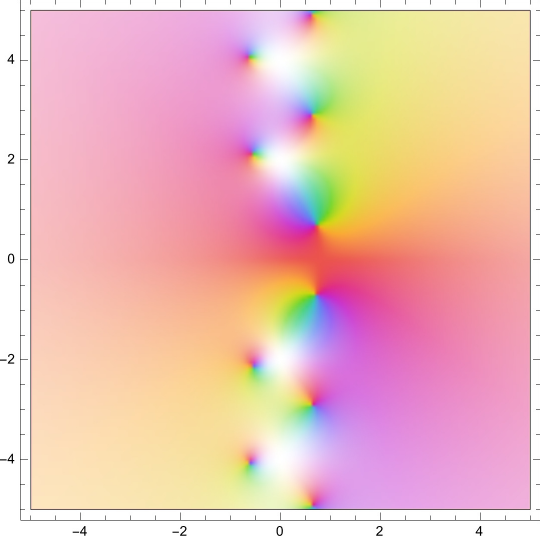}
    \caption{$v_3 =0.00784281 $}
  \end{subfigure}
  \hfill
  \begin{subfigure}{0.3\linewidth}
    \includegraphics[width=\linewidth]{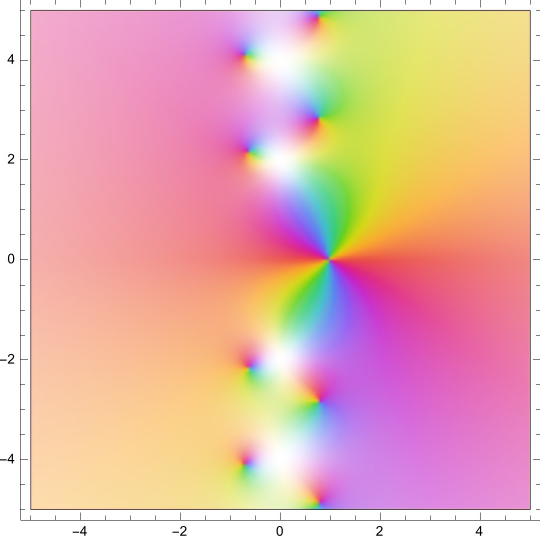}
    \caption{$v_3 = v_3^{({\rm II})}(0.5)$}
   \label{sub:dege2}
  \end{subfigure}
  \hfill
  \begin{subfigure}{0.3\linewidth}
    \includegraphics[width=\linewidth]{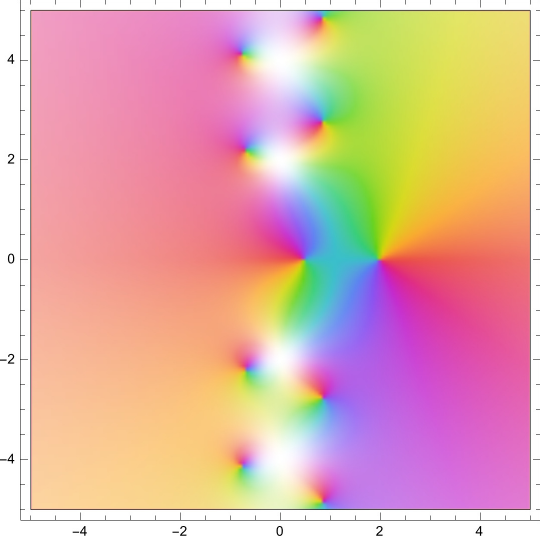}
    \caption{$v_3 = 0.0156856$  }
    \label{subfig:alpha3}
        \label{sub:twozeros2}
  \end{subfigure}
  \caption{Positions of the zeros (black) and poles (white) of $f(\nu,\alpha = 0.5 ,v_3)$ in the  $\nu$-complex plane for different values of $v_3$. For $\abs{v_3} < v_3^{({\rm II})}(0.5) \simeq 0.01176$ (\figref{sub:twozeros1}, \ref{sub:twozeros2} )there are two distinct real zeros that degenerate for $v_3 =  v_3^{({\rm II})}(0.5)$ (\figref{sub:dege1}, \ref{sub:dege2} )  at $\nu^{({\rm II})(0.5)} \simeq 0.974336$. In all the physical region $\abs{v_3}< v_3^{({\rm II})(0.5)}$, there are no real zeros of $f(\nu, 0.5 ,v_3)$. } 
  \label{fig:complexplanef3reals}
\end{figure}

\begin{figure}
     \centering
     \includegraphics{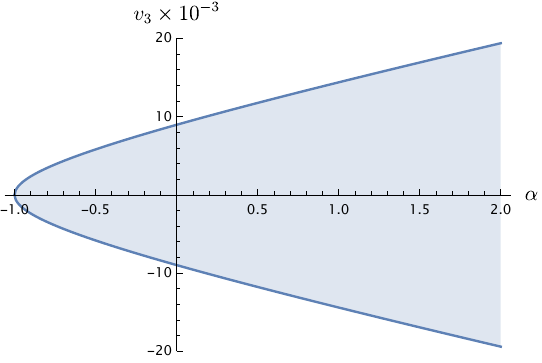}
     \caption{The shaded region correspond to the values of the couplings where the model has a positive spectrum of mesons, while the curve identifies the critical points where the theory has a massless mode. }
     \label{realspectrumreal}
\end{figure}

Let us start by considering real values of the quark masses corresponding to $\alpha \geq -1$. For arbitrary values of $\alpha$ and $v_3\in \bR$, zeros of $f(\nu)$ are all simple. Given any fixed value of $\alpha$, there are two critical values $\pm v_3^{({\rm II})}(\alpha)$,  where the two zeros closest to the real axis degenerate to a double zero 
at $\nu =\pm \nu^{({\rm II})}(\alpha)\in \bR$:
    \be
    \label{doublezeros}f\(\pm \nu^{({\rm II})}(\alpha),\alpha, \pm v_3^{({\rm II})}(\alpha)\) = \partial_\nu f\(\pm \nu^{({\rm II})}(\alpha),\alpha, \pm v_3^{({\rm II})}(\alpha)\) =0 \period 
    \ee
    Solutions to this equation give a curve in \figref{realspectrumreal} in the plane $(\alpha, v_3)$. As shown in \figref{fig:complexplanef3reals}, while  for $\abs{v_3} < v_3^{({\rm II})}(\alpha)$ there are no real zeros of $f(\nu.\alpha,v_3)$, when  $\abs{v_3} > v_3^{({\rm II})}(\alpha)$ exactly two of them lie on the real axis. 
Therefore, the analytical structure of the bound state equation \eqref{eq:eignuspace} implies that the spectrum of mesons is positive in a shaded region of Figure \ref{realspectrumreal} bounded by the curve \footnote{Analytic solutions for the coupled equations \eqref{doublezeros} are not available. However near $\alpha=-1$, one can study the solution $v_3^* = v_3^*(\alpha)$ analytically by linearizing \eqref{doublezeros} in $\nu$ since we know that for small $\alpha\to -1$, both $v^{({\rm II})}_3,\nu^{({\rm  II})}\to 0$. The result of the linear approximation,  $v^*_3(\alpha) \simeq  \pm \frac{\sqrt{\alpha +1}}{8 \sqrt{6} \pi ^2}, \nu^*(\alpha) \simeq \pm\frac{\sqrt{6} \sqrt{\alpha +1}}{\pi }$, exhibit many of the features we demonstrate in the text.} \eqref{doublezeros}.
For any $v_3 \neq 0$, $f(\nu, -1 , v_3 \neq 0 )$ has two real zeros, 
(one of which is always at $\nu = 0$), and therefore the eigenfunction are  not-normalizable. So, for $m^2=0$ the massless meson becomes unstable at $v_3 \neq 0$. So that, for points laying outside of the region bounded by the curve, the eigenfunction are not normalizable and the spectrum becomes tachyonic. 

\begin{figure}[htbp]
\centering
\begin{subfigure}[b]{0.23\linewidth}
\centering
    \includegraphics[width = \linewidth]{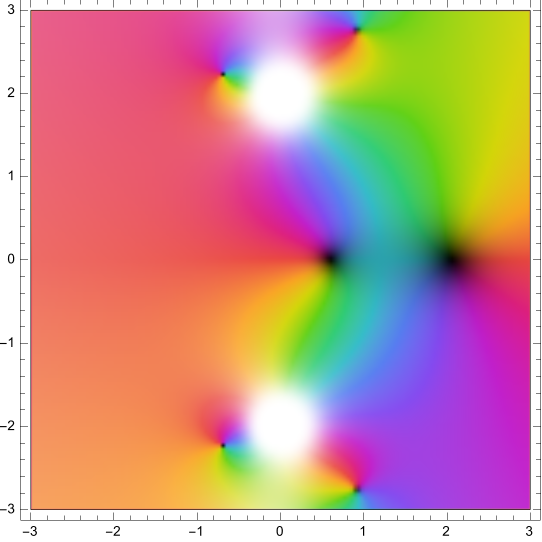}
    \caption{$\theta = 0$}
\end{subfigure}\hfill
    \begin{subfigure}[b]{0.23\linewidth}
    \includegraphics[width = \linewidth]{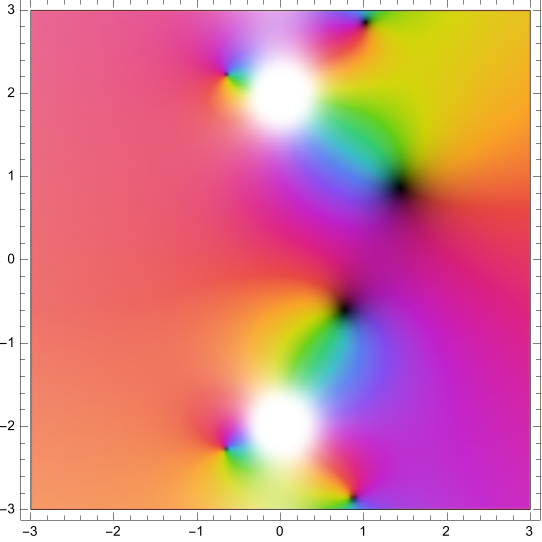}
    \caption{$\theta = 1/3$}
\end{subfigure}\hfill
\begin{subfigure}[b]{0.23\linewidth}
    \includegraphics[width = \linewidth]{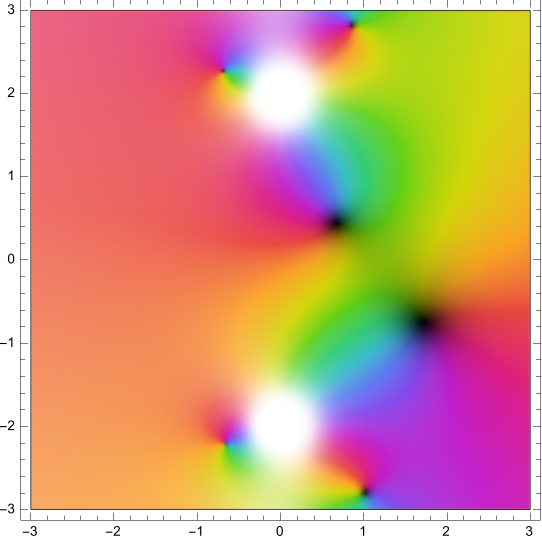}
    \caption{$\theta = 3/5$}
\end{subfigure}\hfill
\begin{subfigure}[b]{0.23\linewidth}
    \includegraphics[width = \linewidth]{figgen/complex_plot_alphav30.pdf}
    \caption{$\theta = 1$}
\end{subfigure}\hfill
\caption{We fix $\overline{\alpha} \simeq 2.12$, its corresponding positive critical point is $v^{({\rm II})}\(\overline{\alpha}\) \simeq 0.2$. The figure illustrates the positions of the zeros (black) of $f\(\nu, \alpha(\theta), v^{({\rm II})}\(\overline{\alpha}\) \)$, for values of $\alpha(\theta)= 2.4 - 0.4 e^{i 2\pi \theta}$ that encircle the point $\overline{\alpha}$. Along this analytical continuation in the complex plane the two zeros are exchanged.  } 
\label{figexchaningzeros}
\end{figure}

It is worth stressing that, although the cubic potential is not bounded from below, in the large $N_c$ approximation, the mesons can have a well-defined positive spectrum.   
\paragraph{Massless meson.}
The critical points $\(v_3^{({\rm II})}(\overline{\alpha}),\overline{\alpha}\)$ solving \eqref{doublezeros}  are square-root branch points of the eigenvalues $\lambda(\alpha,v_3)$, at which the lowest meson becomes massless. One can infer the branch-cut structure by studying the monodromy of the roots under analytical continuation of the quark mass $\alpha$ along a curve that encircles  $\overline{\alpha}$ an odd number of times. As illustrated in \figref{figexchaningzeros}, it exchanges the two simple zeros $\overline{\nu}_\pm$ of $f(\nu,\alpha, v_3)$, which eventually degenerate into $\nu^{(\rm II)}(\overline{\alpha})$. 

This shows that, by tuning the strength of the potential $v_3$, for any real and positive value of the quark masses $\alpha$, one can always produce a massless meson \footnote{This agrees with some preliminary exploration of \cite{Douglas_1994} for some specific values of the coupling. Note also that this meson is not associated with chiral symmetry as in \cite{Ambrosino:2023dik}. }.  If we set $\alpha = -1$, where the quark are massless, the curve \eqref{doublezeros} degenerates into a point $v_3^{({\rm II})} = 0$.

\begin{figure}
\begin{subfigure}{0.45\linewidth}
        \centering
    \includegraphics[width = \textwidth]{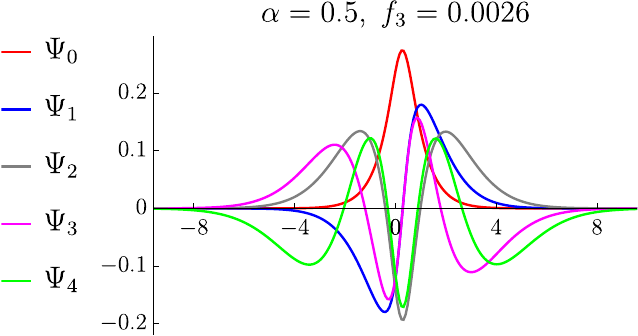}
    \caption{First five eigenfunctions with cubic potential for $v_3 = 0.0026$.}
    \label{fig:f3eigeinfunctions}
\end{subfigure}
\hfill
\begin{subfigure}{0.45\linewidth}
    \centering
    \includegraphics[width = \textwidth]{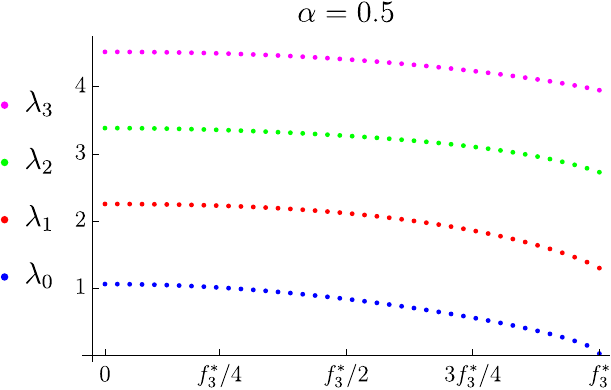}
    \caption{First 4 energy levels for $v_3\in [0,f^*_3]$ spanning the whole physical spectrum. }
    \label{fig:f3levels}
    \end{subfigure}
    \caption{Energy levels and eigenfunctions for $\alpha = 0.5$, $v_3^{(\rm II)}(0.5) = f_3^*\simeq 0.01176$.  }
    \label{fig:numericalf3}
\end{figure}

We tested these results against the numerical solutions of the $\nu$-space eigenproblem \eqref{eq:eignuspace}, as explained in Appendix \ref{appendixnumerical}. The numerical solutions in \figref{fig:numericalf3}   illustrate a typical confining spectrum of mesons with Regge-like trajectories. The ground state becomes massless in correspondance with the critical point identified by the collision of zero of $f(\nu,\alpha,v_3)$.  
\subsubsection[Analytical continuation and tricritical point]{\boldmath Analytical continuation and tricritical point}
\label{tricrit}
In the previous section, we analyzed the spectrum of mesons for real values of the quark masses ($\alpha \geq -1$) and cubic couplings. We showed that at any fixed $v_3>0$, there is a square root branch-cut in the $\alpha$ plane at $\alpha= \alpha^{(\rm II)}(v_3)$, corresponding to a critical point in the first sheet, where the first meson becomes massless. In this section we are interested in the analytic continuation to negative values of $\alpha$. Because of the branch cuts in the $\alpha,v_3\in\bR$ (see \figref{realspectrumreal}), this requires an extra care. Below, we first analytically continue $v_3$ to imaginary values keeping $\alpha > 1$, and we study the region $\alpha<-1$ aferwards.

\begin{figure}
\centering
        \centering
    \includegraphics{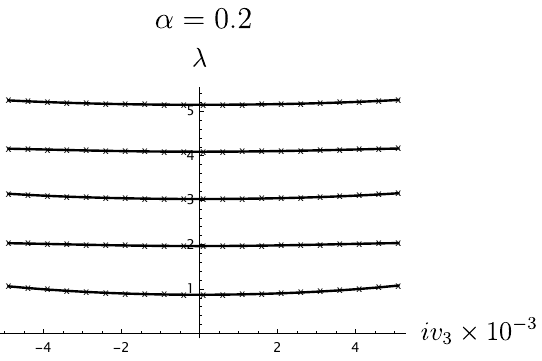}
    \caption{Numerical solutions for the first 5 levels.}
    \label{fig:unbrokenpt}
\end{figure}

\paragraph{Analytical continuation to imaginary couplings.}
At complex values of $v_3\in\bC$, the Hamiltonian is no longer hermitian, and the spectrum will be generically complex. Yet,  if  restrict to purely imaginary values of $v_3$, i.e.\ we take the self-interaction of the $B$-field to be:
\be 
\label{nonhermpot}
V(B) = \frac{1}{8\pi} B^2 + i \abs{v_3} B^3 \comma \ee
the corresponding bound-state equation \eqref{eq:eignuspace}  enjoys $\cP\cT$-symmetry\footnote{The structure of the potential here considered can be regarded as a field theory analogous of the quantum mechanical system $\cH = p^2 + m^2 x^2 + i x^3$ considered e.g.\ in \cite{Bender_1998}. }, since the $B$ field is odd under $\cP\cT$ symmetry as discussed in general in section \ref{analiticalpr}. At fixed $\alpha > -1$, for $v_3 \in i \mathbb{R}$, there are no collision-of-root singularities in the first sheet, and no zeros of $f(\nu,\alpha,v_3)$ on the real line. So we expect for this region of the parameter to have a positive spectrum of mesons. Indeed, by solving numerically the eigensystem at imaginary values of $v_3$ in a neighborhood of the real axis, we find that the spectrum remains real and positive as shown in \figref{fig:unbrokenpt}. This means that $\cP\cT$-symmetry is not spontaneously broken for $\alpha>-1$ at least in the neighborhood of the real axis of $v_3$.  

\begin{figure}
\centering
\begin{subfigure}{0.3\textwidth}
        \centering
    \includegraphics[width = \textwidth]{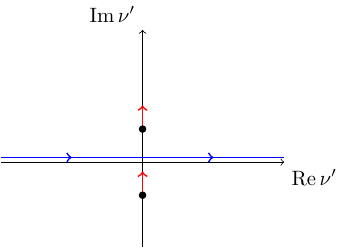}
    \caption{$\alpha > -1$ fixed, and  $v_3 = 0$}
    \label{fig:plot1}
\end{subfigure}
\begin{subfigure}{0.3\textwidth}\hfill
        \centering
    \includegraphics[width = \textwidth]{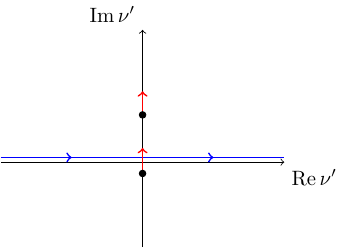}
    \caption{$\alpha > -1$ fixed, $i v_3 < 0$}
    \label{fig:plot2}
\end{subfigure}
\begin{subfigure}{0.3\textwidth}\hfill
        \centering
    \includegraphics[width = \textwidth]{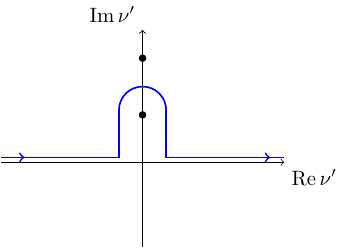}
    \caption{$\alpha < -1$, fixed  $i v_3 < 0$}
    \label{fig:plot3}
\end{subfigure}
\caption{First two roots of $f(\nu',\alpha,v_3)$ upon analytical continuation. In (a) and (b), at fixed $\alpha > -1$, we analytically continue in $v_3$ to $v_3 \in i \bR^+$. The red arrows illustrate the shift of the first two roots under this continuation. In (c), we then plot the result of the analytical continuation from $\alpha > -1 $ to $\alpha < -1$ at fixed value $v_3\in i \bR$. 
The root crosses the real axis exactly at $\alpha = -1$. When the root crosses the real axis, the integration contour is deformed to avoid the singularity. For $i v_3 > 0$ it is analogous but with the root in the upper half plane moving in the lower half plane. }
\label{fig:plot123}
\end{figure}

\paragraph{\boldmath Critical points at $\alpha < -1$.} The analytical continuation to  imaginary values of $v_3$ allow us to also continue the spectrum to  $\alpha < -1$ without crossing any of the branch cuts. More concretely, we first continue to\footnote{$\abs{v_3}$ has to be large enough to guarantee any pinching of root singularity on the contour, as we will discuss below.} $v_3=i \mathbb{R}^{+}$ at $\alpha > -1$, and only then we continue from $\alpha > -1 $ to $\alpha < -1$. In doing that,  exactly one of the roots crosses the real axis, (as $f(\nu,-1, v_3)$ has always a root at $\nu = 0$),  yet, as long as one makes sure that there are no collision of roots on the contour,  one can deform the  $\nu'$ integration contour in \eqref{eq:eignuspace}, to avoid the singular point. This is illustrated in \figref{fig:plot123}.

\begin{figure}
    \centering
    \includegraphics[width = \textwidth]{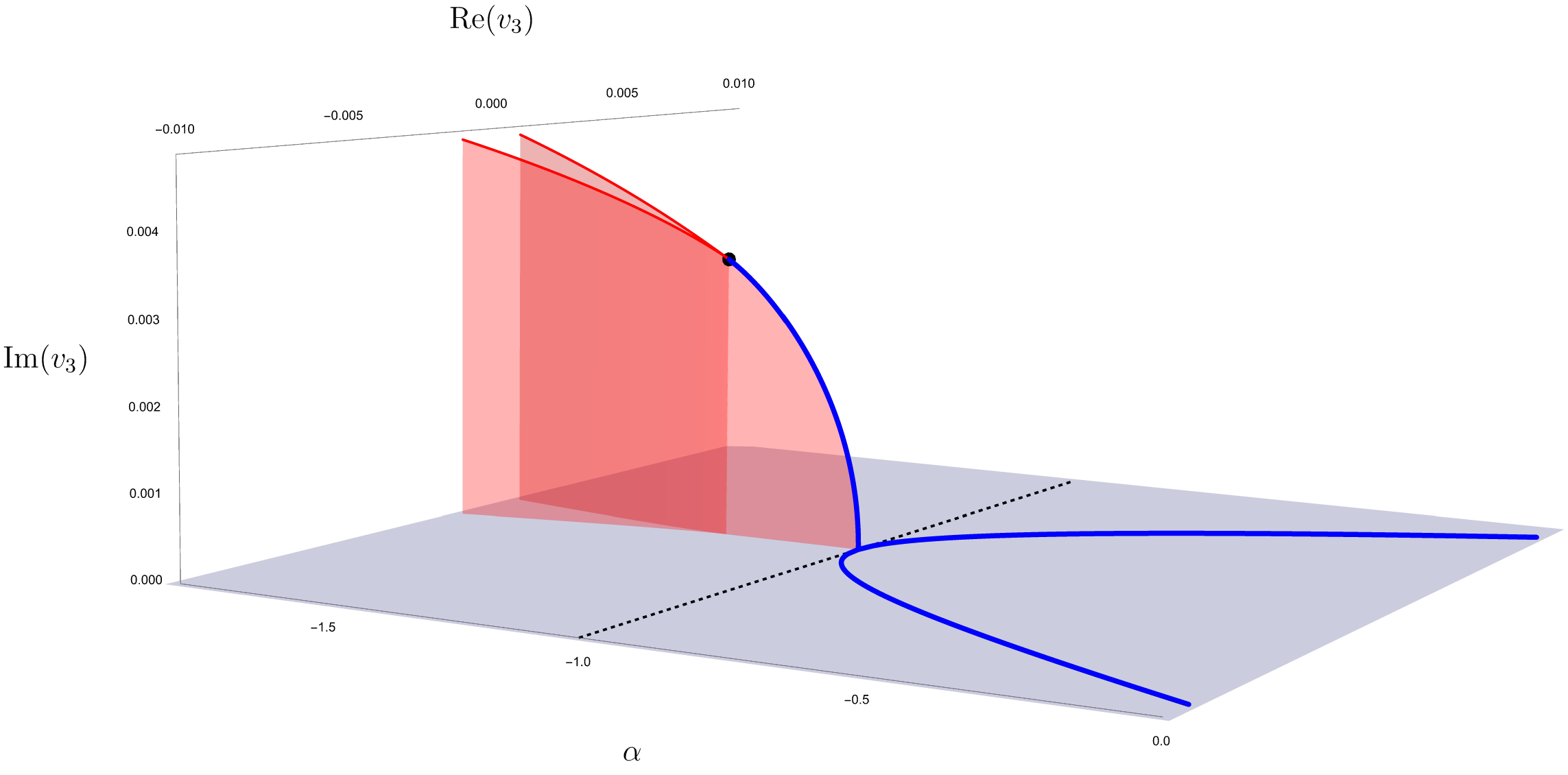}
    \caption{Critical points for $v_3\in\bR^+$ or $v_3 \in i\bR^+$ and $\alpha\in\bR$. For $\alpha > -1$, the blue curves are loci of critical points lying at $\Im(v_3) =0$ and are exactly the one studied in Section \ref{realcouplings} (cfr. \figref{realspectrumreal}). The blue curve at $\alpha < -1$, is located at $\Re(v_3) = 0$ while the red ones correspond to critical points for the complex spectrum. The black dot corresponds to the tricritical point.}
    \label{fig:tricriticalpoints}
\end{figure}

Collision-of-roots singularities happens at $\alpha < -1$, whenever, upon varying $v_3$ the first two roots in \figref{fig:plot123} pinch the contour. This happens, for any $\alpha < -1$, at two isolated points on the imaginary axis $\pm v_3^{(\rm II)}(\alpha)$, determined by the solutions of the equation $f(\nu,\alpha,v_3) = 0$ and $\partial_\nu f(\nu,\alpha,v_3) = 0$ at fixed $\alpha<-1$. Upon decreasing $\alpha$, one finds solutions $v_3^{(\rm II)}(\alpha)\in i \bR$ only for $\alpha > \alpha^{({\rm III })}\simeq  -1.27649$, while for $\alpha > \alpha^{({\rm III })}$ the corresponding solutions $v_3^{(\rm II)}(\alpha)$ split in the complex plane (red curves in \figref{fig:tricriticalpoints}). The collision-of-roots singularities are critical points in the spectrum where the first meson becomes massless. In \figref{fig:tricriticalpoints}, we report the loci of critical points at $\alpha\in\bR$ and $v_3$ in the upper half plane (the lower half plane of $v_3$ is symmetrical). The red curves correspond to the solutions for $\alpha> \alpha^{(\rm III)}$ and are indeed located in a region of the parameter space where the 't Hooft equation is neither hermitian nor $\cP\cT$-symmetric, and the masses of mesons are $\lambda_n \sim M^2_n$ are complex.  The critical points in this region correspond to points where one of the complex eigenvalues turns to zero. 
Any analytical continuation performed along a path on the first sheet of $v_3$ and $\alpha$ that does not intersect any of the blue or red curves and does not ends in the region shaded in red, leads to a discrete and positive spectrum of mesons for $\alpha > \alpha^{(\rm III)}$. Correspondingly, the $\cP\cT$ symmetry is not spontaneously broken in this region of the parameter space.
In Appendix \ref{noncriticalbreaking}, we describe the spectrum also after analytically continuing it around the $\alpha^{\rm (III)}$ in the region shaded in red. There, $\cP\cT$-symmetry is spontaneously broken.

\begin{figure}
\centering
\begin{subfigure}{0.45\textwidth}
       \centering
    \includegraphics[width = \textwidth]{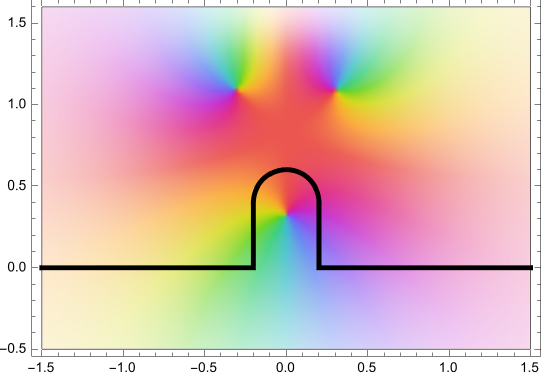}
\end{subfigure}\hfill
\begin{subfigure}{0.45\textwidth}
        \centering
    \includegraphics[width = \textwidth]{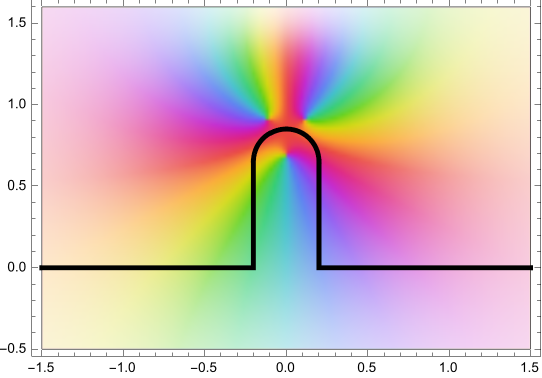}
\end{subfigure}
    \caption{Integration contour in the $\nu'$ complex plane for $\alpha < \alpha^{(\rm III)}$ at fixed $v_3= v_3^{(\rm III)}$. As $\alpha$ approaches the tricritical point, the contour is trapped between three singularities.}
    \label{contour12}
\end{figure}

\paragraph{Tricritical point.}
As illustrated in the previous subsection, the line of critical points in the first sheet of the $(\alpha,v_3)$ plane for $\alpha < -1$, terminates at the special point: 
 \be\label{cubicroot} 
 \alpha^{({\rm III })} \simeq  -1.27649 \comma \quad  v_3^{({\rm III })} \simeq \pm  0.00344328 \, i    \comma
\ee  
where $f(\nu)$ has a triple zero.  Here the contour has a singularity due to the collision of three roots simultaneously as shown in \figref{contour12}.
\begin{figure}
    \centering
    \includegraphics{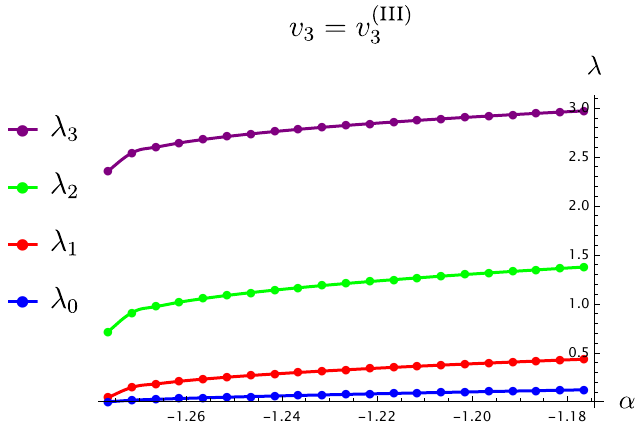}
    \caption{First 4 meson masses as a function of $\alpha$ at fixed $v_3= v_3^{(\rm III)}$. The first two levels turn to zero at $\alpha = \alpha^{(\rm III)}$. }
    \label{tricriticalhermite}
\end{figure}
The monodromy of the roots under analytical continuation in $\alpha$, is exactly the one illustrated in \figref{fig:3rdorderroot}. 
Henceforth, according to the general discussion of section \ref{analiticalpr}, one expects that the theory admits a description in terms of a tricritical CFT where two of the energy levels turn to zero simultaneously as $\lambda \sim \sqrt[3]{\alpha - \alpha^{(\rm III)}}$. In \figref{tricriticalhermite} the result of the numerical spectrum after analytical continuation; we refer to Appendix \ref{appendixnumerical} for a detailed explanation of the method we employed to analytically continue to $\alpha < -1$.

\subsubsection{Singularities beyond the cuts}
\label{singbey}
\begin{figure}\centering
\begin{subfigure}{0.65\textwidth}
    \centering
    \includegraphics[width = \textwidth]{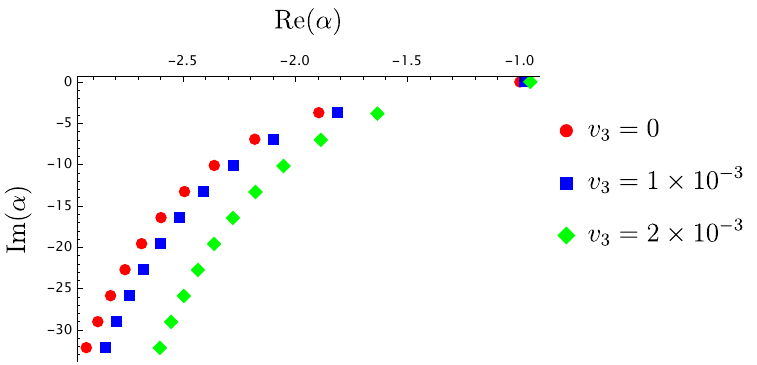}
    \caption{Singularity in the second sheet of the $\alpha$ plane after analytical continuation through the cut along the real axis, for various values of $v_3$.}
    \label{fig:highersheet}
    \end{subfigure}\hfill
\begin{subfigure}{0.27\textwidth}
    \centering
    \includegraphics[width = \textwidth]{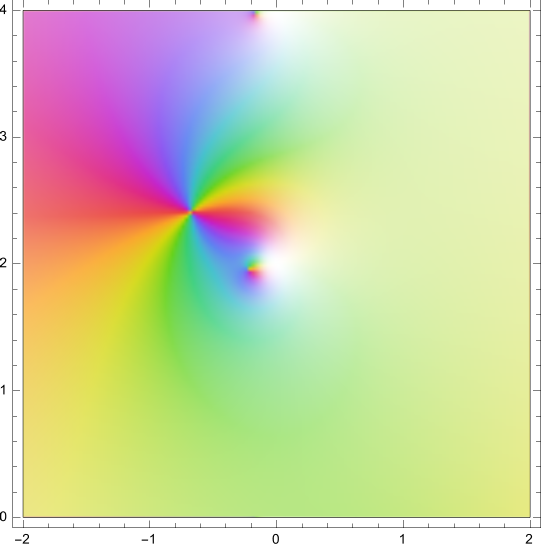}
    \caption{Collision of the first singularity, after analytical continuation with one of the higher.}
    \label{fig:complexplothigherroot}
    \end{subfigure}
    \caption{Singularities in the second sheet of the $\alpha$ plane continued through the cut at fixed $v_3\in\bR$.}
\end{figure}

Each of the critical point we discussed is associated to a branch cut in the complex plane of the coupling. 
As a generalization to what happens in the 't Hooft model \cite{Ambrosino:2023dik}, as we analytically continue the model to complex values of $\alpha$ and $v_3$, we probe a very rich and complicated multi-sheeted structure. 
\paragraph{Critical points in the second sheet.}
For fixed $v_3\in\mathbb{R}$, by analytically continuing in $\alpha$ through  the square root cut, the two roots inevitably cross the integration contour, and the eigenproblem differs by a further contribution due to the contour integration around those singularities, as in \figref{fig:monodromy}. In the second sheet of these cuts, we find infinitely many other critical points where, at complex values of $\alpha$, one of the roots that took a monodromy after analytical continuation collides with one of the higher singularities (e.g.\ \figref{fig:complexplothigherroot}).  On  each of the corresponding sheets, the $k-$th meson becomes massless, rather than the first energy level $\lambda_0$. 
This is a generalization of what already discussed for the 't Hooft model \cite{Ambrosino:2023dik}, but at generic values of $v_3\in\bR$. 

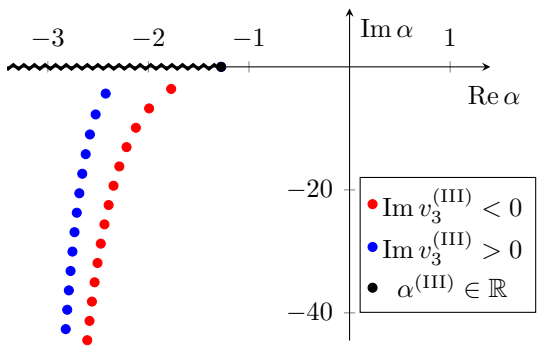
\begin{figure}
\centering
\begin{tikzpicture}
  \begin{axis}[
      xlabel={$\Re \,\alpha $},
      ylabel={$\Im \,\alpha$},
      axis lines=middle,
      enlargelimits,
      width=8cm,
      height=6cm,
      font=\footnotesize,
      smooth,
      xmin=-3,
      xmax=1,
      ymin=-40,
      ymax=5,
      xticklabel style={
        yshift=4ex,
      },
      xlabel style={
        yshift=-3.5ex,
        xshift = 3ex,},
       legend style={at={(axis cs:0.1,-40)},anchor=south west},
       title={ Tricritical points in the complex $\alpha$-plane}
    ]
    \addplot[only marks, mark=*, mark size=1.7, red] table[x index=0, y index=1, col sep=comma] {figgen/solneg.csv};
    \addlegendentry{$\Im\, v^{(\rm III)}_3 < 0$}
    \addplot[only marks, mark=*, mark size=1.7, blue] table[x index=0, y index=1, col sep=comma] {figgen/solpos.csv};
        \addlegendentry{$\Im\, v^{(\rm III)}_3 > 0$} 
       \addplot[only marks, mark=*, mark size=1.7, black] coordinates {(-1.27649,0)};
           \addlegendentry{$ \alpha^{(\rm III)} \in \bR$} 
    \draw [decorate, very thick, decoration={zigzag,segment length=4,amplitude=0.8,post=lineto,post length=0}] (-4,0) -- (-1.27649,0);
\end{axis}
\end{tikzpicture}
\caption{Position of the tricritical points in the complex $\alpha$-plane. Each of them is associated with a distinct value of $v^{(\rm III)}_3$.}
\label{trictriticalpoint}
\end{figure}

\paragraph{Tricritical points in higher sheets.}
A similar argument shows the existence of  infinitely many tricitrical points after the analytic continuation in $\alpha$ through  the cubic root branch cut at \eqref{cubicroot}. On the second sheet relative to this cut, one finds two infinite towers of cubic branch points, each associated with distinct values of $v_3$ and located respectively in the second and third sheet of the $\alpha$-plane. We report them in  \figref{trictriticalpoint}, as determined by the points corresponding to triple zeros of $f(\nu,\alpha,v_3)$ in the complex plane.

The two infinite towers are distinguished  by the sign of the imaginary part of the corresponding $v_3^{(\rm III)}$\footnote{This can also be understood from the square root branch cut in the $v_3$-$\alpha$ plane, present for any value of $\alpha$, in correspondence of which the value of $v_3$ changes sign. }. Each of those tricritical points is associated to a point  where two of the higher meson masses, rather than the first and second,  turn to zero simultaneously. 

\paragraph{\boldmath Second sheet critical points at imaginary $v_3$.}
As one consider the region $v_3\in i \bR$, a new type of critical points in the second sheet arise. Following the same strategy of the previous subsection, it is not difficult to show that for any value of $\alpha\in\bR$, there are infinitely many double zeros  $\pm v^{({\rm II}, k)}_3(\alpha)\in  i\bR$, $k\in \mathbb{N}$, strictly ordered as:
\be 
\label{criticalpointspt}
\abs{v^{({\rm II},0)}_3(\alpha)} \leq \abs{v^{({\rm II},1)}_3(\alpha)} \leq \abs{v^{({\rm II},2)}_3(\alpha)} \leq \cdots \comma
\ee
where $f(\nu,\alpha,v_3)$ has a double zero at $\nu^{({\rm II},k)}(\alpha) \in i \bR$:
\be 
\label{locuscritical}
f\(\pm \nu^{({\rm II},k)}(\alpha),\alpha,\pm v^{({\rm II},k)}_3(\alpha)\) = \partial_\nu f\(\pm \nu^{({\rm II},k)}(\alpha),\alpha,\pm v^{({\rm II},k)}_3(\alpha)\)=0 \comma 
\ee
At the first of those critical points $\pm v_3^{({\rm II},0)}$, the two closest roots of $f(\nu)$ in the upper-half plane (that are always in the strip $\nu \in [0,2i]$) collide. At fixed $\alpha$, none of these collision-of-root point is a singularity in the first sheet. But, if one first analytically continue in $\alpha$ along a closed curve enclosing a the square root branch point, then $v^{({\rm II},k)}_3(\alpha)$ becomes a singular point where the first eigenvalues closes the gap and breaks spontaneously $\cP\cT$-symmetry. This is exactly identical to what discussed in Section 6.3.1 of  \cite{Ambrosino:2023dik} (see Figure 16 there).

\begin{figure}
\centering
\begin{subfigure}{0.6\textwidth}
           \centering
    \includegraphics[width = \textwidth]{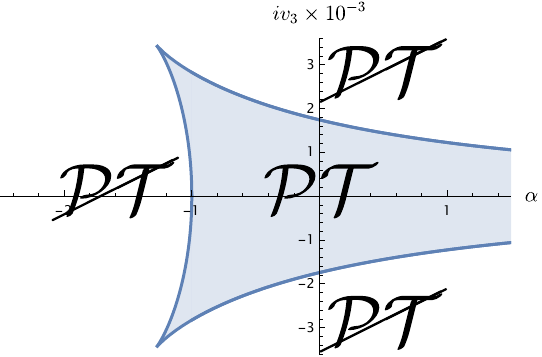}
    \caption{Phase space of the theory for $\alpha > -1$ in the second sheet of the $\alpha$ plane, and $v_3 \in i\bR$. In the shaded region, the model present a physical and confining spectrum of mesons. }
    \label{fig:phasesunbroken} 
\end{subfigure}\hfill
\begin{subfigure}{0.3\textwidth}
           \centering
    \includegraphics[width = \textwidth]{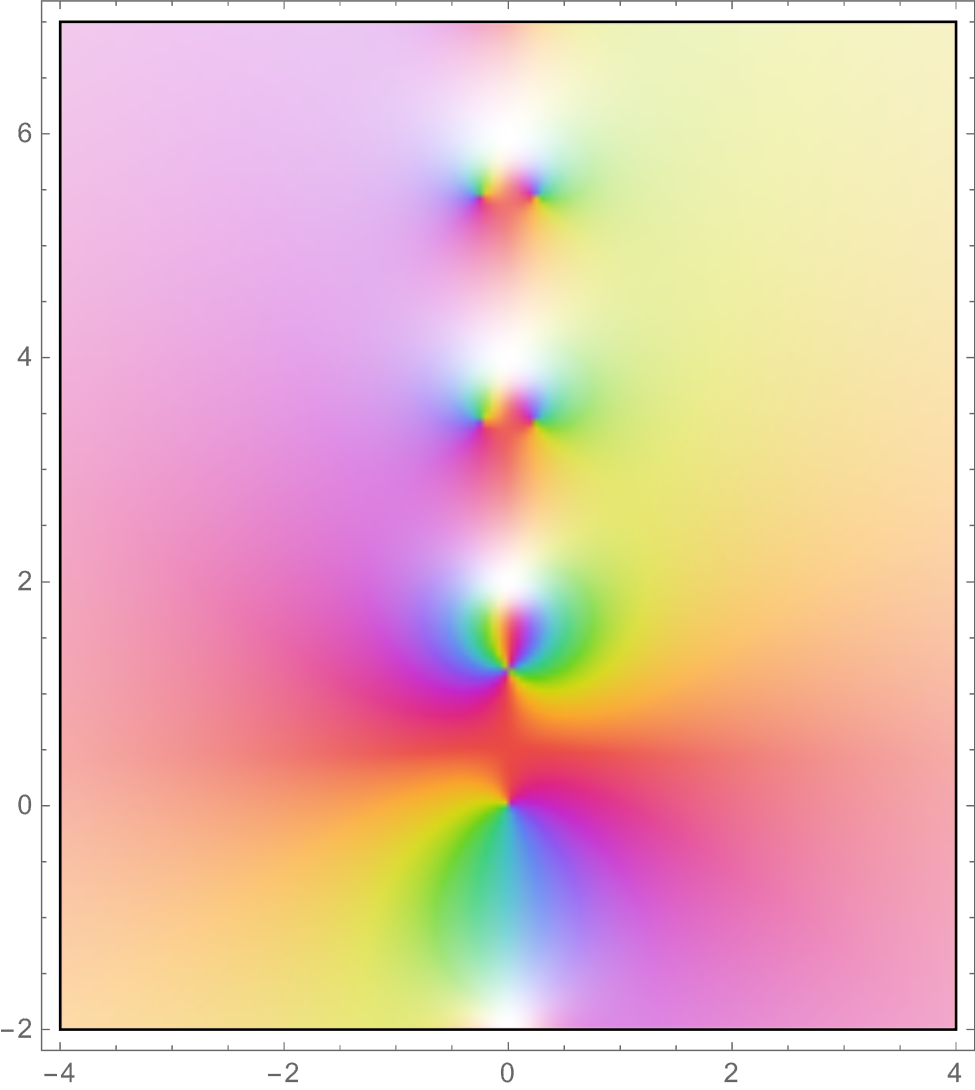}
    \caption{First collision of roots of $f(\nu,\alpha,v_3)$ at $v^{({\rm II},0)}_3$. \\
    }
    \label{fig:phasesunbroken2} 
\end{subfigure}
\caption{Collision of roots for imaginary $v_3$ and corresponding phase-space.}
\end{figure}

Note that the line of critical points at $\alpha< -1$ remains singular also in this second sheet, the resulting phase-space of the theory  is illustrated in \figref{fig:phasesunbroken}, where is is interesting to note that these new locus of critical point intersects the one discussed in the previous subsection exactly at the tricritical point. 

\begin{figure}[!h]\centering
\begin{subfigure}{0.3\textwidth}
        \centering
    \includegraphics[width = \textwidth]{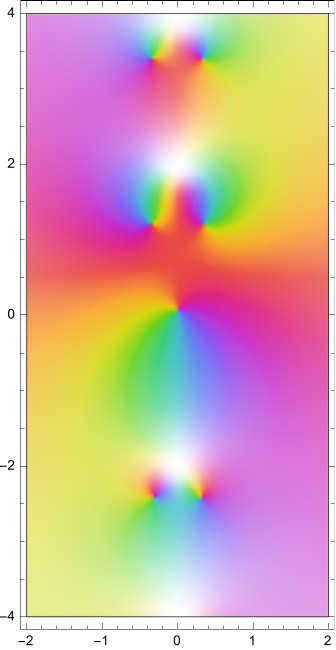}
    \caption{$v_3 > v_3^{(\rm II,0)}$}
\end{subfigure}\hfill
\begin{subfigure}{0.3\textwidth}
        \centering
    \includegraphics[width = \textwidth]{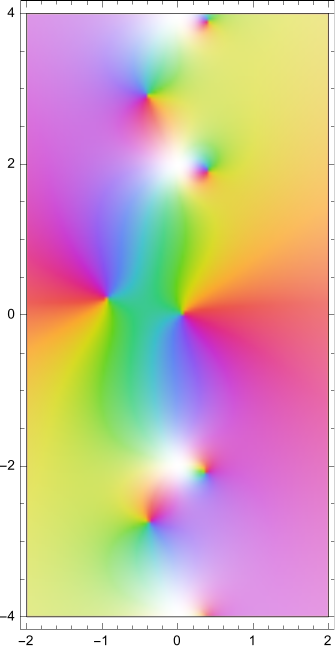}
    \caption{$v_3 e^{i2\pi /5 }$}
\end{subfigure}\hfill
\begin{subfigure}{0.3\textwidth}
        \centering
    \includegraphics[width = \textwidth]{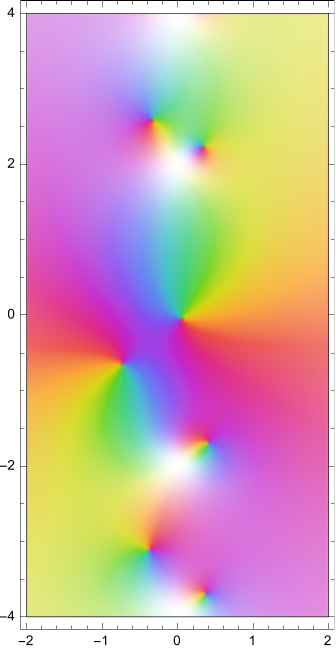}
    \caption{$v_3 e^{i4\pi /5 }$}
\end{subfigure}\hfill\\
\begin{subfigure}{0.3\textwidth}
        \centering
    \includegraphics[width = \textwidth]{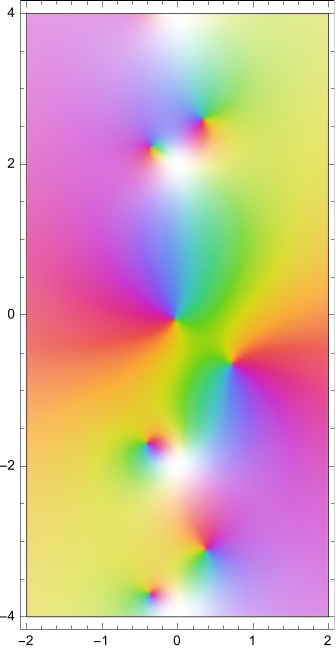}
    \caption{$v_3 e^{i6\pi /5 }$}
\end{subfigure}\hfill
\begin{subfigure}{0.3\textwidth}
        \centering
    \includegraphics[width = \textwidth]{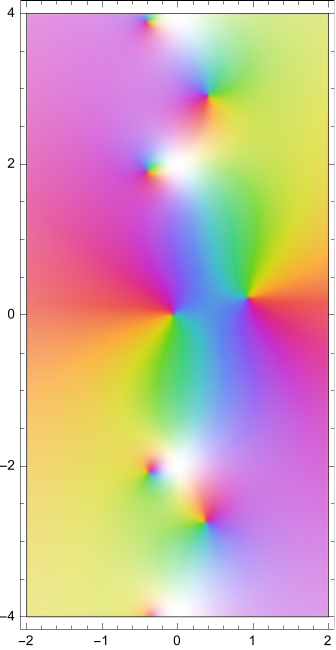}
      \caption{$v_3 e^{i8\pi /5 }$}
\end{subfigure}\hfill
\begin{subfigure}{0.3\textwidth}
        \centering
    \includegraphics[width = \textwidth]{figgen/monov32.pdf}
    \caption{$v_3 e^{2i\pi }$}
\end{subfigure}\hfill
\caption{Monodromy of the roots under analytical continuation in $v_3$ around $v_3^{({\rm II} ,0)}$.}
\label{fig:monov3}
\end{figure}

 This can be generalized to any of the $\nu^{({\rm II}, k)}$.
It turns out that, at any fixed $\alpha$, $v^{({\rm II},0)}_3(\alpha)$ is also a branch point of infinite degree in the $v_3$ plane. Indeed upon analytically continue in $v_3$ around any closed curve enclircling a single time $v^{({\rm II},k)}_3(\alpha)$, one observes a monodromy of the roots of the form $\nu_{i} \to \nu_{i\pm 1}$ where $\nu_i$ is the position of the roots of $f(\nu,\alpha,v_3)$ (with the exception of the one corresponding to $-v^{({\rm II},0)}_3(\alpha)$) and $\pm$ depends on the sign of  $\rm{Re}(\nu_i)$. This is illustrated in \figref{fig:monov3}. Then, it is easy to see that under this analytical continuation, one can always find a sheet where $\nu^{({\rm II}, k)}$ becomes a critical point and the corresponding $k$-th meson closes the gap and breaks spontaneously $\cP\cT$-symmetry. 

\section{Mesons spectrum in the large representation limit}\label{sec:largespin}
In this section we consider a theory with a finite number of colors $N$, but of fermions in a very large representation of the gauge group. This limit was also considered in \cite{Kaushal:2023ezo} for the 't Hooft model. Here we illustrate that  analogous results can be derived also for the generalized QCD case.
Specifically, we consider an $SU(2)$ gauge theory coupled to a single flavor of quarks in the isospin $J$ representation, and consider a scaling limit where $J\to \infty$.  

\paragraph{Effective Hamiltonian for fermions.}
For our scopes, it is best to work with an Hamiltonian formulation of the theory. The starting point is the  generalized YM Lagrangian in lightcone gauge, coupled to quarks in a generic representation of the gauge group. Not not specifying yet $N=2$ the Hamiltonian is:
\be 
\cL = \frac{N}{4\pi} {\rm tr} \, B \,\partial_-\, A_+  - \frac{Ng^2}{4\pi} \sum_{n=2}^{\infty} {\rm tr } \, v_n B^n + \ovps_a \(i \slashed{\partial} - m_a\)  \psi_a - {\rm tr } \,  A^A_+ \( \ovps_a \, \gamma_- \, T^A\, \psi_a\)\comma
\ee
and we take the generators $T^A\comma A=1,\cdots N^2-1$ in a generic large dimensional representation of $SU(N)$ . 
Integrating out the $A_+$ field in the path integral leads to the insertion of a delta function distribution setting
\be
\frac{N}{4\pi} \partial_- B^A - \ovps_a \, \gamma_-\, T^A \, \psi_a = 0\comma
\ee
and leads to
\be 
\cL = - \frac{Ng^2}{4\pi} \sum_{n=2}^{\infty} {\rm tr}_{\rm adj} \,   \(\frac{4\pi}{N}\)^n v_n \( \partial^{-1}_-\ovps_a \, \gamma_- \, T\, \psi_a\)^n + {\rm tr}\,  \ovps_a\, \(i \slashed{\partial} - m_a\) \, \psi_a \comma
\ee
where with  ${\rm tr}_{\rm adj}$ we mean that the trace has to be taken in the adjoint representation:
\be 
{\rm tr}_{\rm adj} \( \partial^{-1}_-\ovps_a \, \gamma_- \, T\, \psi_a\)^n = {\rm tr}[F^{A_1}\cdots F^{A_N} ]  \, \prod_{i=1}^n\( \partial^{-1}_-\ovps_a \, \gamma_- \, T^{A_i}\, \psi_a\)\comma\qquad  A_i = +, - , 3\comma
\ee
where $F^A$ are the generators in the adjoint representation of $SU(N)$. 

The component $\psi_+$ is non-dynamical because the field has no time derivatives in the lightcone gauge Lagrangian; it is fixed by the equations of motion to:
\be 
\psi_+ = \frac{i m}{2} \partial^{-1}_- \, \psi_-\period
\ee
Henceforth, one reduces the Lagrangian \eqref{lagrangian} to the following effective action for the only dynamical quark component $\psi_-$:
\be \label{fermionlagrangian}
\cL_{\rm{eff}} = - \frac{Ng^2}{4\pi} \sum_{n=2}^{\infty} {\rm tr}_{\rm adj} \, \(\frac{4\pi}{N}\)^n v_n \( \partial^{-1}_-\ovps_a \, \gamma_- \, T\, \psi_a\)^n + {\rm tr}  \, \ovps_a\, \(i \slashed{\partial} - m_a\) \, \psi_a \comma
\ee
The standard quantization:
\be
[\pi_-(x), \psi_-(y)] = \delta(x_--y_-)\, , \qquad \pi(x) = \pdv{\cL_{\text{eff}}}{\psi_-(x)} = i \sqrt{2} \, \psi^\dag_+(x) \comma
\ee
leads to the effective Hamiltonian density:
\be \label{effectiveham}
\cH_{\rm{eff}} = - \frac{Ng^2}{4\pi} \, \sum_{n=2}^{\infty}   \(\frac{4\pi}{N}\)^n v_n  {\rm tr}_{\rm adj} \( \partial^{-1}_-\ovps_a \, \gamma_- \, T\, \psi_a\)^n -  i\frac{m^2}{4} {\rm tr}\,\psi^\dag_- \partial^{-1}_- \psi_- \comma
\ee
The quarks depend on the two lightcone coordinates $\psi(y) = \psi(y_-, y_+)$, while in the potential only derivatives with respect to the $y_-$ coordinate appear. To simplify the notation, we will momentarily drop the explicit ``$-$" label of the coordinates, with the understanding that inside the argument of the fermions $\psi(y) \equiv \psi(y_-, y_+)$, and  everywhere else $y \equiv y_-$. Using the identities in Appendix A of \cite{Kaushal:2023ezo}, each of the quark bilinears that appears in \eqref{effectiveham} can readily expressed as\footnote{A similar rewriting was also employed in this context in \cite{Pesando:1994cw}.}:
\be
 \partial^{-1}_{x} \psi^\dag_-(x) \psi_-(x) =  \int\dd{y}\partial^{-1}_{y} \delta(x -y) \psi^\dag_-(y) T^a \psi_-(y) = \frac{1}{2}\int\dd{y} \sgn(x-y) \psi^\dag_-(x) T^a \psi_-(y)\period
\ee
Then, the effective Hamiltonian is ($y_1 := x, y_{0} = y_{n}$):
\be \begin{split}\label{eff}
H_{\rm{eff}} 
&= - \frac{g^2}{4\pi}\sum_{n=2}^{\infty}{\rm tr}\( 2\pi\)^nv_n\(\ts\prod_{k= 1}^n \int\dd{y_k}\sgn(y_k - y_{k-1})\psi^\dag_{-}(y_k) T^{A_k}\psi_{-}(y_k)\) \Tr[F^{A_1}\cdots F^{A_n}] \\ 
&-{\rm tr} \, i\frac{m_a^2}{8} \int\dd{x}\dd{y} \psi_{-}^\dag (x)  \sgn(x- y)   \psi_{-} (y) 
\end{split}
\ee
\paragraph{Large representation limit. }
Let us specialize to the group $SU(2)$. In the spin $J$ representation, we have
\be \label{matrixelements}\begin{split}\psi^\dag_- T^+ \psi_-(x) &= \sum_{m=-J}^J\sqrt{(J-m)(J+m+1)}\, \psi^*_{m+1,-}(x) \psi^{m}_-(x)\comma \\
\psi^\dag_- T^- \psi_-(x) &=  \sum_{m=-J}^{J}\sqrt{(J+m)(J-m+1)}\, \psi^*_{m,-}(x) \psi^{m+1}_-(x)\comma\\
\psi^\dag_- T^3 \psi_-(x) &= \sum_{m=-J}^{J} m\, \psi^*_{m-}(x) \psi^{m}_-(x) \period
\end{split}\ee
Following the approach of \cite{Kaushal:2023ezo}, in the large $J$ limit $\frac{m}{J}$ is approximated by a continuous variable $\frac{m}{J} = \cos\theta$. 
Rather than carrying a representation index $m$, in this limit, the quarks depend explicitly on the continuous variable $\theta$:
\be \psi_-(\theta, x) := \sqrt{J}\, \psi^m_-(x)\comma \ee
where, the additional factor $\sqrt{J}$  guarantees the correct scaling of the canonical quantization in the limit $J\to \infty$:
\be 
\{ \psi^\dag_-(\theta,x) \, ,\,  \psi_-(\theta',y) \} = J \, \(\frac{1}{J \sin\theta} \delta(\theta-\theta') \)\, \delta(x-y)
\ee
Accordingly, we must replace the trace over the representation by its continuous limit:
\be
\sum_{m = -J}^{J} \to J \int_0^\pi\dd{\theta} \sin\theta  \comma
\ee
Using the continuous fields $\psi_-(\theta,x)$, one derives \cite{Kaushal:2023ezo} the following large $J$ scaling of \eqref{matrixelements}:
\be\begin{split}
\label{elementlarge}
  &\psi^\dag_-(x)\, T^\pm \, \psi_-(x) \to  J \int_0^\pi\dd{\theta}\sin^2\theta \, \psi^*_-(\theta,x) \psi_-(\theta,x) \comma\\
  &\psi^\dag_-(x)\, T^3 \, \psi_-(x) \to J \int_0^\pi\dd{\theta}\sin\theta \cos\theta\, \psi^*_{-}(\theta,x) \psi_-(\theta,x)\ .
\end{split}\ee
Now, let us assume that the potential truncates at a certain order $\bar{n} \in 2\bN$ (the reason for fixing $\bar{n}$ even will be clear in a moment), and that the none of $v_{n}$ scales with $J$.  Then, because of the overall trace,  the Hamiltonian is order $J$, and admits a double-scaling limit at $J\to \infty$ where the combination \be\label{doublescaling} 4 (2\pi)^{\bar{n}-1} g^2 J^{\bar{n}} = \lambda \ee is taken fixed. Taking $v_n = \cO(J^0)$ suppresses any term $V(B)$ with $n< \bar{n}$\footnote{\label{footnotegen} Any of them can be retained by imposing a scaling limit on the $v_n = \cO\(\ts J^{\bar{n}-n}\)$ at large $J$. Since the  generalization to this case is straighforward, here we consider the simplest case where only the highest power contributes at large spin.}. Under this conditions, in the large spin limit, \eqref{eff} is:
\be \begin{split}
H_{\rm{eff}} =  - \frac{v_{\bar{n}}\lambda}{8} &{\rm tr}\,   \(\textstyle \prod_{k= 1}^{\bar{n}} \int\dd{y_k} \sgn(y_k - y_{k-1}) \psi^\dag_{a,-}(y_k) \, T^{A_k}\, \psi_{a,-}(y_k)\) {\rm tr}\,[F^{A_1} \cdots F^{A_{\bar{n}}}] \\ 
&-{\rm tr}\, i\frac{m_a^2}{4} \int \dd{x}\dd{y} \psi^\dag_{a,-} (x)  \sgn(x- y)   \psi_{a,-} (y)  \comma
\end{split}
\ee
where we intend that all the representation indexes are contracted through integrals over the respective $\theta$ variables and that we take the trace thereof  ($\theta_1 = \theta_n$) to ensure gauge invariance. 
For the special case of $SU(2)$, only traces of an even number of adjoint generators  are non-vanishing\footnote{We normalize the generators in the adjoint representation such that the killing form is $\kappa^{ab} = \begin{psmallmatrix} 0 & 1/2 &0\\ 1/2 & 0 &0 \\ 0 & 0 &1 
\end{psmallmatrix}$. }. Henceforth, odd powers in the potential of $V(B)$ do not contribute (this is the reason for setting $\bar{n}\in 2\bN$) to in an $SU(2)$ theory. 

As in \cite{Kaushal:2023ezo}, we introduce the bi-local field:
\be \label{defM}
\bra{y,\phi} \bs{M}  \ket{x,\theta} = M(x,y;\theta,\phi) = \psi_-(\theta,x) \psi^*_-(\phi,y) \period
\ee
We will also use  the notation $\bs{M}(x,y)$ whenever we want to specify the positions while keeping the short hand matrix notation. This notation is convenient for expressing contraction of bilinear fields in the continuum limit $\int\dd{\cos \theta} A(x,y,\theta,\phi) B(z,t,\theta,\lambda)$ in the short-hand matrix multiplication notation $\bs{A}(x,y)\, \bs{B}(z,t)$.
After we have substituted the adjoint representation traces, the interaction term of $\cH_{\rm eff}$, keeping only the symmetric terms in $\theta$, and indicating the integration measure as  $\dd{\bs{y}}\dd{\bs{\theta}}  = \frac{1}{\bar{n}!}\prod_{i=1}^{\bar{n}}\dd{\theta_i} \sin\theta_i \dd{y_i}$, takes the form :
\small
\be \begin{split}
&\int\dd{\bs{y}}\dd{\bs{\theta}} \prod_{i=1}^{\bar{n}} \sgn{(y_k - y_{k-1})} \Bigg[ M(y_1, y_2;\theta_1,\theta_2) \widehat{M}(y_2, y_3,\theta_2,\theta_3) M(y_3,y_4,\theta_3,\theta_4) \widehat{M}(y_4,y_5;\theta_4,\theta_5) \cdots \\ & \hspace{12cm} +{\rm perm.\ } \Bigg]\comma
 \end{split}\ee
 \normalsize
 where $ \widehat{M}(y_i, y_j,\theta_i,\theta_j) := M(y_i, y_j,\theta_i,\theta_j) \cos(\theta_i - \theta_j)$. 
These quantities are conveniently expressed through the matrix notation introduced in \eqref{defM}:
\be 
{\rm tr}\int \prod_{i=1}^{\bar{n}}\(\dd{y_i}\sgn{(y_k - y_{k-1})}\) \(\bs{M}(y_1,y_2)\,\bs{\widehat{M}}(y_2,y_3) \bs{M}(y_3,y_4) \cdots \) \period
\ee 

Then, for an $SU(2)$ theory, at large spin, the Hamiltonian  \eqref{effectiveham} is:
\be \begin{split}
H_{\rm eff} = & \frac{i m^2}{8}\int \dd{x}\dd{y} \sgn(x-y)M(x,y;\theta,\theta) + \\ &+ \frac{\lambda v_{\bar{n}}}{8} \int \prod_{i=1}^{\bar{n}}\(\dd{y_i}\sgn{(y_k - y_{k-1})}\) \,{\rm tr} \, \(\bs{M}(y_1,y_2) \bs{\widehat{M}}(y_2,y_3) \bs{M}(y_3,y_4) \cdots  + {\rm perm.\  }\)\comma
\end{split}\ee
 \paragraph{Bound state equation. }
The fermions satisfy the same anti-commutation relation of \cite{Kaushal:2023ezo}, and therefore their same considerations apply here\footnote{Adding a potential for the gauge part of the action, does not modify the quantization of the fermions. }: in particular the bilinear quantum operators $\bs{M}$ satisfy a $W_\infty^{2J + 1}$ algebra (cf.\ \S 3.2, Appendix C,D of \cite{Kaushal:2023ezo} and references therein). It has been shown in \cite{DHAR_1993,DHAR_1994, Kaushal:2023ezo} that, in the case of 't Hooft model at large $N$ or large $J$, the fluctuation along the co-adjoint orbit around a classical value of $\bs{M}$ can be identified with the meson wavefunction. The equation of motion for those fluctuations give the t' Hooft equation determining the spectrum of the mesons. At large spin, this is a very efficient way to obtain the bound state equation, as this method does not rely on planar perturbation theory. We illustrate this in Appendix \ref{appboundstatelargespin} where we compute the equation of motions for the fluctuation $\bs{W}$ around the classical solution:
 \be \begin{split}\label{almostthoofttext}
 &-i \partial_+ \, W(p,q;\theta,\phi) - \frac{m^2}{4} \( \frac{1}{p}+ \frac{1}{ q} \) \, W(p,q;\theta,\phi)\\
 &- \frac{\lambda (-i)^n n v_n}{4 \pi} \int_0^{p-q}\dd{k} I_n(k,q) \(W(k, q-k,\theta,\phi) \cos(\theta-\phi) - W(p,q;\theta,\phi)\) =0 
 \period\end{split}\ee

In order to get an equation for a gauge-invariant quantity, we take the trace over the representation indexes:
\be
\phi(x) \equiv \phi(x,1-x) := \int\dd{\cos\theta}W (x,1-x;\theta,\theta)\comma\qquad x := \frac{p}{r}\comma \quad q:= p-r \period
\ee
Recall the the functions $\phi(x)$ retain a residual dependence on the $x_+$ component. Going into its Fourier space $\phi(x; p_+) \equiv \phi(x)$, and restoring the $-$ label on the momenta, \eqref{almostthooft} takes the form:
\be 
4 r_+ r_- \phi(x) = m^2 \( \frac{1}{x}+ \frac{1}{1-x}\) \phi(x) - \frac{\lambda}{\pi} \int_0^{1}\dd{y} K(y , x;1-x,1-x) \(\phi(y)- \phi(x)  \)
\ee
that, is exactly equivalent to the generalized Bethe-Salpeter equation \eqref{bseq} \cite{Douglas_1994}, with $K$ and $\Gamma$ computed using the potential $V(B) = B^{\bar{n}}$. 

We can also obtain arbitrary potentials in the large representation limit (see also footnote \ref{footnotegen}), if we scale the coefficients $v_n$ in the potential with appropriate powers of $J$. Specifically, once a double scaling limit of the YM coupling $g^2$ and $J$ is fixed as in \eqref{doublescaling}: $4 (2\pi)^{\bar{n}-1}\, g^{2} J^{\bar{n}} = \lambda$,  imposing that of each of the other coefficients scales as $v_m = \cO(J^{\bar{n}-m})$ at large $J$ guarantees that all the combinations \be 4 (2\pi)^{m-1}\ g^2 v_m J^m = \lambda_m\ee are finite and fixed in the limit $J\to \infty$. 
Henceforth, at large spin $J$, an arbitrary potential (that now depends also explicitly on $J$), $V(B,J)$ leads to a mass spectrum $\mu^2$ of mesons determined by the same generalized 't Hooft equation \eqref{thoofteq}:
\be
\label{thoofteqlargesin}
    2\pi \mu^2 \phi(x) =  \(m^2 - \sum^\infty_{n=1} \lambda_{2n}\, \alpha_{2n}\) \(\frac{1}{x} + \frac{1}{1-x}\) \phi(x) - \fint_0^1\dd{y}\,\frac{\sum_{n=1}^{\infty} \lambda_{2n}\, p_{2n}(\Lambda)}{(x-y)^2} \, \phi(y) \comma
\ee
where the summation is only over the even coefficients as a consequence of fixing the specific case of the $SU(2)$ algebra.

This implies that the all the results we have derived for the meson spectra at large $N$ are equally valid in this large representation limit. This generalizes the result of \cite{Kaushal:2023ezo}, obtained for the specific case of standard QCD$_2$, corresponding to setting $ v_n = \frac{1}{8\pi} \delta_{n,2}$. In particular, also in this limit, the theory exhibit the integrable structure. \section{Conclusion}\label{sec:conclusion}
In this paper, we studied the meson spectrum of the generalized QCD$_2$ at large $N_c$. We demonstrated that the integral equation determining the meson spectrum can be recast into the TQ-Baxter equation. The end result \eqref{eq:TQBaxter} turned out to be remarkably simple and universal; the dependence on the potential of the generalized Yang-Mills $V(B)$ is contained entirely in the form of the transfer matrix $T(\nu)$ that can be computed explicitly for any given potential. Based on the reformulation, we analyzed the spectrum as a function of the coefficients of $V(B)$ and uncovered a multi-sheeted structure with infinitely many branch points, each of which signals the emergence of a massless meson. These results extend previous findings for the 't Hooft model  \cite{Ambrosino:2023dik, Fateev_2009} to much a broader class of theories. There are numerous future directions worth exploring, many of which have already been mentioned in our previous paper \cite{Ambrosino:2023dik}. Below we list a few more which were not mentioned there:
\begin{itemize}
\item We showed that the spectrum of the generalized QCD$_2$ contains infinitely many branch points in the complex plane of quark masses. As is the case with the 't Hooft model, it is likely that these branch points correspond to physical critical points of the theory at which the infrared phase is described by a conformal field theory. It would be interesting to see if these branch points persist at finite $N_c$ and, if so, to understand what their precise CFT description is by extending the analysis of \cite{Delmastro:2021otj,Delmastro:2022prj}.
\item As already mentioned multiple times, it would be interesting to study theories with quarks in the adjoint representation of the gauge group and analyze the spectrum of glueballs and confining fluxtubes. Developing analytic approaches to these theories is a challenging yet important direction. Another direction worth pursuing is to perform the numerical analysis for the generalized Yang-Mills coupled to adjoint quarks, extending the existing results in the literature for the adjoint QCD$_2$ \cite{Bhanot:1993xp, Demeterfi:1993rs, Katz:2013qua, Katz:2014uoa, Dempsey:2021xpf, Dempsey_2023, Dempsey:2023fvm}.
\item It would be interesting to find examples of four-dimensional gauge theories for which the meson spectrum can be computed exactly. One promising target is $\mathcal{N}=4$ supersymmetric Yang-Mills (SYM) theory on the Coulomb branch. Although $\mathcal{N}=4$ SYM is conformal, going to the Coulomb branch allows us introduce massive excitations i.e.~W-bosons. When the symmetry breaking pattern is $U(N)\to U(N-1)\times U(1)$, these W-bosons transform under the fundamental representation of the residual $U(N-1)$ gauge group and form stable mesonic bound states in the large $N$ limit. As shown in \cite{Caron-Huot:2014gia} based on a judicious use of the Regge theory, a part of the spectrum of these bound states can be determined from the cusp anomalous dimension, which in turn can be computed exactly using the planar integrability of $\mathcal{N}=4$ SYM \cite{Beisert:2010jr}. More recently, direct evidence for the integrability on the Coulomb branch was obtained in \cite{Ivanovskiy:2024vel}. It would be interesting to explore this direction further and determine the full spectrum of these mesonic bound states more directly from integrability. The first step may be to derive the TQ-Baxter equation governing the spectrum of these bound states at weak 't Hooft coupling. Also interesting would be to find a similar setup in the fishnet limit of $\mathcal{N}=4$ SYM \cite{Zamolodchikov:1980mb,Gurdogan:2015csr,Caetano:2016ydc,Loebbert:2020hxk,Loebbert:2020tje}.
\item There is another potential point of contact between the analysis of this paper and four-dimensional gauge theories. As discussed in Section 2.2 of \cite{Alday:2007mf}, the dynamics in the large spin sector of four-dimensional Yang-Mills theory with (massless) dynamical quarks can be described by an effective two-dimensional QCD with higher derivative corrections, when the gauge coupling is sufficiently small (but finite). Thus, it is interesting to extend our analysis and make predictions on the large-spin dynamics of four-dimensional gauge theories.  
\end{itemize}
	\subsection*{Acknowledgement} We thank Bruno Balthazar, Aleksey Cherman, Gabriel Cuomo, Diego Delmastro, Sergei Dubovsky, John Donahue, Matijn Fran\c cois, Jaume Gomis, Zohar Komargodski,  Sergei Lukyanov, Marcos Mari\~no, Giuseppe Mussardo, Alessio Miscioscia, Alba Grassi, Yifan Wang, Alexander Zamolodchikov for useful discussions.
	
	\appendix
 
\section{Recurrence relation for integrals}
\label{app:rec}
In this Appendix we provide a more in-depth proof for the formula \eqref{recintegral}. The starting point is the observation that the principal part integrals:
\be 
   J_{n+2} =   \fint_0^1 \frac{\(\frac{y}{1-y}\)^{\frac{i \nu }{2}} \(\log \(\frac{1-y}{y}\)\)^{n}}{(x-y)^2}\comma
\ee
can be computed as Fourier transforms of distributions, once represented through rapidity variables $\theta = \frac{1}{2}\log(\frac{y}{1-y})$:
\be \begin{split}
   J_{n+2} &= \infint \dd{\theta} e^{i\theta \nu}\( -2 \frac{\cosh^2(\theta)} { \sinh^2(\theta-\varphi)}    \) (2\theta)^{n} = \(\frac{2}{i}\)^n  \dv[n]{\nu} J_2 \\
   &=  \(\frac{2}{i}\)^n  \dv[n]{\nu}  \left[\(\frac{\(\frac{x}{1-x}\)^{\frac{i \nu }{2}}}{\frac{2}{\pi}x(1-x)} \) \, \cI_2 \right] \\
   &= \(\frac{2}{i}\)^n \(\frac{\(\frac{x}{1-x}\)^{\frac{i \nu }{2}}}{\frac{2}{\pi}x(1-x)} \)  \dv[n]{\nu} \cI_2  + \sum_{p=0}^{n-1} (-1)^{n-p+1} \binom{n}{p}  \log^{n-p}\(\frac{x}{1-x}\)\, J_{p}
   \end{split}
\ee
where $\cI_2$ is the one defined in \eqref{integrals}. This can be 
equivalently expressed as:
\be \label{almostdone}
\(\frac{2}{i}\)^n \dv[n]{\nu} \cI_2 = \(\frac{\(\frac{x}{1-x}\)^{\frac{i \nu }{2}}}{\frac{2}{\pi}x(1-x)} \)^{-1}\sum_{p=0}^{n} (-1)^{n-p} \binom{n}{p}  \log^{n-p}\(\frac{x}{1-x}\)\, J_{p+2}\comma
\ee
but, using that:
\be 
\fint_0^1 \frac{\(\frac{1-y}{y}\)^{\frac{i \nu }{2}} \(\log \(\frac{1-y}{y}\)-\log \(\frac{1-x}{x}\)\)^{n}}{(x-y)^2} = \sum_{p=0}^{n} (-1)^{n-p} \binom{n}{p}  \log^{n-p}\(\frac{1-x}{x}\)\, J_{p+2}\comma
\ee
from \eqref{almostdone} we obtain \eqref{recintegral}:
\be 
\(\frac{2}{i}\)^n \dv[n]{\nu} \cI_2 = \cI_{n+2} \period
\ee

\section{On conjectural formula for the asymptotic expansion}
\label{app:spectraldet}
Here we present an explicit form of the conjectural formula for the asymptotic expansion of the spectrum:
\be
\begin{split}
   & n = 2 \lambda_n  - \frac{2 \alpha  \log (2\lambda_n)}{\pi ^2} + \frac{\alpha^2}{\pi ^4 \lambda_n } - \frac{3}{4} - \frac{2 \alpha  \log (4\pi e^{\gamma_E})}{\pi ^2} +\frac{1}{\pi^3}\,\cI(\alpha,v_3)\\&+\frac{1}{2\pi^6\lambda^2_n}\(\alpha^3 + (-1)^n\pi^2 (\alpha+1) + (-1)^n v_3^2 \left(C_1 + \frac{288 \log ^2(\lambda_n)}{\pi ^2}+ C_2\log (\lambda_n ) \right) \) + \mathcal{O}(\lambda^{-3}_n)\comma
\end{split}
\end{equation}
where the constants are given by
\begin{equation}
\begin{split}
 C_1 &=\displaystyle -288 (2\alpha+3)+\frac{(\alpha+1)}{\pi^2} \Big[144 (2 \gamma_E ^2+1+2 \log ^2(2)+\log (4)+\\
&\qquad\qquad +\gamma_E(2+4 \log (2 \pi ))+2 \log (\pi ) (1+\log (4 \pi )\Big]\comma\\
C_2 &= \frac{288 (2 \gamma_E +1+\log (4)+2 \log (\pi )) }{\pi ^2} \period
\end{split}
\ee 
\small
\be \begin{split}
\label{intdef}
\cI(\alpha,v_3) &= \Re \infint\dd{\nu} \Bigg[\frac{\alpha^2 \(\sinh(\pi\nu) - \pi\nu \)}{\frac{\pi\nu}{2} \cosh^2\(\frac{\pi\nu}{2}\)  f(\nu)}+ \frac{24 \pi ^2 \alpha  v_3 \left(\pi ^2 \nu ^2-2 \cosh (\pi  \nu )+2 \pi  \nu  \coth \left(\frac{\pi  \nu }{2}\right)-2\right)} { \frac{\pi\nu}{2} \cosh^2\(\frac{\pi\nu}{2}\)  f(\nu)} + \\ 
&\qquad\qquad\quad\quad \quad  + \frac{72 \pi ^4 v_3^2 \left(-2 \pi ^2 \nu ^2 \cosh (\pi  \nu ) +\cosh (2 \pi  \nu )-1\right) \text{csch}^4\left(\frac{\pi  \nu }{2}\right) }{ \frac{\pi\nu}{2} \cosh^2\(\frac{\pi\nu}{2}\)  f(\nu) } \Bigg] \period
\end{split}
\ee\normalsize
\section{Derivation of the bound state equation at large representation}
\label{appboundstatelargespin}
In this appendix we follow closely \cite{DHAR_1993,DHAR_1994, Kaushal:2023ezo}; we do not repeat many of the technical points that are carefully explained in Appendix D of \cite{Kaushal:2023ezo} to which we refer the reader for more details. 

The starting point is the action for a field for $\bs{M}$, that can be borrowed from the standard theory of quantization of the coadjoint orbits by adding a {\it kinetic term} $S_{\rm kin}(\bs{M})$\footnote{The details of this are not important for our discussion, we refer the interested reader to \cite{Kaushal:2023ezo}. } to the Hamiltonian $\cH_{\rm eff}$:
\be \begin{split}\label{action}
&S\left[\bs{M}\right] = S_{\rm kin}(\bs{M}) - \frac{i m^2}{8}\int\dd{x^+}\dd{y_1^-} \tr  \bs{S} \bs{M} \quad  \\ &+\frac{\lambda v_n}{8} \int\dd{x^+}\int \prod_{i=1}^{\bar{n}}\(\dd{y_i}\sgn{(y_k - y_{k-1})}\) \,{\rm tr} \, \(\bs{M}(y_1,y_2) \bs{\widehat{M}}(y_2,y_3) \bs{M}(y_3,y_4) \cdots  + {\rm perm.\  }\)\comma
\end{split}\ee
where $S(x,y,\theta,\phi) = \sgn(x_- - y_-) \frac{1}{\sin(\theta)} \delta(\theta - \phi)$. This generalizes the action used in \cite{Kaushal:2023ezo} only in the last term in \eqref{action}.

We chose a coadjoint orbit with vanishing $U(1)$ charge, of whose a representative is $M_0(p,q,\theta,\phi) = 2\pi \Theta(p_-) \delta(p_- - q_-) \frac{1}{\sin\theta}\delta(\theta - \phi)$, where \be M(p,q,\theta,\phi) = \infint \dd{x_-}\dd{y_-} e^{i p x_-} e^{i q y_-} M(x,y,\theta,\phi)\period\ee Fluctuation $\bs{W}$ around the coadjoint orbit are be parametrized as:  (eq.\ 3.26 of \cite{Kaushal:2023ezo}):
\be \label{variation}
\bs{M}= \bs{M_0} + \frac{1}{\sqrt{J}} [\bs{W}, \bs{M_0}] - \frac{1}{2J} [\bs{W}, [\bs{W},\bs{M_0}]] + \cO(\ts J^{-3/2}) \period
\ee
The action \eqref{action} is order $J$  , and therefore  we must take the  $\cO(\ts 1/J)$ term under the variation \eqref{variation} to get the finite leading order action for the fluctuations. The first two terms of  \eqref{action}, are directly borrowed the result from Appendix D of \cite{Kaushal:2023ezo}:
\be \begin{split}
\delta S_{\rm kin} &= i  \frac{1}{J} \int\dd{x^+}\int_{p,q >0}\dd{p}\dd{q} {\rm tr} \, \bs{W}^{+-}(p,q) \partial_+\bs{W}^{-+}(q,p)\comma\\
\delta S_{\rm mass} &= - \frac{m^2}{4J}\int_{k,q >0}\dd{p}\dd{q} \( \frac{1}{k} + \frac{1}{p} \) \tr \bs{W}^{+-}(k,p) \bs{W}^{ -+}(p,k) \comma
\end{split}
\ee
where, for clarity of comparison, we use the notation\footnote{The same of \cite{Kaushal:2023ezo} up to $+(-)\leftrightarrow p(n)$. } : $W^{ \pm \pm}(p,q) = W(\pm p,\pm q)$. The interaction term of $H_{\rm eff}$ receives three contributions (up to permutations of $\theta_i$ ) at order $\cO(1/J)$. Schematically:
\be 
{\rm I}: \hspace{1 mm } \bs{M_0} \frac{1}{2J} \, \bs{\widehat{W W}} \,\(\underbrace{\bs{M_0}\, \bs{\widehat{M_0}}\cdots }_{(\bar{n}-2) \, {\rm terms}}\)\comma \quad {\rm II}: \hspace{1 mm } \widehat{\bs{M_0}} \frac{1}{2J} \, \bs{W W} \,\(\underbrace{\bs{M_0}\, \bs{\widehat{M_0}}\cdots }_{(\bar{n}-2) \, {\rm terms}}\) \simeq {\rm I}\comma
\ee
the ${\rm II}$ and the ${\rm I }$ are equal contributions up to permutations of the $\theta_i$s; a third, inequivalent contribution is:
\be 
{\rm III}: \hspace{1 mm} \frac{1}{\sqrt{J}} \bs{W} \frac{1}{\sqrt{J}} \bs{\widehat{W}} \,\(\underbrace{\bs{M_0}\, \bs{\widehat{M_0}}\cdots }_{(\bar{n}-2) \, {\rm terms}}\)
\ee
All this contribution to the variation of \eqref{action} can be computed as a generalization of Appendix D of \cite{Kaushal:2023ezo}. The contributions {\rm I, II} are evaluated to\footnote{The form of the integral over the $k_i$ can be readily obtained (for instance) from a direct generalizing eq.\  D.17 of \cite{Kaushal:2023ezo}: in each of the two integral the sign can be recognized to be $\sgn(k)$, and $1/k^2$ (Fourier transform of $\abs{x}$) is replaced for each term of  our problem with $\frac{i}{(k_{i-1}-k_i)}\frac{i}{(k_i-k_{i+1})}$  (Fourier transform of $\sgn(x)$), with $k_0 = k$, $k_{\bar{n}} = q$}:
\small
 \be 
-i \int_{p,q>0}\dd{p}\dd{q}\int_0^{p-q}\dd{k} \int\dd[\bar{n}-1]{\bs{k}} \frac{-i\sgn(k_1)}{2\pi (k- k_1)}\,  \frac{-i\sgn(k_2)}{2\pi (k_1 - k_2)} \cdots \frac{-i\sgn(k_{\bar{n}-1})}{2\pi (k_{\bar{n}-1} - k_1)} {\rm tr}\,\bs{W}^{-+}(p,q) \bs{W}^{+-}(q,p)
 \ee\normalsize
all the $\bar{n}+1$ integrations over $\dd{p}\dd{q}$ but the two remaining above over $\dd{p}\dd{q}$ are removed by the $\bar{n}-3$ $\delta(\theta_i - \theta_{i+1})$ inside each of the $\bs{M_0}$, this also cancel all the $\cos(\theta_i - \theta_{i+1})$ trigonometric factors. the integral over the $k_i$ is eactly the \textit{master integral} \eqref{masterintegral} introduced in section \ref{sec:thoofteq}:
 \be 
\int_{p,q>0}\dd{p}\dd{q}\int_0^{p-q}\dd{k} I_n(k,q) {\rm tr}\, \bs{W}^{-+}(p,p) \bs{W}^{+-}(q,p)\period
 \ee
Analogously, the third contribution is evaluated as:
\be 
\int_{p,q>0}\dd{p}\dd{q}\int_0^{p-q}\dd{k} I_n(k,q) {\rm tr}\, \left[\bs{W}^{+-}(q-k,q) \bs{\widehat{W}}^{-+}(p,q) + \bs{W}^{-+}(k,q-k) \bs{\widehat{W}}^{+-}(q,p)\right] \period
\ee
Once we sum over all the permutations of  ${\rm I}$, ${\rm II}$ and ${\rm III}$, taking into account the combinatorial factor $\frac{1}{n!}$, we obtain the following expression for the interaction term of $\cH_{\rm eff}$ at order $\cO(\ts 1/J)$:
\be \begin{split}
\frac{-\lambda}{4J}v_n(-i)^{\bar{n}} \bar{n}  \int_{q,p>0}\dd{p}\dd{q} \int_0^{p-q}\dd{k} I_n(k,q) \, {\rm tr}\, \bigg[ &\( \bs{\widehat{W}}^{+-}(q-k,k) - \bs{W}^{+-}(q,p) \) \bs{W}^{-+}(p,q) + \\
&\(\bs{\widehat{W}}^{-+}(k,q-k) - \bs{W}^{-+}(p,q) \) \bs{W}^{+-}(q,p) \bigg]
\end{split}\ee
Varying the action w.r.t.\ $W^{+-}(p,q;\theta,\phi)$ we obtain the following equation of motions:
\be \begin{split}\label{almostthooft}
&-i \partial_+ \, W^{-+}(p,q;\theta,\phi) - \frac{m^2}{4} \( \frac{1}{p}+ \frac{1}{ q} \) \, W^{-+}(p,q;\theta,\phi)\\
&- \frac{\lambda (-i)^n n v_n}{4 \pi} \int_0^{p-q}\dd{k} I_n(k,q) \(W^{-+}(k, q-k,\theta,\phi) \cos(\theta-\phi) - W^{-+}(p,q;\theta,\phi)\) =0 
\period\end{split}\ee
In the main text we report \eqref{almostthooft} by replacing $W^{-+} \mapsto W$ in
\eqref{almostthoofttext} to simplify the notation.

\section{Numerical methods for analytical continuation}
\label{appendixnumerical}
In this appendix we illustrate the numerical method we employ to solve numerically the 't Hooft equation in $\nu$-space \eqref{eq:eignuspace} in the complex space of the couplings. As discussed in the main text (cfr.\ Section  \ref{analiticalpr}), the integral eigenvalue problem exhibits a multi-sheeted sctructure in the complex plane, arising from collision-of-roots singularities inside the kernel of \eqref{eq:eignuspace} due to the poles of $\Psi(\nu)$. Compared to the same integral equation in $x$-space, the problem in Fourier space, is particularly convenient to solve numerically even in the complex space of the couplings. To solve this eigenproblem numerically, we employ an orthonormal basis of Hermite polynomials:
\be 
\bra{n}\ket{x} = H_n(x) = \frac{ (-1)^n e^{\frac{x^2}{2}}}{\sqrt{ 2^n \pi^{1/2} n!}}  \dv[n]{x} e^{-x^2}\comma \qquad  \bra{m}\ket{n} = \delta_{nm}\comma
\ee
that share many properties with the eingefunction $\Psi_n(x)$ having exactly $n$ zeros and definite parity in $x\in (-\infty, + \infty)$.

In this basis the generalized 't Hooft equation \eqref{eq:eignuspace}reduces to the eingeproblem for the following ``Hamiltonian'':
\be \label{matrixham}
\bra{m}\cH \ket{n} = \cH_{mn} = \infint\dd{x}\infint\dd{y}  \frac{H_m(x) H_n(y)}{f(x)} \frac{\pi(x-y)}{2\sinh(\frac{\pi}{2}(x-y))}\comma
\ee
that can be straightforwardly solved upon truncation of the basis. The eigenvalues $e_k$ of $\cH_{nm}$ are related to the masses of the mesons $\lambda_k$ by inversion: $\lambda_k= e_k^{-1}$.

This numerical method is particularly akin to analytical continuation. Indeed, whenever upon analytically  continuation of the couplings $\alpha, \bs{v}$, one (or more) of the zeros of $f(\nu)$ crosses the integration contour, one can just modify the Hamiltonian $\cH_{mn}$ simply by including the corresponding residue terms. 

To exemplify, consider the case of $V(B) = \frac{1}{8\pi} B^2 + i \abs{v_3} B^3$. Fixed any $v_3>0$, for $\alpha > -1$ no zeros of $f(\nu)$ lie on the integration contour, while upon analytically continue in $\alpha$ to values $\alpha < -1$, exactly one of the zero crosses the real axis, and therefore the contour has to be deformed to avoid this singular point as in \figref{fig:contourapp}.  Then,  the $\dd{x}$ integral in \eqref{matrixham} will pick a further contribution due a residue term\footnote{Note that the sign of the residue comes from the contour orientation in \figref{fig:contourapp}}:
\be 
\cH_{mn} \to \cH_{mn} - 2\pi i H_m(x_0) \Res\left[{f^{-1}(x,\alpha,v_3)}, x = x_0 \right]\infint\dd{y}  H_n(y)\frac{\pi(x_0-y)}{2\sinh(\frac{\pi}{2}(x_0-y))}\comma
\ee
where $x_0 = x_0(\alpha,v_3)$ is the position of the zero of $f(x,\alpha,v_3)$ that has crossed the integration contour.
\begin{figure}
    \centering
    \includegraphics{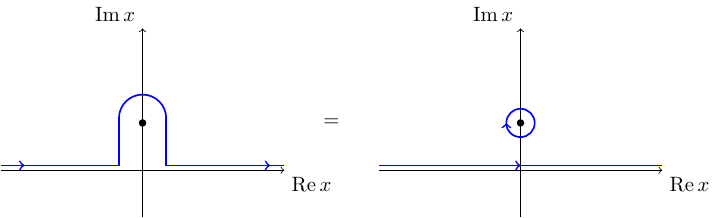}
    \caption{The integration contour in the $x$-plane of \eqref{matrixham} is deformed under analytical continuation from $\alpha>-1$ to $\alpha < -1$ at fixed $-i v_3 >0$.}
    \label{fig:contourapp}
\end{figure}

Analogous manipulations make straightforward the numerical analysis of \eqref{eq:eignuspace} among the various sheets in the couplings-space. In practice, already truncating the basis at $n_{max}\sim 30$, reproduces the first 4 eigenvalues of \cite{Fateev_2009} up to an accuracy of $10^{-5}$, that used a basis of $400$ elements to diagonalize the  $x$-space for teh simplest case of $\bs{v}= \bs{0}$ and $\alpha = 0$. This is yet another illustration of the efficiency of solving the $\nu$-space problem, rather than the original one in $x$-space, also at the level of numerical methods. 
\section{\boldmath Non-critical spontaneous symmetry breaking of $\cP\cT$-symmetry}
\label{noncriticalbreaking}
\begin{figure}
    \centering
    \includegraphics[width=0.5\linewidth]{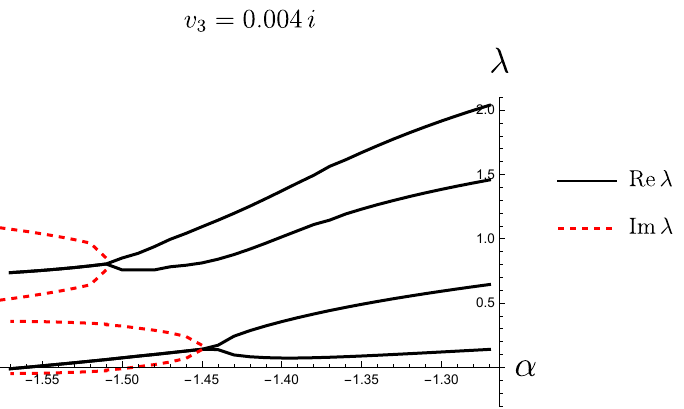}
    \caption{First four eigenvalues for $v_3 = 0.004 i$ varying $\alpha$. The non-critical $\cP\cT$ symmetry is associated to the first two eigenvalues becoming complex-conjugated pairs without turning to zero. To facilitate reading this plot, the imaginary parts of the eigenvalues are displayed shifted by the corresponding real values at the degenerated point.}
    \label{noncriticassb}
\end{figure}
In this appendix, we study the meson spectrum associated with a cubic generalized QCD potential, $V(B) = \frac{1}{8\pi} B^2 + v_3 B^3$, in the region of the parameter space obtained by analytically continuing around the cubic-root branch cut located at $\alpha = \alpha^{(\rm III)}$. We perform this analytical continuation along a path lying at imaginary values of $v_3$. This guarantees that the $\cP\cT$ symmetry is preserved all along the path. As already stated in the main text, whenever $\cP\cT$ symmetry is not spontaneously broken by the eigenfunctions, the spectrum is real. This is the case for values of $\alpha > \alpha^{(\rm{ III})}$ and  $i \mathbb{R} \ni v_3 > v_3^{(\rm II)}$, i.e.\ in the entire region bounded from  below by the critical line in \figref{fig:tricriticalpoints}. Another possibility is that $\cP\cT$ symmetry could be broken \textit{spontaneously} by the eigenfunctions, i.e.\ \textit{some} of the eigenfunctions and eigenvalues become complex conjugate pairs. As we have already explained in Section \ref{singbey} (cfr. \figref{fig:phasesunbroken}) in our model this happens in correspondence with the critical points at $v_3^{(\rm II)}$. Crossing the branch cut associated to any of the critical points leads into a spontaneously broken phase of $\cP\cT$-symmetry. At this special point the massless state become doubly degenerated with a non-normalizable eigenstate\footnote{The argument follows from the analogous discussion in section 6.3.1 of \cite{Ambrosino:2023dik}. }. At fixed $\alpha$, the pair becomes complex conjugated as we cross the branch cut beyond $v_3 < v_3^{(\rm{II} )}(\alpha)$. 

These critical points shall be regarded as associated to the spontaneous symmetry breaking of $\cP\cT$ symmetry, so that the \textit{critical} $\cP\cT$ breaking is nothing but a usual incarnation of the Landau paradigm. 
\begin{figure}
    \centering
    \includegraphics[width=0.7\linewidth]{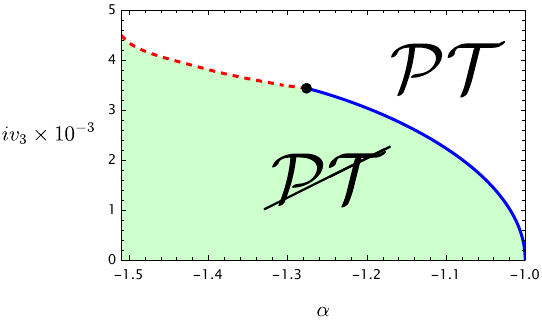}
    \caption{Critical and non-critical $\cP\cT$ symmetry breaking in the $(\alpha,  v_3\in i \bR)$ plane. In the region shaded in green $\cP\cT$ symmetry is spontaneously broken by the eigenfunctions and the spectrum is populated by complex-conjugate eigenvalues. The blue line is the line of critical points where the first meson mass turns to zero, as computed analytically from the singularities of $f(\nu)$. The dashed red line is the line of spontaneous non-critical $\cP\cT$ breaking determined numerically. The dot is located at the tricritical transition.}
    \label{ssbphase}
\end{figure}
More subtle is what happens beyond the tricritical point $(\alpha^{(\rm III)}, v^{(\rm III)}_3)$, where the line of critical points is split to complex values of $v_3$ and does not lie anymore at $\cP\cT$-symmetric values of the parameters. Henceforth, $\cP\cT$ symmetry cannot be broken spontaneously at a critical points: it is rather broken\footnote{Yet, one should expect that ssb breaking of $\cP\cT$-symmetry must happen at some point in the parameter space: as if this would not be the case, one could find a continuous path, from the $\cP\cT$-symmetric phase at $(\alpha > \alpha^{(\rm II)}, v_3 > v_3^{(\rm II)})$ to the ssb one at values $v_3 < v^{(\rm II)}_3$ without crossing any branch cut. }\textit{non-critically} \cite{Lencses:2024wib,Lencses:2023evr}. 
 Non-critical $\cP\cT$-breaking happens when two of the meson masses degenerate and become complex conjugate. In this case the transition to a spontaneously broken phase of $\cP\cT$-symmetry is not associated with a critical point as there are no massless modes in the spectrum\footnote{Interestingly, the  same phenomenon was also recently illustrated for the non-unitary version of tricritical Ising model (the minimal model M(2,7) ) in \cite{Lencses:2024wib,Lencses:2023evr}. Generalized YM shares many features with the the multicritical minimal models of the Ising or YL series.}. The typical situation is reported in \figref{noncriticassb}. At the non-critical breaking point, the lowest two eigenvalues (and eigenfunctions) degenerate and become complex conjugate. Upon delving deepen in the ssb broken phase, more and more states become degenerate and and turn into complex conjugate pairs, exhibiting non-critical $\mathcal{PT}$-breaking.

 These non-critical transitions cannot be diagnosed by the singularities of the  (inverse of)  tranfer matrix $f(\nu)$. Yet one can compute numerically the eigenvalues for a large range of $\alpha$ and $v_3$ to scan\footnote{In practice, one fixes $v_3$ and scans over all the $\alpha$'s until the first two level become complex conjugate. Repating this over $v_3$ leads to the figure.} where the first two levels  undergo a non-critical phase transition. This results in the phase diagram reported in \figref{ssbphase}. This completes the phase space of  \figref{fig:tricriticalpoints}.


\pdfbookmark[1]{\refname}{references.bib}
\bibliographystyle{JHEP}
\bibliography{references}

\providecommand{\href}[2]{#2}\begingroup\raggedright\begin{thebibliography}{10}

\bibitem{Ambrosino:2023dik}
F.~Ambrosino and S.~Komatsu, {\it {2d QCD and Integrability, Part I: 't Hooft model}},  \href{http://arxiv.org/abs/2312.15598}{{\tt arXiv:2312.15598}}.

\bibitem{Fateev_2009}
V.~A. Fateev, S.~L. Lukyanov, and A.~B. Zamolodchikov, {\it {On mass spectrum in 't Hooft's 2D model of mesons}},  {\em J. Phys. A} {\bf 42} (2009) 304012, [\href{http://arxiv.org/abs/0905.2280}{{\tt arXiv:0905.2280}}].

\bibitem{Delmastro:2021otj}
D.~Delmastro, J.~Gomis, and M.~Yu, {\it {Infrared phases of 2d QCD}},  {\em JHEP} {\bf 02} (2023) 157, [\href{http://arxiv.org/abs/2108.02202}{{\tt arXiv:2108.02202}}].

\bibitem{Delmastro:2022prj}
D.~Delmastro and J.~Gomis, {\it {RG flows in 2d QCD}},  {\em JHEP} {\bf 09} (2023) 158, [\href{http://arxiv.org/abs/2211.09036}{{\tt arXiv:2211.09036}}].

\bibitem{Douglas_1994}
M.~R. Douglas, K.~Li, and M.~Staudacher, {\it {Generalized two-dimensional QCD}},  {\em Nucl. Phys. B} {\bf 420} (1994) 118--140, [\href{http://arxiv.org/abs/hep-th/9401062}{{\tt hep-th/9401062}}].

\bibitem{Litvinov:2024riz}
A.~Litvinov and P.~Meshcheriakov, {\it {Meson mass spectrum in QCD$_2$ 't Hooft's model}},  \href{http://arxiv.org/abs/2409.11324}{{\tt arXiv:2409.11324}}.

\bibitem{Dalley:1992yy}
S.~Dalley and I.~R. Klebanov, {\it {String spectrum of (1+1)-dimensional large N QCD with adjoint matter}},  {\em Phys. Rev. D} {\bf 47} (1993) 2517--2527, [\href{http://arxiv.org/abs/hep-th/9209049}{{\tt hep-th/9209049}}].

\bibitem{Kutasov:1993gq}
D.~Kutasov, {\it {Two-dimensional QCD coupled to adjoint matter and string theory}},  {\em Nucl. Phys. B} {\bf 414} (1994) 33--52, [\href{http://arxiv.org/abs/hep-th/9306013}{{\tt hep-th/9306013}}].

\bibitem{Boorstein:1993nd}
J.~Boorstein and D.~Kutasov, {\it {Symmetries and mass splittings in QCD in two-dimensions coupled to adjoint fermions}},  {\em Nucl. Phys. B} {\bf 421} (1994) 263--277, [\href{http://arxiv.org/abs/hep-th/9401044}{{\tt hep-th/9401044}}].

\bibitem{Bhanot:1993xp}
G.~Bhanot, K.~Demeterfi, and I.~R. Klebanov, {\it {(1+1)-dimensional large N QCD coupled to adjoint fermions}},  {\em Phys. Rev. D} {\bf 48} (1993) 4980--4990, [\href{http://arxiv.org/abs/hep-th/9307111}{{\tt hep-th/9307111}}].

\bibitem{Demeterfi:1993rs}
K.~Demeterfi, I.~R. Klebanov, and G.~Bhanot, {\it {Glueball spectrum in a (1+1)-dimensional model for QCD}},  {\em Nucl. Phys. B} {\bf 418} (1994) 15--29, [\href{http://arxiv.org/abs/hep-th/9311015}{{\tt hep-th/9311015}}].

\bibitem{Smilga:1994hc}
A.~V. Smilga, {\it {Instantons and fermion condensate in adjoint QCD in two-dimensions}},  {\em Phys. Rev. D} {\bf 49} (1994) 6836--6848, [\href{http://arxiv.org/abs/hep-th/9402066}{{\tt hep-th/9402066}}].

\bibitem{Lenz:1994du}
F.~Lenz, M.~A. Shifman, and M.~Thies, {\it {Quantum mechanics of the vacuum state in two-dimensional QCD with adjoint fermions}},  {\em Phys. Rev. D} {\bf 51} (1995) 7060--7082, [\href{http://arxiv.org/abs/hep-th/9412113}{{\tt hep-th/9412113}}].

\bibitem{Katz:2013qua}
E.~Katz, G.~Marques~Tavares, and Y.~Xu, {\it {Solving 2D QCD with an adjoint fermion analytically}},  {\em JHEP} {\bf 05} (2014) 143, [\href{http://arxiv.org/abs/1308.4980}{{\tt arXiv:1308.4980}}].

\bibitem{Katz:2014uoa}
E.~Katz, G.~Marques~Tavares, and Y.~Xu, {\it {A solution of 2D QCD at Finite $N$ using a conformal basis}},  \href{http://arxiv.org/abs/1405.6727}{{\tt arXiv:1405.6727}}.

\bibitem{Cherman:2019hbq}
A.~Cherman, T.~Jacobson, Y.~Tanizaki, and M.~\"Unsal, {\it {Anomalies, a mod 2 index, and dynamics of 2d adjoint QCD}},  {\em SciPost Phys.} {\bf 8} (2020), no.~5 072, [\href{http://arxiv.org/abs/1908.09858}{{\tt arXiv:1908.09858}}].

\bibitem{Komargodski:2020mxz}
Z.~Komargodski, K.~Ohmori, K.~Roumpedakis, and S.~Seifnashri, {\it {Symmetries and strings of adjoint QCD$_{2}$}},  {\em JHEP} {\bf 03} (2021) 103, [\href{http://arxiv.org/abs/2008.07567}{{\tt arXiv:2008.07567}}].

\bibitem{Dempsey:2021xpf}
R.~Dempsey, I.~R. Klebanov, and S.~S. Pufu, {\it {Exact symmetries and threshold states in two-dimensional models for QCD}},  {\em JHEP} {\bf 10} (2021) 096, [\href{http://arxiv.org/abs/2101.05432}{{\tt arXiv:2101.05432}}].

\bibitem{Popov:2022vud}
F.~K. Popov, {\it {Supersymmetry in QCD2 coupled to fermions}},  {\em Phys. Rev. D} {\bf 105} (2022), no.~7 074005, [\href{http://arxiv.org/abs/2202.04017}{{\tt arXiv:2202.04017}}].

\bibitem{Dempsey_2023}
R.~Dempsey, I.~R. Klebanov, L.~L. Lin, and S.~S. Pufu, {\it {Adjoint Majorana QCD$_{2}$ at finite N}},  {\em JHEP} {\bf 04} (2023) 107, [\href{http://arxiv.org/abs/2210.10895}{{\tt arXiv:2210.10895}}].

\bibitem{Dempsey:2023fvm}
R.~Dempsey, I.~R. Klebanov, S.~S. Pufu, and B.~T. S\o{}gaard, {\it {Lattice Hamiltonian for Adjoint QCD$_2$}},  \href{http://arxiv.org/abs/2311.09334}{{\tt arXiv:2311.09334}}.

\bibitem{Dempsey:2024ofo}
R.~Dempsey, I.~R. Klebanov, S.~S. Pufu, and B.~T. S\o{}gaard, {\it {Small Circle Expansion for Adjoint QCD$_2$ with Periodic Boundary Conditions}},  \href{http://arxiv.org/abs/2406.17079}{{\tt arXiv:2406.17079}}.

\bibitem{Dubovsky:2018dlk}
S.~Dubovsky, {\it {A Simple Worldsheet Black Hole}},  {\em JHEP} {\bf 07} (2018) 011, [\href{http://arxiv.org/abs/1803.00577}{{\tt arXiv:1803.00577}}].

\bibitem{Donahue:2019adv}
J.~C. Donahue and S.~Dubovsky, {\it {Confining Strings, Infinite Statistics and Integrability}},  {\em Phys. Rev. D} {\bf 101} (2020), no.~8 081901, [\href{http://arxiv.org/abs/1907.07799}{{\tt arXiv:1907.07799}}].

\bibitem{Donahue:2019fgn}
J.~C. Donahue and S.~Dubovsky, {\it {Classical Integrability of the Zigzag Model}},  {\em Phys. Rev. D} {\bf 102} (2020), no.~2 026005, [\href{http://arxiv.org/abs/1912.08885}{{\tt arXiv:1912.08885}}].

\bibitem{Donahue:2022jxu}
J.~C. Donahue and S.~Dubovsky, {\it {Quantization of the zigzag model}},  {\em JHEP} {\bf 08} (2022) 047, [\href{http://arxiv.org/abs/2202.11746}{{\tt arXiv:2202.11746}}].

\bibitem{Asrat:2022aov}
M.~Asrat, {\it {(1+1)D QCD with heavy adjoint quarks}},  {\em Phys. Rev. D} {\bf 107} (2023), no.~10 106022, [\href{http://arxiv.org/abs/2212.02162}{{\tt arXiv:2212.02162}}].

\bibitem{Beisert:2010jr}
N.~Beisert et~al., {\it {Review of AdS/CFT Integrability: An Overview}},  {\em Lett. Math. Phys.} {\bf 99} (2012) 3--32, [\href{http://arxiv.org/abs/1012.3982}{{\tt arXiv:1012.3982}}].

\bibitem{Kaushal:2023ezo}
A.~Kaushal, N.~S. Prabhakar, and S.~R. Wadia, {\it {Meson spectrum of $\text{SU}(2)$ QCD$_{1+1}$ with Quarks in Large Representations}},  \href{http://arxiv.org/abs/2307.15015}{{\tt arXiv:2307.15015}}.

\bibitem{Bourget:2018obm}
A.~Bourget, D.~Rodriguez-Gomez, and J.~G. Russo, {\it {A limit for large $R$-charge correlators in $\mathcal{N}=2$ theories}},  {\em JHEP} {\bf 05} (2018) 074, [\href{http://arxiv.org/abs/1803.00580}{{\tt arXiv:1803.00580}}].

\bibitem{Arias-Tamargo:2019xld}
G.~Arias-Tamargo, D.~Rodriguez-Gomez, and J.~G. Russo, {\it {The large charge limit of scalar field theories and the Wilson-Fisher fixed point at $\epsilon=0$}},  {\em JHEP} {\bf 10} (2019) 201, [\href{http://arxiv.org/abs/1908.11347}{{\tt arXiv:1908.11347}}].

\bibitem{Arias-Tamargo:2019kfr}
G.~Arias-Tamargo, D.~Rodriguez-Gomez, and J.~G. Russo, {\it {Correlation functions in scalar field theory at large charge}},  {\em JHEP} {\bf 01} (2020) 171, [\href{http://arxiv.org/abs/1912.01623}{{\tt arXiv:1912.01623}}].

\bibitem{Badel:2019oxl}
G.~Badel, G.~Cuomo, A.~Monin, and R.~Rattazzi, {\it {The Epsilon Expansion Meets Semiclassics}},  {\em JHEP} {\bf 11} (2019) 110, [\href{http://arxiv.org/abs/1909.01269}{{\tt arXiv:1909.01269}}].

\bibitem{Watanabe:2019pdh}
M.~Watanabe, {\it {Accessing large global charge via the $\epsilon$-expansion}},  {\em JHEP} {\bf 04} (2021) 264, [\href{http://arxiv.org/abs/1909.01337}{{\tt arXiv:1909.01337}}].

\bibitem{Giombi:2020enj}
S.~Giombi and J.~Hyman, {\it {On the large charge sector in the critical O(N) model at large N}},  {\em JHEP} {\bf 09} (2021) 184, [\href{http://arxiv.org/abs/2011.11622}{{\tt arXiv:2011.11622}}].

\bibitem{Caetano:2023zwe}
J.~a. Caetano, S.~Komatsu, and Y.~Wang, {\it {Large Charge 't Hooft Limit of $\mathcal{N}=4$ Super-Yang-Mills}},  \href{http://arxiv.org/abs/2306.00929}{{\tt arXiv:2306.00929}}.

\bibitem{Cuomo:2022xgw}
G.~Cuomo, Z.~Komargodski, M.~Mezei, and A.~Raviv-Moshe, {\it {Spin impurities, Wilson lines and semiclassics}},  {\em JHEP} {\bf 06} (2022) 112, [\href{http://arxiv.org/abs/2202.00040}{{\tt arXiv:2202.00040}}].

\bibitem{Aharony:2022ntz}
O.~Aharony, G.~Cuomo, Z.~Komargodski, M.~Mezei, and A.~Raviv-Moshe, {\it {Phases of Wilson Lines in Conformal Field Theories}},  {\em Phys. Rev. Lett.} {\bf 130} (2023), no.~15 151601, [\href{http://arxiv.org/abs/2211.11775}{{\tt arXiv:2211.11775}}].

\bibitem{Aharony:2023amq}
O.~Aharony, G.~Cuomo, Z.~Komargodski, M.~Mezei, and A.~Raviv-Moshe, {\it {Phases of Wilson Lines: Conformality and Screening}},  \href{http://arxiv.org/abs/2310.00045}{{\tt arXiv:2310.00045}}.

\bibitem{Rodriguez-Gomez:2022xwm}
D.~Rodriguez-Gomez and J.~G. Russo, {\it {Wilson loops in large symmetric representations through a double-scaling limit}},  {\em JHEP} {\bf 08} (2022) 253, [\href{http://arxiv.org/abs/2206.09935}{{\tt arXiv:2206.09935}}].

\bibitem{Beccaria:2022bcr}
M.~Beccaria, S.~Giombi, and A.~A. Tseytlin, {\it {Wilson loop in general representation and RG flow in 1D defect QFT}},  {\em J. Phys. A} {\bf 55} (2022), no.~25 255401, [\href{http://arxiv.org/abs/2202.00028}{{\tt arXiv:2202.00028}}].

\bibitem{Rodriguez-Gomez:2022gif}
D.~Rodriguez-Gomez and J.~G. Russo, {\it {Defects in scalar field theories, RG flows and dimensional disentangling}},  {\em JHEP} {\bf 11} (2022) 167, [\href{http://arxiv.org/abs/2209.00663}{{\tt arXiv:2209.00663}}].

\bibitem{CarrenoBolla:2023sos}
I.~Carre\~no Bolla, D.~Rodriguez-Gomez, and J.~G. Russo, {\it {RG flows and stability in defect field theories}},  {\em JHEP} {\bf 05} (2023) 105, [\href{http://arxiv.org/abs/2303.01935}{{\tt arXiv:2303.01935}}].

\bibitem{THOOFT1974461}
G.~{'t Hooft}, {\it A two-dimensional model for mesons},  {\em Nuclear Physics B} {\bf 75} (1974), no.~3 461--470.

\bibitem{Bender_1998}
C.~M. Bender and S.~Boettcher, {\it Real spectra in non-hermitian hamiltonians having pt symmetry},  {\em Physical Review Letters} {\bf 80} (jun, 1998) 5243--5246.

\bibitem{Bender_2005}
C.~M. Bender, {\it Introduction to pt-symmetric quantum theory},  {\em Contemporary Physics} {\bf 46} (jul, 2005) 277--292.

\bibitem{Bender:2004sa}
C.~M. Bender, D.~C. Brody, and H.~F. Jones, {\it {Extension of PT symmetric quantum mechanics to quantum field theory with cubic interaction}},  {\em Phys. Rev. D} {\bf 70} (2004) 025001, [\href{http://arxiv.org/abs/hep-th/0402183}{{\tt hep-th/0402183}}]. [Erratum: Phys.Rev.D 71, 049901 (2005)].

\bibitem{Lencses:2022ira}
M.~Lencs\'es, A.~Miscioscia, G.~Mussardo, and G.~Tak\'acs, {\it {Multicriticality in Yang-Lee edge singularity}},  {\em JHEP} {\bf 02} (2023) 046, [\href{http://arxiv.org/abs/2211.01123}{{\tt arXiv:2211.01123}}].

\bibitem{Lencses:2023evr}
M.~Lencs\'es, A.~Miscioscia, G.~Mussardo, and G.~Tak\'acs, {\it {$ \mathcal{PT} $ breaking and RG flows between multicritical Yang-Lee fixed points}},  {\em JHEP} {\bf 09} (2023) 052, [\href{http://arxiv.org/abs/2304.08522}{{\tt arXiv:2304.08522}}].

\bibitem{Lencses:2024wib}
M.~Lencs\'es, A.~Miscioscia, G.~Mussardo, and G.~Tak\'acs, {\it {Ginzburg-Landau description for multicritical Yang-Lee models}},  \href{http://arxiv.org/abs/2404.06100}{{\tt arXiv:2404.06100}}.

\bibitem{Eden:1966dnq}
R.~J. Eden, P.~V. Landshoff, D.~I. Olive, and J.~C. Polkinghorne, {\em {The analytic S-matrix}}.
\newblock Cambridge Univ. Press, Cambridge, 1966.

\bibitem{Pesando:1994cw}
I.~Pesando, {\it {Generalized QCD in two-dimensions via the bilocal method}},  {\em Mod. Phys. Lett. A} {\bf 9} (1994) 2927--2936, [\href{http://arxiv.org/abs/hep-th/9408018}{{\tt hep-th/9408018}}].

\bibitem{DHAR_1993}
A.~DHAR, G.~MANDAL, and S.~R. WADIA, {\it W$\infty$ coherent states and path-integral derivation of bosonization of non-relativistic fermions in one dimension},  {\em Modern Physics Letters A} {\bf 08} (Dec., 1993) 3557–3568.

\bibitem{DHAR_1994}
A.~Dhar, G.~Mandal, and S.~R. Wadia, {\it String field theory of two dimensional qcd: a realization of w$\infty$ algebra},  {\em Physics Letters B} {\bf 329} (June, 1994) 15–26.

\bibitem{Caron-Huot:2014gia}
S.~Caron-Huot and J.~M. Henn, {\it {Solvable Relativistic Hydrogenlike System in Supersymmetric Yang-Mills Theory}},  {\em Phys. Rev. Lett.} {\bf 113} (2014), no.~16 161601, [\href{http://arxiv.org/abs/1408.0296}{{\tt arXiv:1408.0296}}].

\bibitem{Ivanovskiy:2024vel}
V.~Ivanovskiy, S.~Komatsu, V.~Mishnyakov, N.~Terziev, N.~Zaigraev, and K.~Zarembo, {\it {Vacuum Condensates on the Coulomb Branch}},  \href{http://arxiv.org/abs/2405.19043}{{\tt arXiv:2405.19043}}.

\bibitem{Zamolodchikov:1980mb}
A.~B. Zamolodchikov, {\it {'FISHNET' DIAGRAMS AS A COMPLETELY INTEGRABLE SYSTEM}},  {\em Phys. Lett. B} {\bf 97} (1980) 63--66.

\bibitem{Gurdogan:2015csr}
O.~G\"urdo\u{g}an and V.~Kazakov, {\it {New Integrable 4D Quantum Field Theories from Strongly Deformed Planar $\mathcal N = $ 4 Supersymmetric Yang-Mills Theory}},  {\em Phys. Rev. Lett.} {\bf 117} (2016), no.~20 201602, [\href{http://arxiv.org/abs/1512.06704}{{\tt arXiv:1512.06704}}]. [Addendum: Phys.Rev.Lett. 117, 259903 (2016)].

\bibitem{Caetano:2016ydc}
J.~a. Caetano, O.~G\"urdo\u{g}an, and V.~Kazakov, {\it {Chiral limit of $ \mathcal{N} $ = 4 SYM and ABJM and integrable Feynman graphs}},  {\em JHEP} {\bf 03} (2018) 077, [\href{http://arxiv.org/abs/1612.05895}{{\tt arXiv:1612.05895}}].

\bibitem{Loebbert:2020hxk}
F.~Loebbert, J.~Miczajka, D.~M\"uller, and H.~M\"unkler, {\it {Massive Conformal Symmetry and Integrability for Feynman Integrals}},  {\em Phys. Rev. Lett.} {\bf 125} (2020), no.~9 091602, [\href{http://arxiv.org/abs/2005.01735}{{\tt arXiv:2005.01735}}].

\bibitem{Loebbert:2020tje}
F.~Loebbert and J.~Miczajka, {\it {Massive Fishnets}},  {\em JHEP} {\bf 12} (2020) 197, [\href{http://arxiv.org/abs/2008.11739}{{\tt arXiv:2008.11739}}].

\bibitem{Alday:2007mf}
L.~F. Alday and J.~M. Maldacena, {\it {Comments on operators with large spin}},  {\em JHEP} {\bf 11} (2007) 019, [\href{http://arxiv.org/abs/0708.0672}{{\tt arXiv:0708.0672}}].

\end{thebibliography}\endgroup

	\end{document}